\documentclass[12pt]{iopart}

\usepackage{epsfig}

\usepackage{iopams}

\begin{document}

\title[Testing non-standard cosmological models with supernovae]{Testing non-standard cosmological models with supernovae}
\author{Dirk Puetzfeld$^\star$, Xuelei Chen$^\ddag$}
\address{$^\star$ Department of Physics and Astronomy, Iowa State University, Ames, IA 50011, USA}
\address{$^\ddag$ Kavli Institute for Theoretical Physics, University of California, Santa Barbara, CA 93106, USA}
\ead{dpuetz@iastate.edu, dp@thp.uni-koeln.de, xuelei@itp.ucsb.edu}

\begin{abstract}
In this work we study the magnitude-redshift relation of a non-standard cosmological model. The model under consideration was firstly investigated within a special case of metric-affine gravity (MAG) and was recently recovered via different approaches by two other groups. Apart from the usual cosmological parameters for pressure-less matter $\Omega_{\rm m}$, cosmological constant/dark energy $\Omega_{\lambda}$, and radiation $\Omega_{\rm r}$ a new density parameter $\Omega_\psi$ emerges. The field equations of the model reduce to a system which is effectively given by the usual Friedmann equations of general relativity, supplied by a correction to the energy density and pressure in form of $\Omega_\psi$, which is related to the non-Riemannian structure of the underlying spacetime. We search for the best-fit parameters by using recent SN Ia data sets and constrain the possible contribution of a new dark-energy like component at low redshifts, thereby we put an upper limit on the presence of non-Riemannian quantities in the late stages of the universe. In addition the impact of placing the data in redshift bins of variable size is studied. The numerical results of this work also apply to several anisotropic cosmological models which, on the level of the field equations, exhibit a similar scaling behavior of the density parameters like our non-Riemannian model. 
\end{abstract}

\submitto{\CQG}
\maketitle

\section{Introduction\label{Introduction_section}}

Today the observational situation in cosmology is very promising. With several independent cosmological tests at hand the so-called standard model of cosmology \cite{KolbTurner,PadmanabhanAstro3} has emerged and passed most of these tests. While one of the main benefits of the FLRW (Friedmann-Lema\^{\i}tre-Robertson-Walker) model is given by its simplicity, recent observations of type Ia supernovae \cite{Perlmutter}-\cite{Riess2} made clear that we only know little about the dominating energy density component of the universe, which enters the general relativistic description in form of a cosmological constant. Such a component is nowadays usually termed dark energy.

Since the cosmological constant seems to be inevitable for the description of the SN Ia data within the FLRW model it is interesting to figure out whether such a concept is still needed within an alternative cosmological model. From the viewpoint of a non-Riemannian gravity theory it would be very satisfactory if such a dark energy component could be ascribed to a deviation from the usual Riemannian geometry. 

In this work we analyze the SN Ia data within a cosmological model \cite{Puetzfeld1,Puetzfeld2} which is based on a non-Riemannian gravity theory namely metric-affine gravity (MAG) \cite{PhysRep}. As we show in the following the field equations of the model turn out to be very similar to the usual Friedmann equations. We carry out a numerical analysis on the basis of the new field equations. Thereby we put a quantitative constraint on the contribution of the new density parameter $\Omega_{\psi}$, which is linked to the non-Riemannian structures of the underlying gravity theory and which contributes to the total energy density in the universe.  

The plan of the paper is as follows. In section \ref{EXTWEYL_magnitude_redshift_relation_section} we derive the magnitude-redshift relation from the field equations found by Obukhov \etal in \cite{Obukhov}. With this relation at hand we perform fits to the combined SN Ia data sets of Wang \cite{Wang} and Tonry \etal \cite{Tonry} in section \ref{Numerical_results_section}. In addition we study the impact of placing the data sets in redshift bins of variable sizes. We compare our findings with the results from independent age estimates  of the universe and work out the deceleration history of our model. Finally, we draw our conclusion in section \ref{CONCLUSION_section}. In \ref{TRIPLET_KAPITEL} we briefly introduce the field equations and geometrical quantities of metric-affine gravity (MAG) and provide an introduction to the so-called triplet ansatz of MAG. Readers who are not familiar with MAG might also want to consult \cite{PhysRep} for a comprehensive review. In \ref{OPERATIONS_SECTION} we fix our conventions and in \ref{NATURAL_UNITS} we provide an overview over the units used throughout the preceding sections.

\section{Magnitude-redshift relation within a non-standard cosmological model\label{EXTWEYL_magnitude_redshift_relation_section}}

\subsection{Lagrangian}

In \cite{Obukhov} Obukhov \etal considered a cosmological model within the triplet regime of MAG, cf.\ \ref{TRIPLET_KAPITEL}. In contrast to the triplet ansatz in vacuum, they considered a dilational hyperfluid model \cite{Obukhov2} with vanishing spin-current and vanishing shear-current. In addition only a constrained version of the triplet Lagrangian was investigated. Using the notation from (\ref{general_v_mag}) the following constants in the Lagrangian are assumed to vanish 
\begin{eqnarray}
\rho=1, \quad \lambda=0,\quad a_{1},\dots ,a_{3}=0,\quad b_{1,2,3,5}=0, \quad c_{2},\dots ,c_{4}=0, \nonumber \\ 
w_1, \dots w_7=0, \quad z_1,\dots,z_3,z_5,\dots,z_9=0.  \label{EXT_Obukhov_coupling_constant_choice}
\end{eqnarray}
Hence within the original ansatz the usual cosmological constant is set to zero. Subsequently the Lagrangian of this model is given by:
\begin{equation}
\hspace{-1.5cm} V_{{\rm Obukhov}}=\frac{1}{2\kappa }\biggl[ -a_{0}R^{\alpha \beta }\wedge \eta_{\alpha \beta }+b_{4}\, Q_{\alpha \beta }\wedge \,^{\star } \,^{(4)}Q^{\alpha \beta } \biggr] -\frac{1}{2\rho }\, z_{4}\, R^{\alpha \beta }\wedge \,^{\star}\,^{(4)}Z_{\alpha \beta }.\\ 
\label{Obukhov_Lagrangian}
\end{equation}
On the Lagrangian level the model represents a special case of a model which was recently proposed by one of the authors in \cite{Puetzfeld1,Puetzfeld2}, as can easily be seen by setting $a_1,\dots,a_6=0$ in equation (1) of \cite{Puetzfeld1}. Note that also Babourova and Frolov investigated a very similar model in \cite{Babourova}. Obukhov \etal constrained their analysis to the case in which the hypermomentum is purely dilational, i.e.\ proportional to its trace part (cf.\ equation (9.6) of \cite{Obukhov}). 

\subsection{Field equations\label{Field_eq_in_ext_model_section}}

By making a triplet ansatz for torsion and nonmetricity, cf.\ equations (\ref{triplet_allg})--(\ref{nonmet_triplet}), and by using the usual Robertson-Walker line element
\begin{equation}
ds^2=-dt^2+S(t)^2\left(\frac{dr^2}{1-kr^2}+r^2 d \theta^2 +r^2 \sin^2(\theta)d\phi^2\right), \label{RW_line_element}
\end{equation}
the general MAG field equations (\ref{zeroth})--(\ref{second}) reduce to the set 
\begin{eqnarray}
\frac{\dot{S}^{2}}{S^{2}}+\frac{k}{S^{2}} &=&\frac{\kappa }{3}\left[ \mu + \frac{\kappa }{48a_{0}}\left( 1-\frac{3a_{0}}{b_{4}}\right) \frac{\psi ^{2}}{S^{6}}\right] ,  \label{EXT_obukhov_triplet1} \\
2\frac{\ddot{S}}{S}+\frac{\dot{S}^{2}}{S^{2}}+\frac{k}{S^{2}} &=&-\kappa \left[ p+\frac{\kappa }{48a_{0}}\left( 1-\frac{3a_{0}}{b_{4}}\right) \frac{\psi ^{2}}{S^{6}}\right],  \label{EXT_obukhov_triplet2}
\end{eqnarray}
as displayed in equations (9.8) and (9.9) of \cite{Obukhov}. Here $a_{0}$ and $b_{4}$ are the coupling constants from (\ref{Obukhov_Lagrangian}) and $\psi $ denotes an integration constant entering the solution for the Weyl 1-form $Q$ which is given by\footnote{Note that we changed some of the variable names of \cite{Obukhov} in order to match our notation in \cite{Puetzfeld1,Puetzfeld2}.} 
\begin{equation}
Q=-\frac{\kappa \psi }{8b_{4}}S^{-3}dt.  \label{EXT_obukhov_triplet_1_form}
\end{equation}
As one can easily see the field equations are the usual Friedmann equations, without cosmological constant, with an additional contribution to the energy and pressure from the dilation current. The non-Riemmannian quantities in this model, i.e.\ torsion and nonmetricity, {\it die out} as the universe expands. 

On the level of the field equations this model proves to be compatible with the one proposed in \cite{Puetzfeld1} if we make the choice $a_6=-a_4$ for the coupling constants in the Lagrangian in equation (1) of \cite{Puetzfeld1}. Additionally, one can show that also the model of Babourova and Frolov \cite{Babourova} yields the same set of field equations if one performs the following substitutions for the variables in section 7 of \cite{Babourova}:
\begin{eqnarray}
a\rightarrow S,{\quad }\varepsilon =\varepsilon _{\gamma }a^{-3\left(1+\gamma \right) } \rightarrow \mu ,\quad \varepsilon _{\rm v}\rightarrow 0,\quad \ae \rightarrow \kappa ,\quad \gamma \rightarrow w,  \nonumber \\
\phantom{dumdedidle}\frac{\alpha }{4\lambda ^{2}m^{4}} \rightarrow \frac{3a_{0}-b_{4}}{48a_{0}b_{4}},\quad JN\rightarrow \psi .
\label{EXT_transformation_obukhov_babourova}
\end{eqnarray}
In the next section we outline the derivation of the magnitude-redshift relation within this model. In contrast to the original model in \cite{Obukhov} we explicitly allow for a cosmological constant, which corresponds to an additional term $-\lambda/3$ on the lhs of (\ref{EXT_obukhov_triplet1}) and an extra $-\lambda$ on the lhs of (\ref{EXT_obukhov_triplet2}).

\subsection{Magnitude-redshift relation\label{EXT_obukhov_magnitude_redshift_relation}}

By defining a new constant $\upsilon :=\frac{\kappa ^{2}}{144 a_0}\left( 1-\frac{3a_{0}}{b_{4}}\right) $ we can rewrite (\ref{EXT_obukhov_triplet1}) according to 
\begin{equation}
1+\frac{k}{S^{2}H^{2}}-\frac{\lambda}{3}=\frac{\kappa }{3 H^2}\mu +\upsilon \frac{\psi ^{2}}{S^{6} H^2}\quad \Rightarrow \quad \Omega _{k}+\Omega_{\lambda}+\Omega _{w}+\Omega _{\psi }=1.
\label{EXT_obuk_relation_between_density_parameters}
\end{equation}
Here we introduced the following density parameters $\Omega _{k}:=-\frac{k}{H^{2}S^{2}}$, $\Omega_\lambda=\frac{\lambda}{3 H^2}$, $\Omega _{w}:=\frac{\kappa }{3 H^2}\mu $, $\Omega _{\psi}:=\upsilon \frac{\psi ^{2}}{S^{6} H^2}$ in the last step. We use the index $w$ since we did not fix the underlying equation of state, i.e.\ $p=w\mu$. It is interesting to note that the new density parameter $\Omega_{\psi}$ redshifts with $z^6$, a behavior which is also known from anisotropic models, see \cite{KamionTurn,Khalatnikov}, e.g. Denoting present day values of quantities by an index ``0'' we can rewrite the Hubble rate in terms of the density parameters and the redshift:
\begin{eqnarray}
H^{2}&=&\frac{\kappa }{3}\mu -\frac{k}{S^{2}}+\frac{\lambda}{3}+\upsilon \frac{\psi ^{2}}{S^{6}}  \nonumber \\
&=&H_{0}^{2}\left[ \Omega _{w0}\left( 1+z\right) ^{3\left( 1+w\right)}+\Omega _{k0}\left( 1+z\right) ^{2}+\Omega_{\lambda 0}+\Omega _{\psi 0}\left( 1+z\right) ^{6}\right]   \nonumber \\
&\stackrel{(\ref{EXT_obuk_relation_between_density_parameters})}{=}&H_{0}^{2}\left( 1+z\right) ^{2}\left\{ \phantom{\frac{}{}}\Omega _{w0}\left[ \left(1+z\right) ^{1+3w}-1\right] +\Omega _{\lambda 0}\left[ \left( 1+z\right) ^{-2}-1\right] \right. \nonumber \\
&\phantom{=}&\left.+\Omega _{\psi 0}\left[ \left( 1+z\right) ^{4}-1\right] +1  \phantom{\frac{}{}} \right\}. \label{EXT_obuk_Hubble_rate_with_density}
\end{eqnarray}
Hence the luminosity distance within this model becomes
\begin{equation}
d_{\texttt{\tiny \rm \rm luminosity}}=S_{0}\,\,\left( 1+z\right) \,\Theta \left[ \left(H_{0}S_{0}\right) ^{-1}\int_{0}^{z}F\left[ \tilde{z}\right] d\tilde{z}\right]
.  \label{EXT_obuk_general_luminosity}
\end{equation}
With $F[\tilde{z}]:=H_0/H$ and the function in front of the integral is given by 
\begin{equation}
\Theta \lbrack x]:=\left\{\begin{tabular}{lll}
$\sin \left( x\right) $ &  & $k=+1$ \\ 
$x$ & for & $k=0$ \\ 
$\sin $h$\left( x\right) $ &  & $k=-1$
\end{tabular}
\right. .\label{theta_definition}
\end{equation}
If we make use of the definition of the $\Omega _{k}$ density parameter we end up with
\begin{eqnarray}
\hspace{-2cm}d_{\texttt{\tiny \rm \rm luminosity}}\left( z,H_{0},\Omega _{w0},\Omega _{\lambda 0},\Omega _{\psi 0},w\right) \nonumber \\
\hspace{-1.5cm}=\,\frac{\left( 1+z\right) }{H_{0}\sqrt{\left| 1-\Omega _{w0}-\Omega _{\lambda 0}-\Omega _{\psi
0}\right| }}\,\Theta \left[ \sqrt{\left| 1-\Omega _{w0}-\Omega _{\lambda 0}-\Omega _{\psi0}\right| }\int_{0}^{z}F\left[ \tilde{z}\right] d\tilde{z}\right] .
\label{EXT_obuk_luminosity_with_omega_and_z}
\end{eqnarray}
The magnitude-redshift relation reads
\begin{equation}
m\left( z,H_{0},\Omega _{w0},\Omega _{\lambda 0},\Omega _{\psi 0},w,M\right) =M+5\log \left(\frac{d_{\texttt{\tiny \rm \rm luminosity}}}{{\rm length}}\right)+25.
\label{EXT_obuk_magnitude_redshift}
\end{equation}

\section{Numerical results\label{Numerical_results_section}}

In this section we present the numerical results obtained by fitting the magnitude-redshift relation in (\ref{EXT_obuk_magnitude_redshift}) to two recent data sets. We start with a collection of the different available data sets of type Ia supernovae. Combinations and subsets of these data sets were also used by other teams to determine the cosmological parameters.

\subsection{Data sets\label{DATA_sets_section}}

\begin{figure}[ht]

\setlength{\unitlength}{1mm}
\vspace{-8cm}
\begin{center}
\begin{tabular}{l}
\begin{picture}(120,80)
\epsfig{file=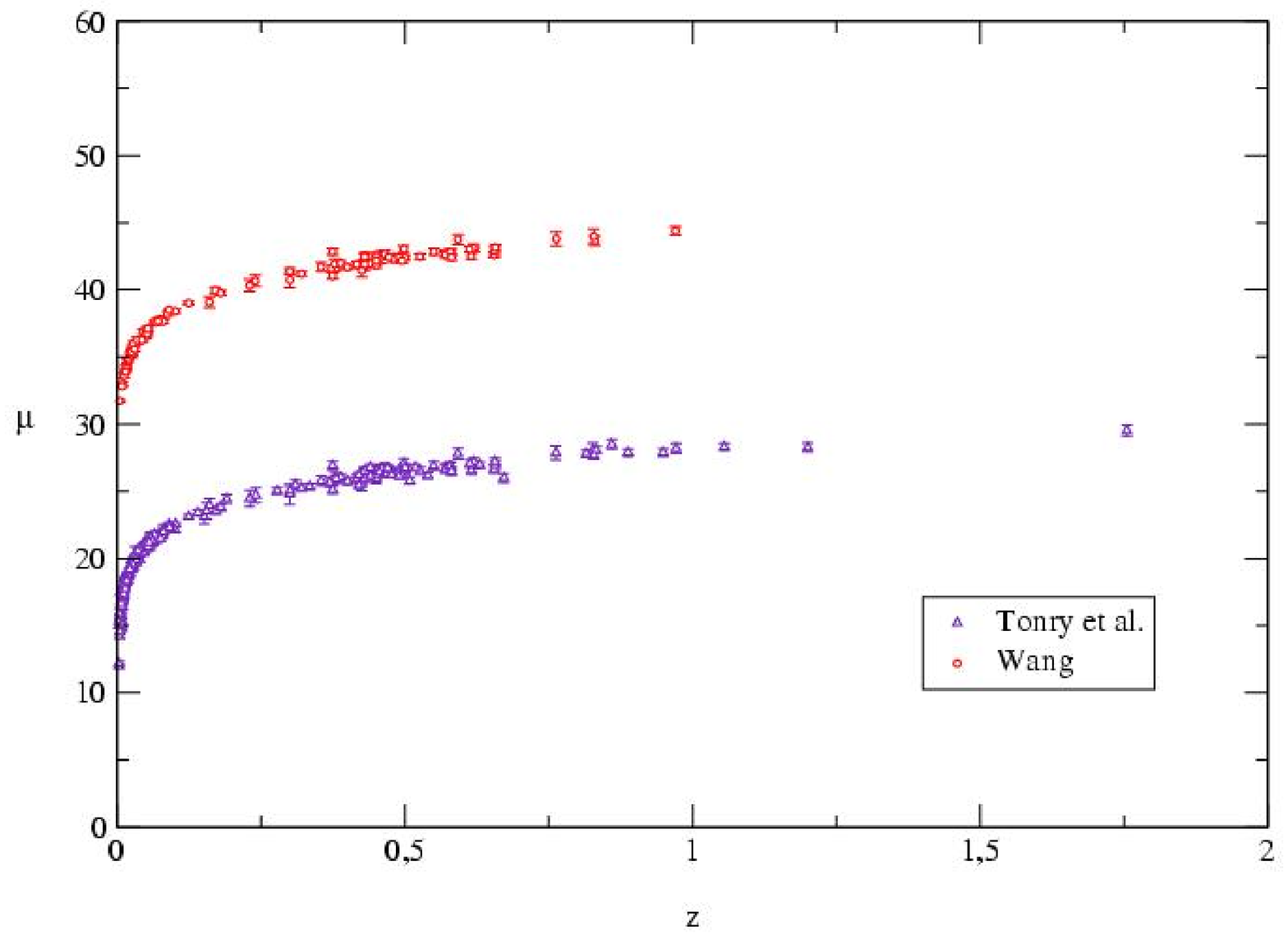,angle=-90,width=12cm}
\end{picture}\\
\begin{picture}(120,80)
\epsfig{file=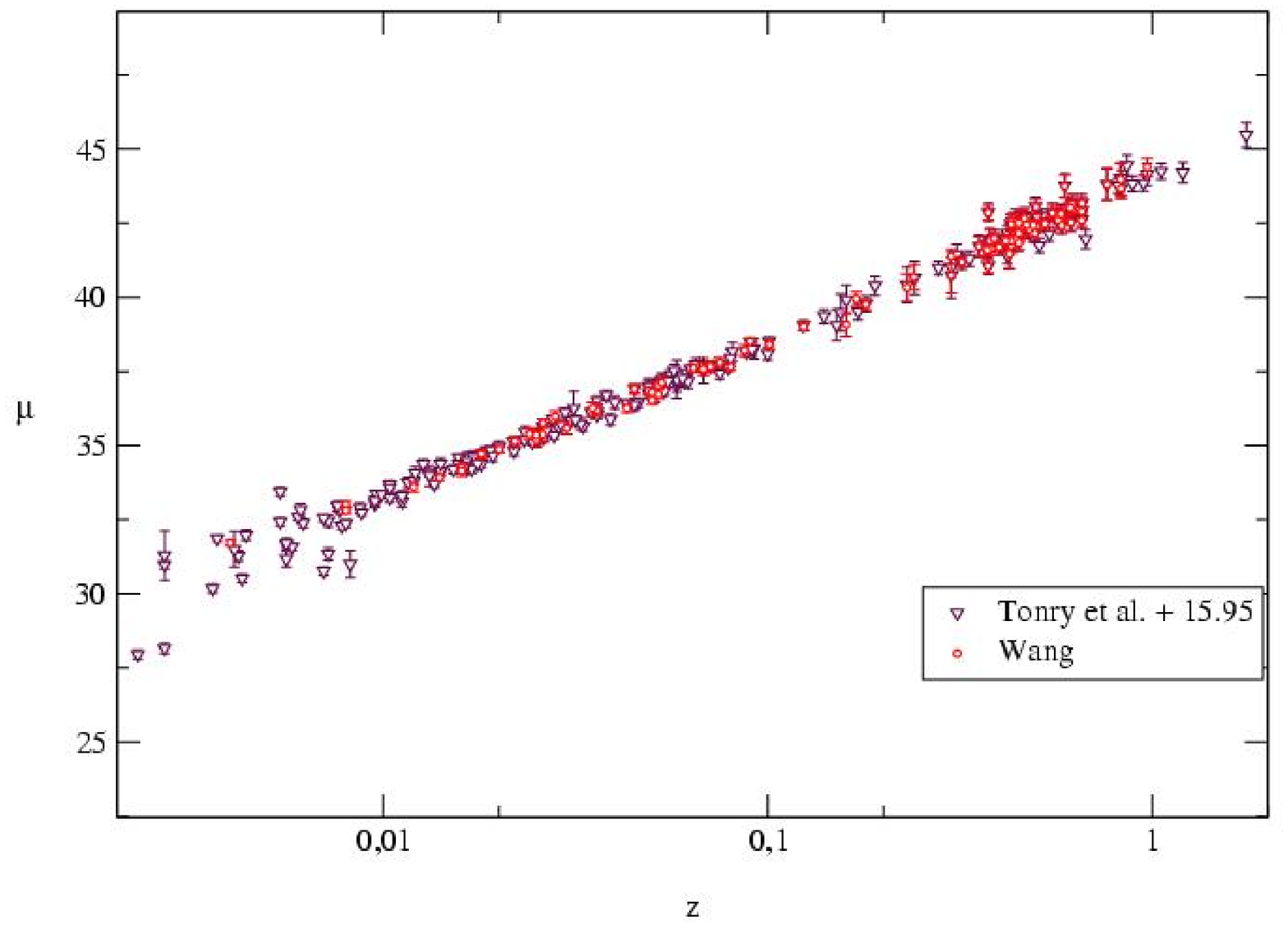,angle=-90,width=12cm}
\end{picture}
\end{tabular}
\end{center}
\vspace{8cm}
\caption[Data sets]{Unbinned data sets of Wang and Tonry et al.\ containing 92 and 230 SNe of type Ia. All of the data points of Wang are included in the data set of Tonry et al. In the lower plot we adjusted the offset between the two data sets by averaging over the common data points in both sets.} 
\label{FIG_data_sets}
\end{figure}

In table \ref{tabelle_2} we collected the number of supernovae and the references which actually contain the data.
\begin{table}
\caption{SN Ia data sets.}
\label{tabelle_2}
\begin{indented}
\item[]\begin{tabular}{@{}lllllll}
\br
Symbol & $\sharp$ SN & Reference & Comments &  &  &  \\ \mr
I & $18$ & p. 571, \cite{Hamuy, Perlmutter} & Cal\'{a}n/Tololo survey &  &  &  \\ 
II & $42$ & p. 570, \cite{Perlmutter} & Supernova Cosmology Project &  &  & 
\\ 
III & $10$ & p. 1021, \cite{Riess2} & High-z Supernova Search Team &  &  & 
\\ 
IV & $10$ & p. 1020, \cite{Riess2} & Same as III but MLCS method &  &  &  \\ 
V & $1$ & \cite{Riess1} & Farthest SN Ia observed to date &  &  &  \\ 
VI & $27$ & p. 1035, \cite{Riess2} & Low-redshift MLCS/template &  &  & \\
VII & $230$ & pp. 33-40,  \cite{Tonry} & Most recent compilation of SN Ia & & & \\\br
\end{tabular}
\end{indented}
\end{table}
The data sets of the different groups are not directly comparable. Perlmutter \etal \cite{Perlmutter} provide the effective magnitude $m_{\texttt{\tiny \rm \rm B}}^{\texttt{\tiny \rm \rm eff}} $ in the B band, while Riess \etal \cite{Riess2} use the so-called distance modulus $\mu $\footnote{Not to be confused with the energy-density within the field equations.}. As shown by Wang in \cite{Wang} it is possible to find a relation between this two data sets by comparing the data of 18 SNe Ia published by both groups. The definition of the magnitude as given in equation (\ref{EXT_obuk_magnitude_redshift}) is compatible with the definition used by Perlmutter \etal, it is related to the definition of Riess \etal by 
\begin{equation}
m=M+\mu =M+5\log d_{\texttt{\tiny \rm \rm luminosity}}+25=\mathcal{M}+5 \log H_{0}d_{\texttt{\tiny \rm \rm luminosity}}.  \label{definition_of_the_distant_modulus}
\end{equation}
As shown in \cite{Wang} we have to choose $M=-19.33\pm 0.25$ in order to transform the different data sets into each other. This value corresponds to the MLCS method applied by Riess \etal. In the following we make use of two data sets\footnote{From here on we use the shortcuts W92 and T230 for the data sets of Wang and Tonry \etal.}, namely the the one of Wang, which contains 92 data points and can be viewed as a compilation of the sets I, II, IV, and VI from table \ref{tabelle_2} in which some outliers were removed and the set VII of Tonry \etal which contains 230 data points and represents the most recent collection of SN Ia \footnote{The additional SN Ia set from the Deep Survey \cite{Barris} was published while this work was nearly finished. Hence it will be considered in future work.}. In order to obtain comparable results we add $15.95$ to the distance modulus in set VII. This constant offset was obtained by averaging over the common points in both sets. The two data sets are displayed in figure \ref{FIG_data_sets}. 

\subsection{Fitting method}\label{NUMERICAL_fitting_method_section}

Since we want results which are comparable to the analysis of the combined data set by Wang in \cite{Wang}, we are going to minimize  
\begin{equation}
\chi^2:=\sum\limits_{i=1}^{92/230}\frac{\left[\mu_i^{\rm theory}\left(z_i | {\tiny \rm parameters} \right)-\mu_i^{\rm measured}\right]^2}{\sigma^2_{\mu \,\, i}+\sigma^2_{mz \,\, i}}, \label{chi_square_fit_function}
\end{equation}
in order to obtain the best-fit parameters \cite{Recipes,Bevington,Martin} for the magnitude-redshift relation displayed in (\ref{EXT_obuk_magnitude_redshift}). Here $\mu_i^{\rm theory}$ denotes the distance modulus at a certain redshift $z_{i}$ as defined in (\ref{definition_of_the_distant_modulus}). The error of the measured $\mu_i^{\rm measured}$ is given by $\sigma^2_{\mu \,\, i}$. The dispersion in the distance modulus $\sigma_{mz}$ due to the dispersion in the galaxy redshift, $\sigma_z$, can be calculated iteratively via
\begin{equation}
\sigma_{mz}:=\frac{5}{ln10}\left[\frac{1}{d_{\rm luminosity}}\frac{\partial d_{\rm luminosity}}{\partial z}\right] \sigma_z \label{Sigma_z_as_in_Wang}
\end{equation}
according to Wang, cf.\ equation (13) of \cite{Wang}.

\subsection{Best-fit parameters\label{NUMERICAL_Best_fit_parameters_section}}

\begin{figure}
\setlength{\unitlength}{1mm}
\vspace{-5cm}
\begin{tabular}{lll}
\vspace{-1cm}\phantom{4242}\begin{picture}(60,55)
\epsfig{file=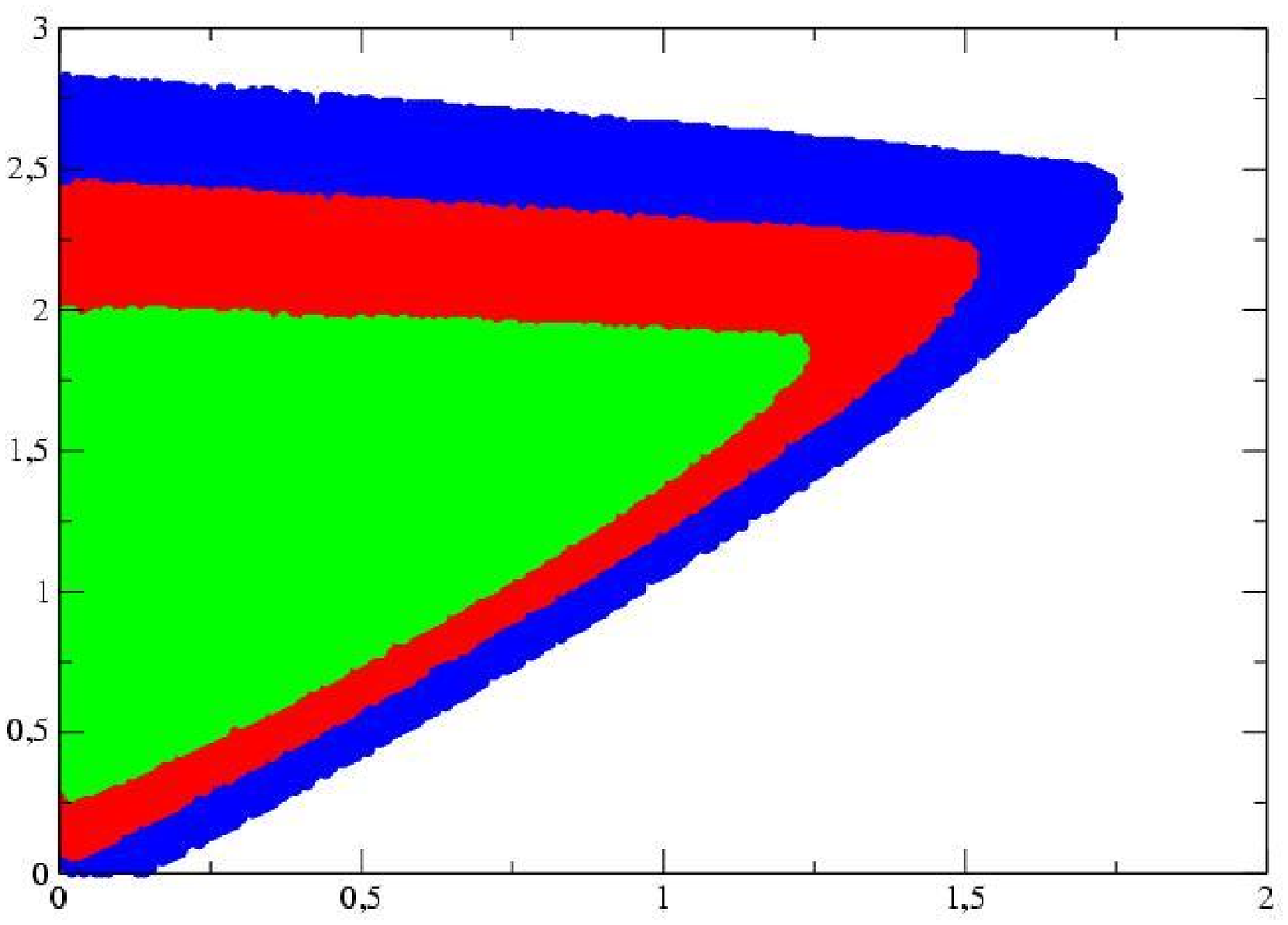,width=5cm,angle=-90}
\put(-35,-46){\tiny{$\Omega_{{\rm m} 0}$}}
\put(-65,-26){\tiny{$\Omega_{{\lambda} 0}$}}
\end{picture}&\phantom{42}&
\begin{picture}(60,55)
\epsfig{file=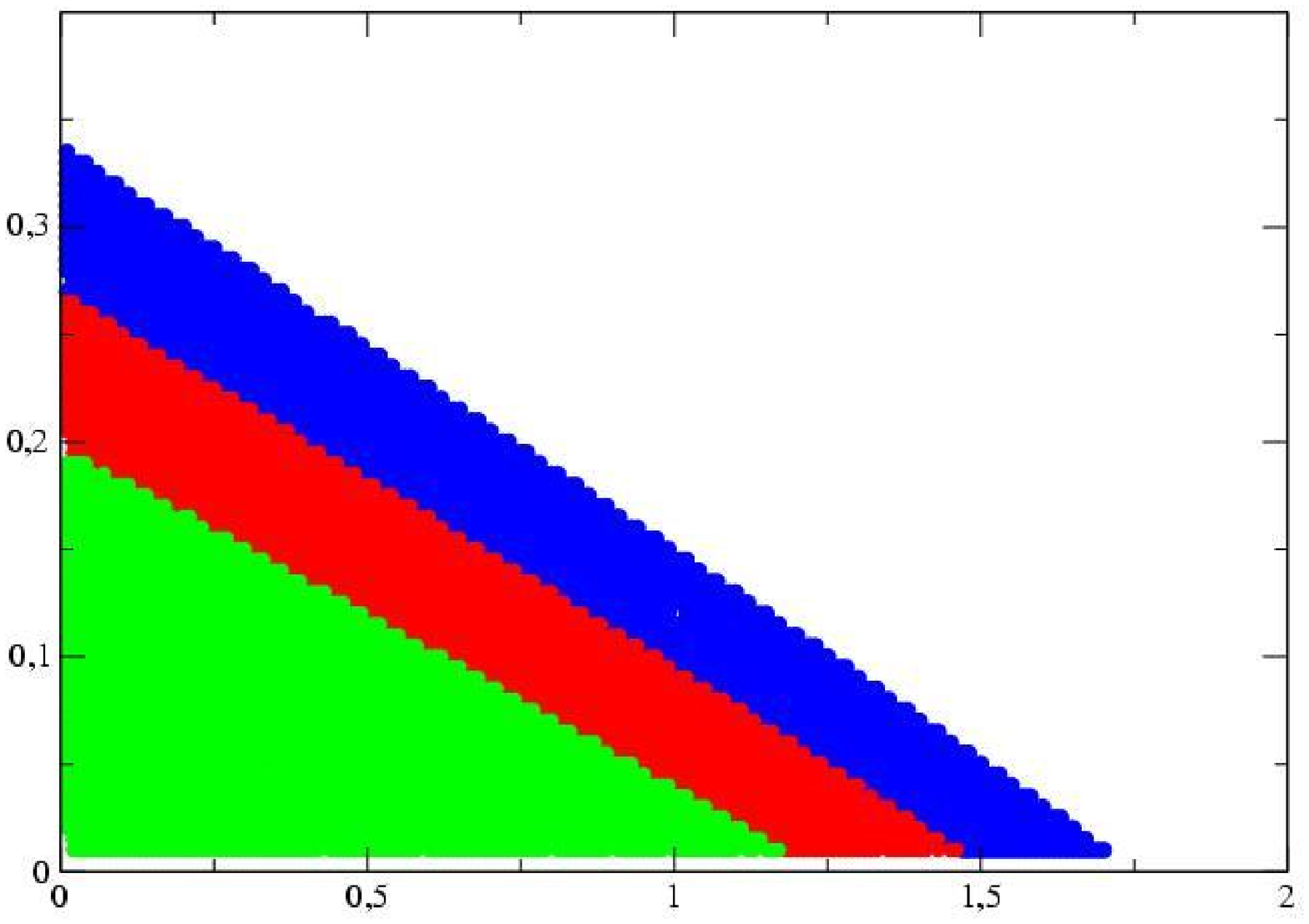,width=5cm,angle=-90}
\put(-35,-46){\tiny{$\Omega_{{\rm m} 0}$}}
\put(-65,-26){\tiny{$\Omega_{{\psi} 0}$}}
\end{picture}\\
\vspace{-1cm}\phantom{4242}\begin{picture}(60,55)
\epsfig{file=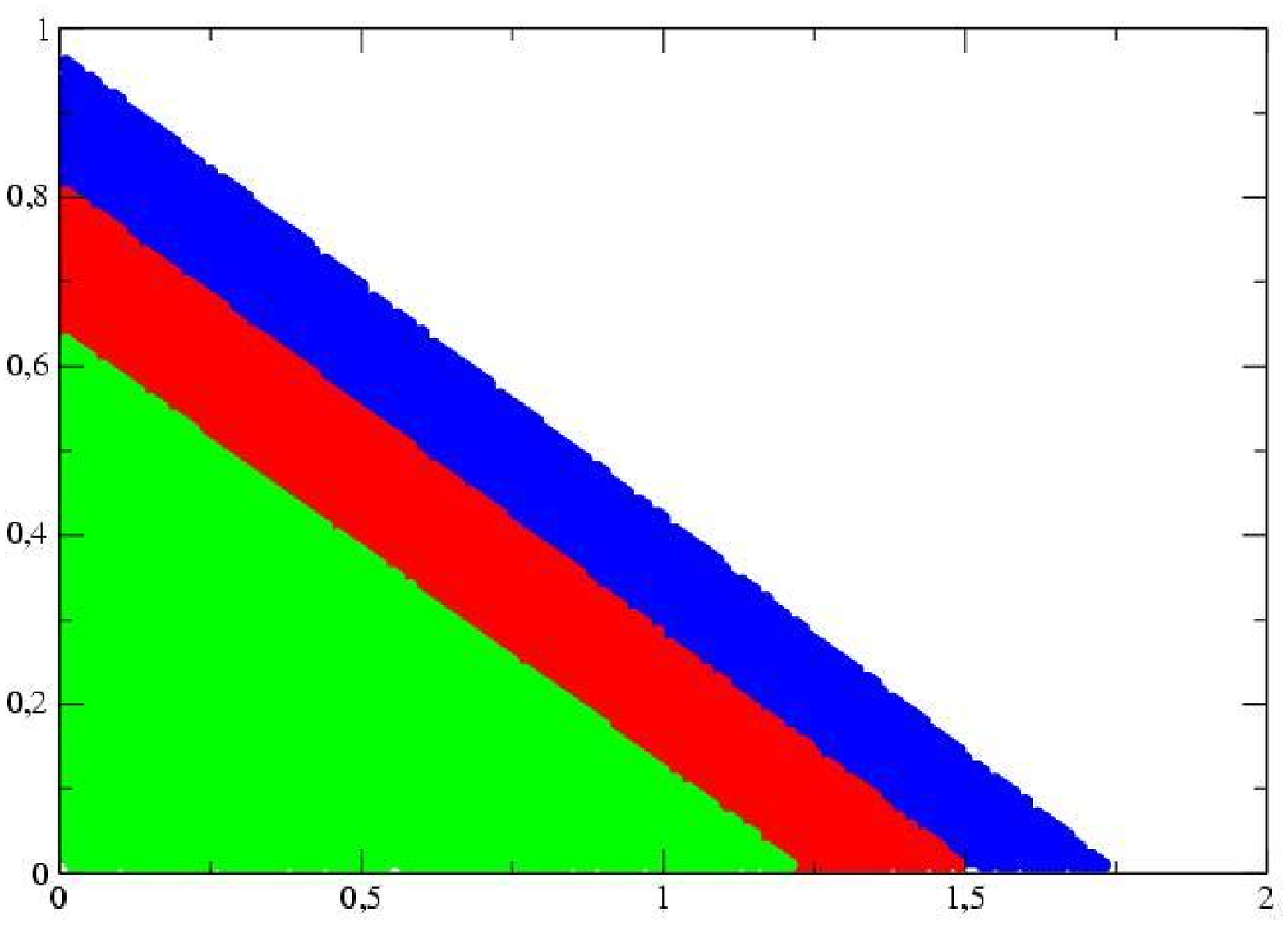,width=5cm,angle=-90}
\put(-35,-46){\tiny{$\Omega_{{\rm m} 0}$}}
\put(-65,-26){\tiny{$\Omega_{{\rm r} 0}$}}
\end{picture}&&
\begin{picture}(60,55)
\epsfig{file=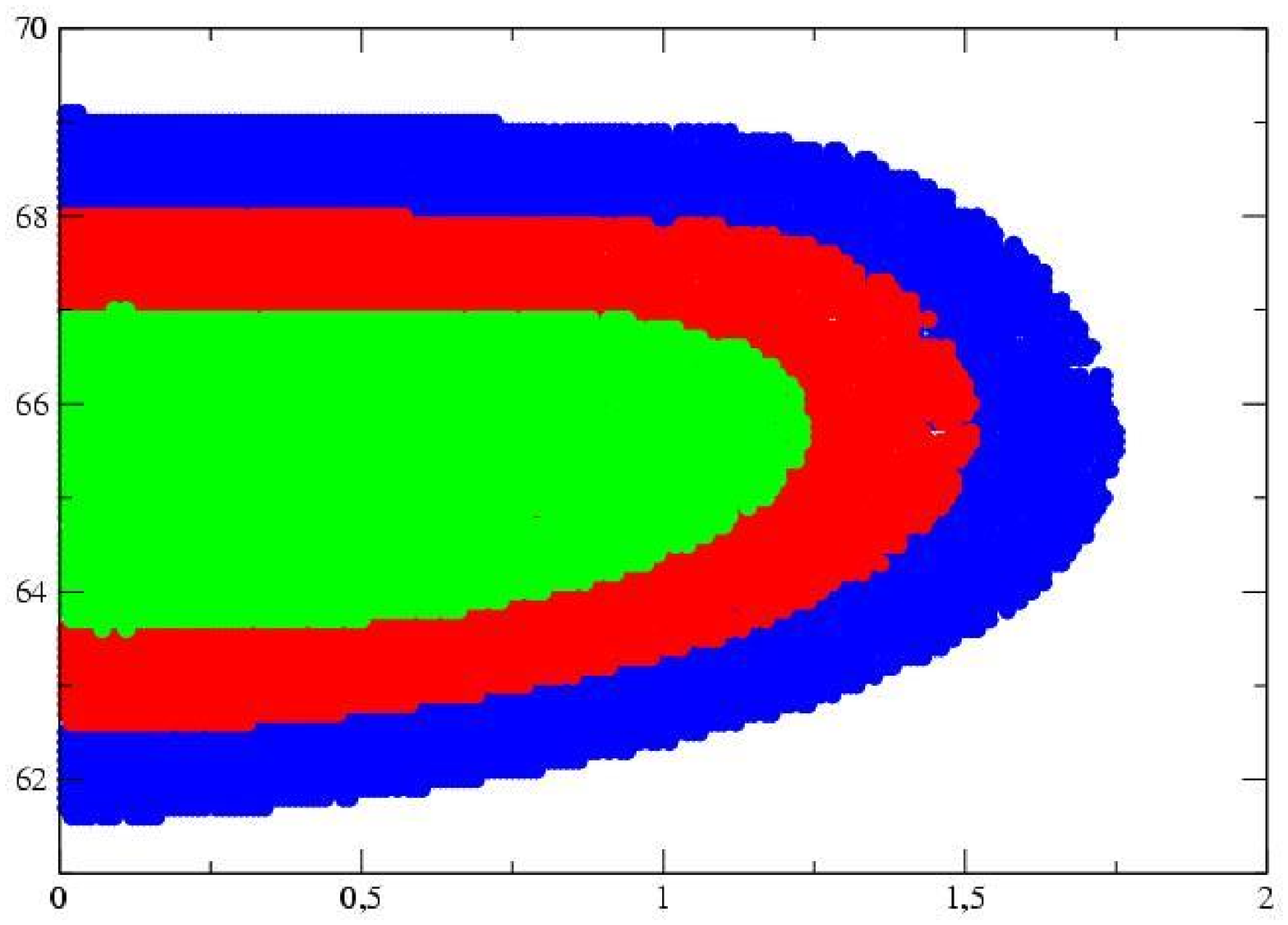,width=5cm,angle=-90}
\put(-35,-46){\tiny{$\Omega_{{\rm m} 0}$}}
\put(-65,-26){\tiny{$H_0$}}
\end{picture}\\
\vspace{-1cm}\phantom{4242}\begin{picture}(60,55)
\epsfig{file=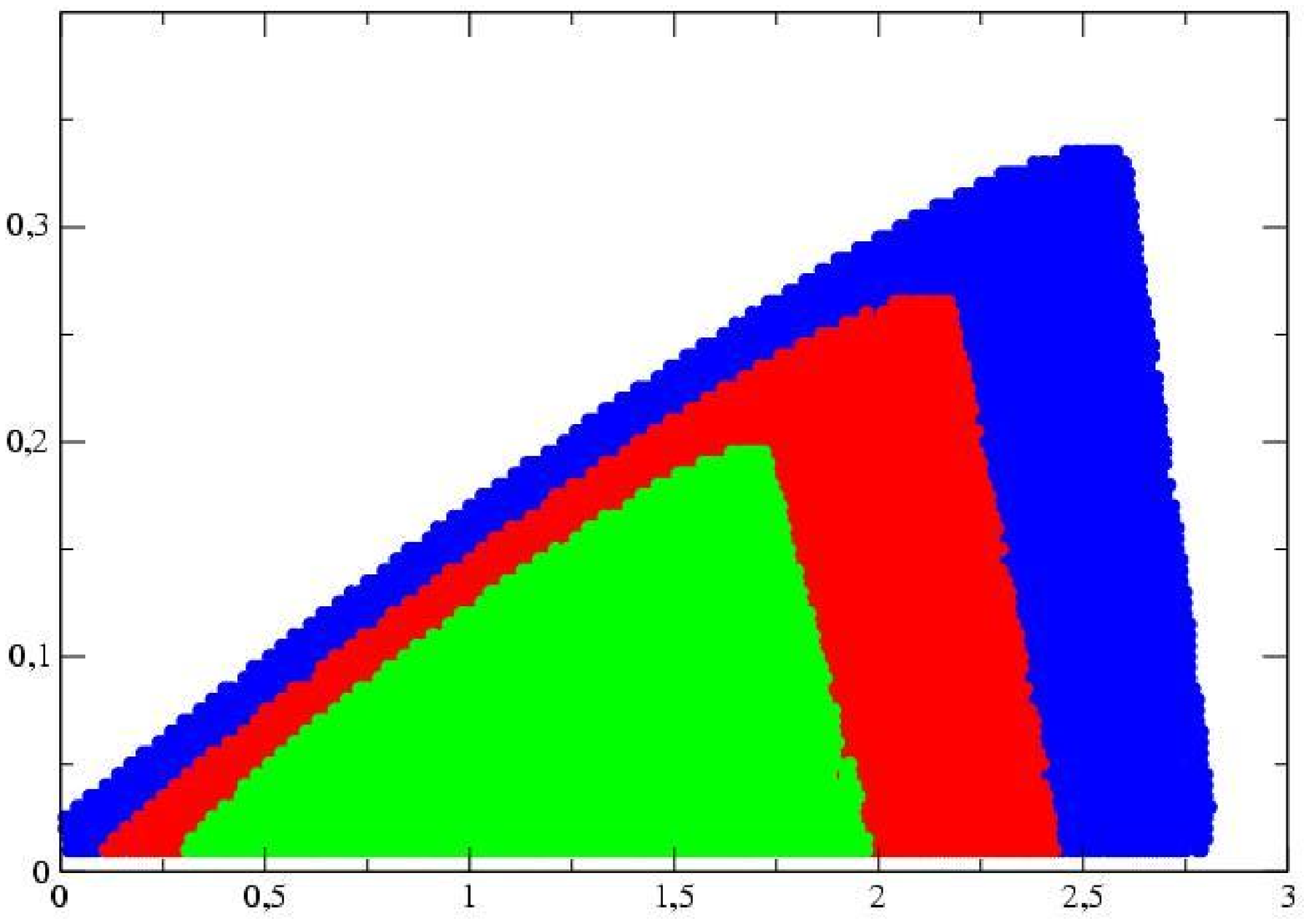,width=5cm,angle=-90}
\put(-35,-46){\tiny{$\Omega_{{\lambda} 0}$}}
\put(-65,-26){\tiny{$\Omega_{{\psi} 0}$}}
\end{picture}&&
\begin{picture}(60,55)
\epsfig{file=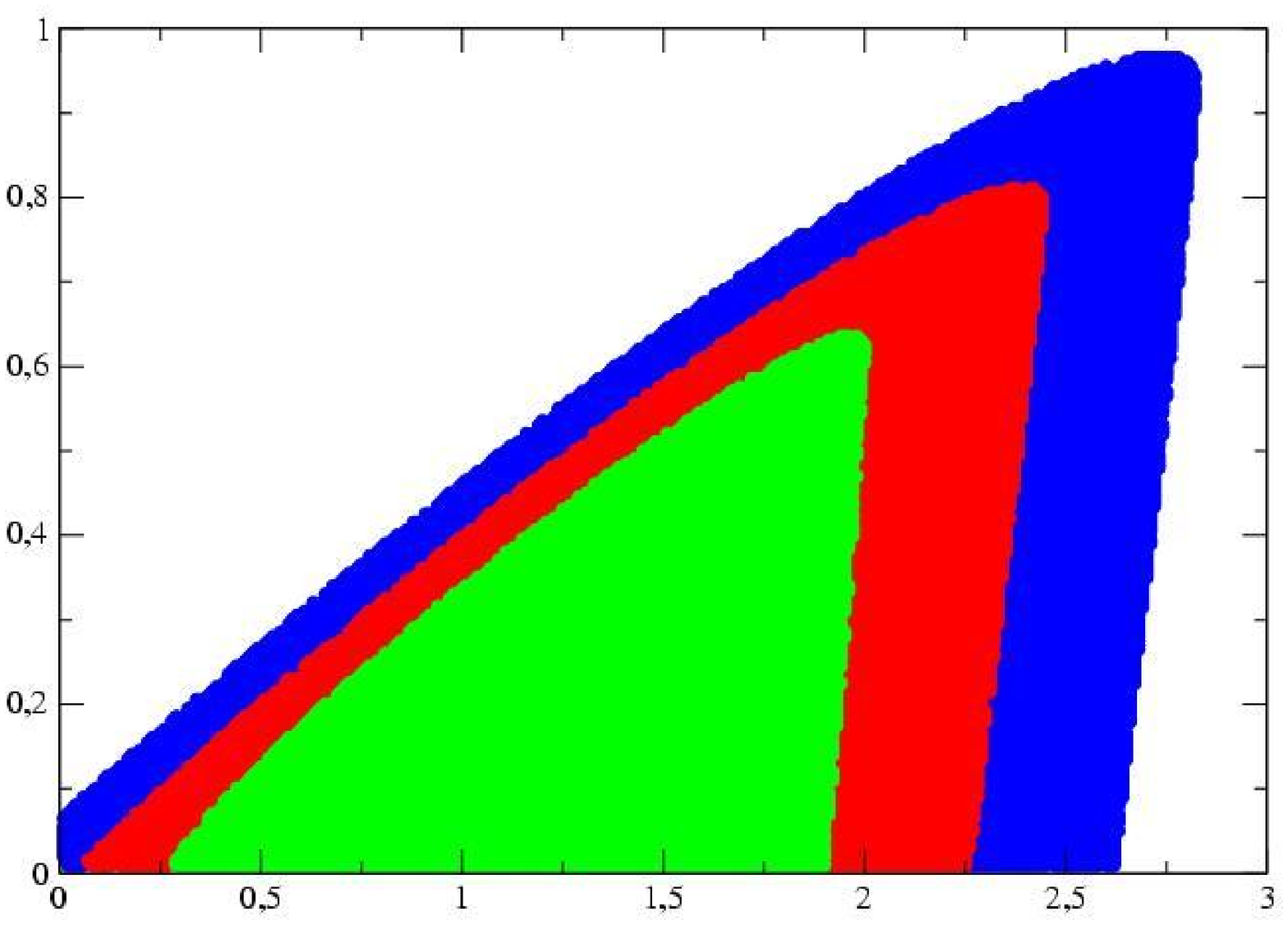,width=5cm,angle=-90}
\put(-35,-46){\tiny{$\Omega_{{\lambda} 0}$}}
\put(-65,-26){\tiny{$\Omega_{{\rm r} 0}$}}
\end{picture}\\
\vspace{-1cm}\phantom{4242}\begin{picture}(60,55)
\epsfig{file=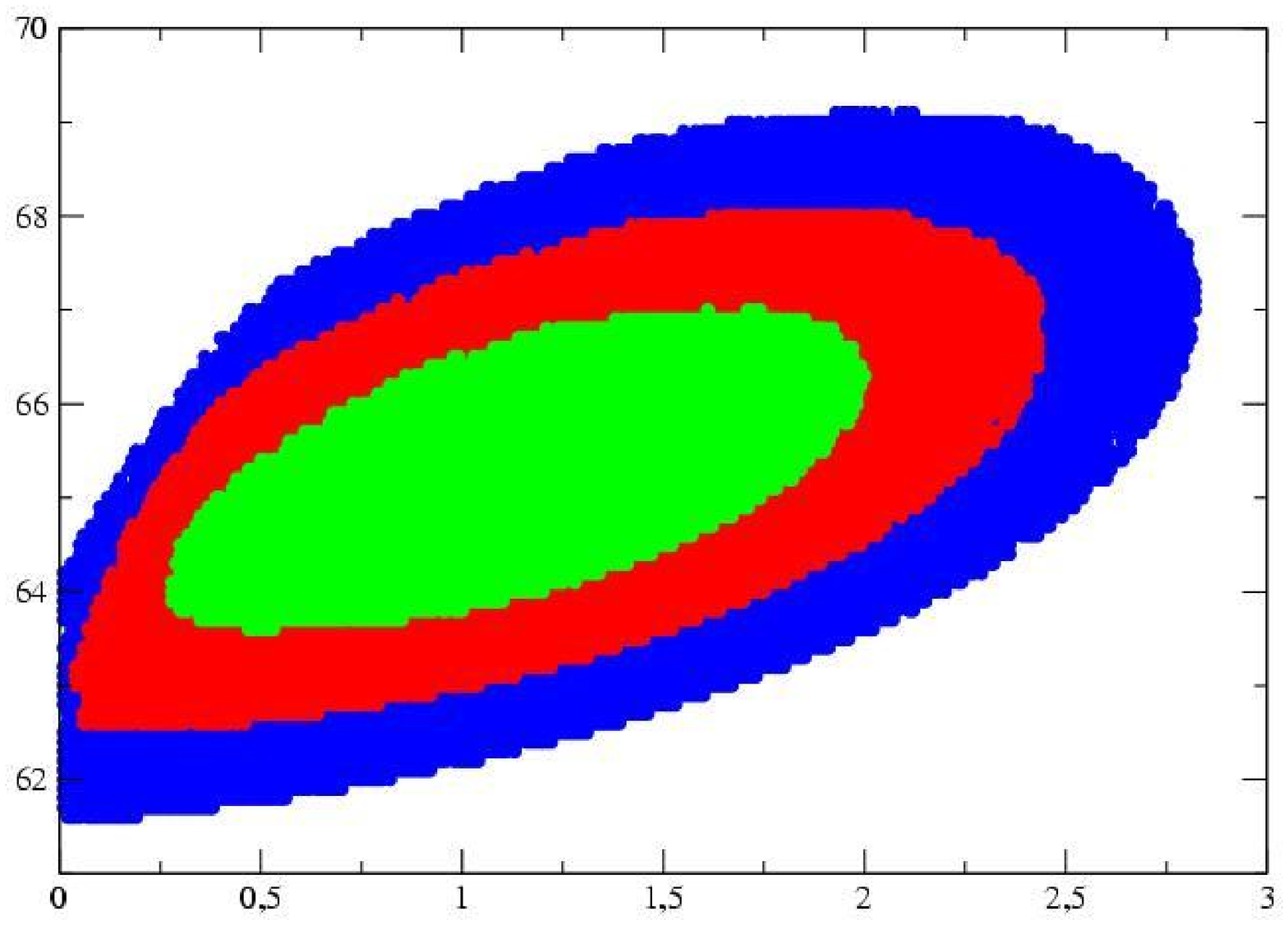,width=5cm,angle=-90}
\put(-35,-46){\tiny{$\Omega_{{\lambda} 0}$}}
\put(-65,-26){\tiny{$H_0$}}
\end{picture}&&
\begin{picture}(60,55)
\epsfig{file=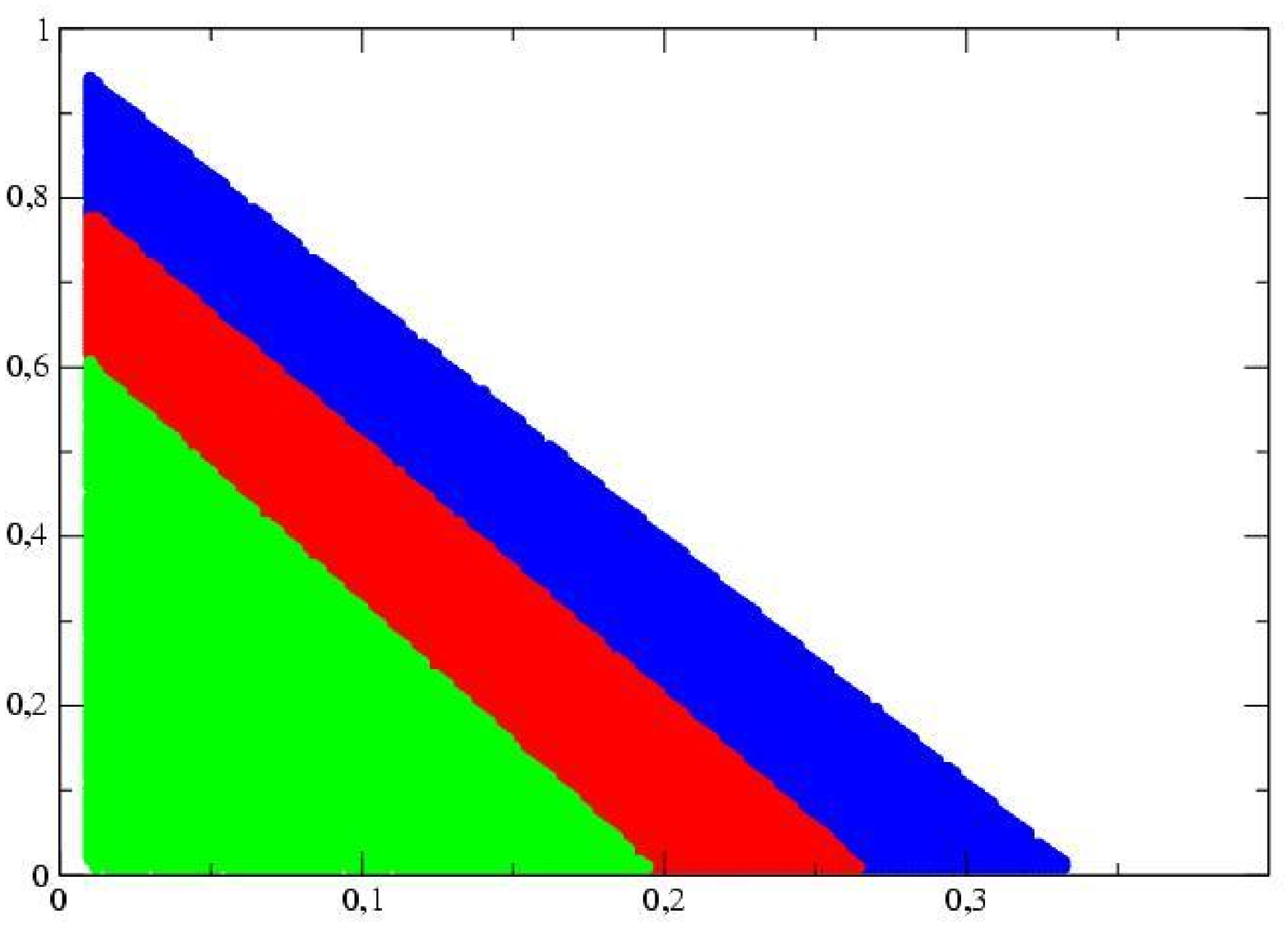,width=5cm,angle=-90}
\put(-35,-46){\tiny{$\Omega_{{\psi} 0}$}}
\put(-65,-26){\tiny{$\Omega_{{\rm r} 0}$}}
\end{picture}\\
\vspace{-1cm}\phantom{4242}\begin{picture}(60,55)
\epsfig{file=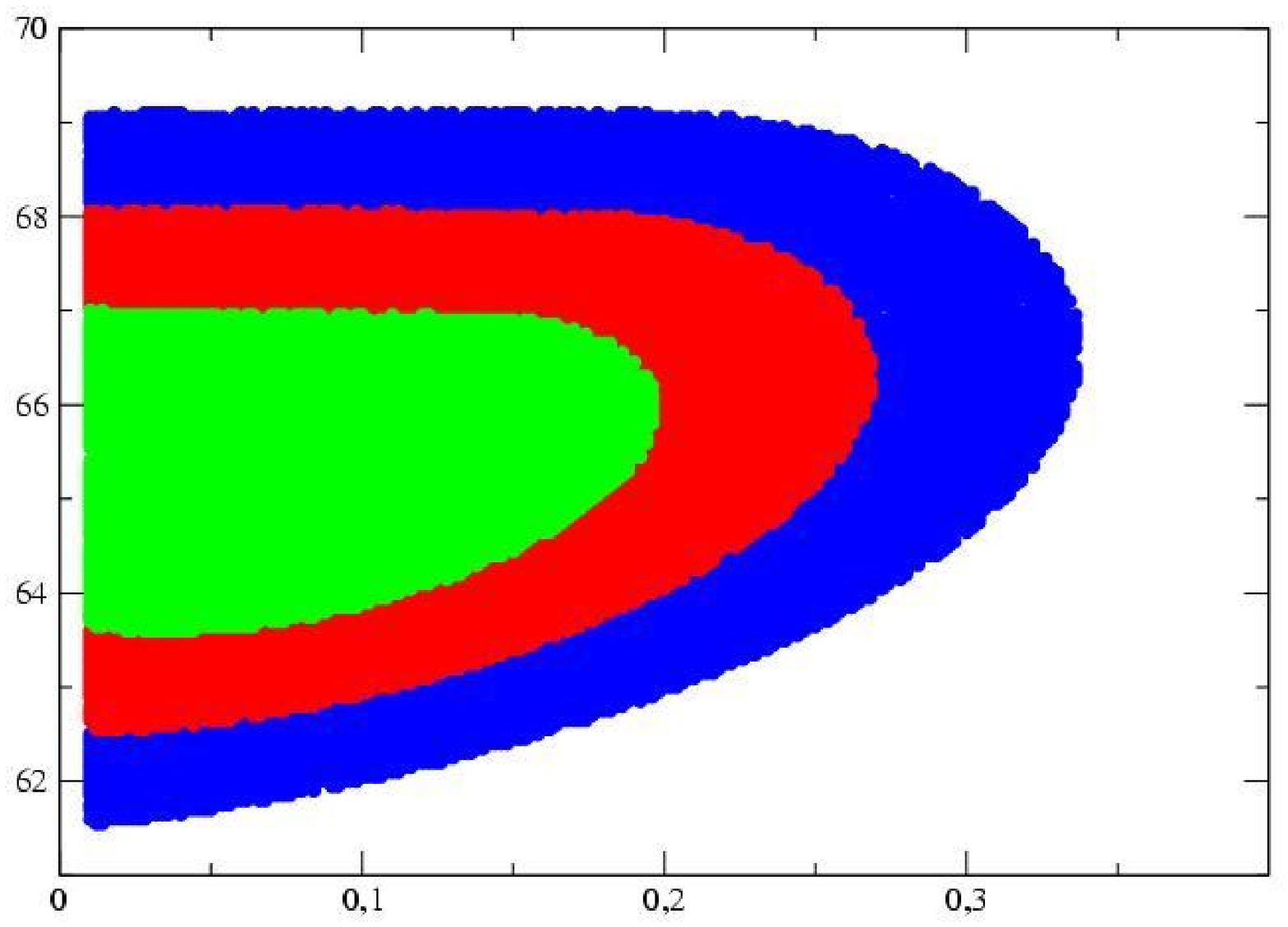,width=5cm,angle=-90}
\put(-35,-46){\tiny{$\Omega_{{\psi} 0}$}}
\put(-65,-26){\tiny{$H_0$}}
\end{picture}&&
\begin{picture}(60,55)
\epsfig{file=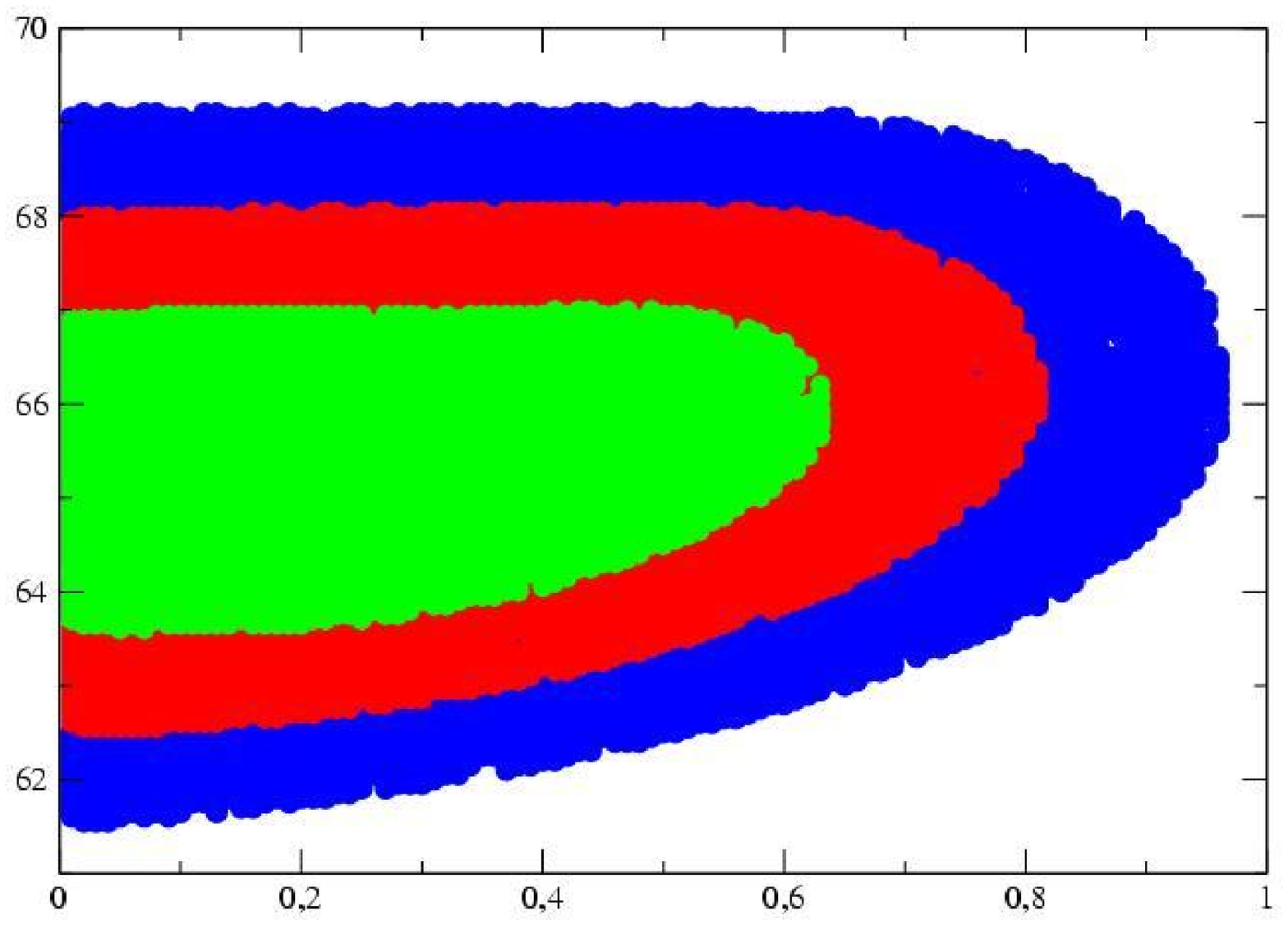,width=5cm,angle=-90}
\put(-35,-46){\tiny{$\Omega_{{\rm r} 0}$}}
\put(-65,-26){\tiny{$H_0$}}
\end{picture}\\\vspace{0.5cm}
\end{tabular}
\vspace{5cm}
\caption[Confidence contours for different parameter planes]{Confidence contours for all parameter planes of the Weyl-Cartan model for the unbinned data set containing 92 SN Ia.}
\label{FIG_triplet_contours_lambda}
\end{figure}

\begin{figure}
\setlength{\unitlength}{1mm}
\vspace{-5cm}
\begin{tabular}{lll}
\vspace{-1cm}\phantom{4242}\begin{picture}(60,55)
\epsfig{file=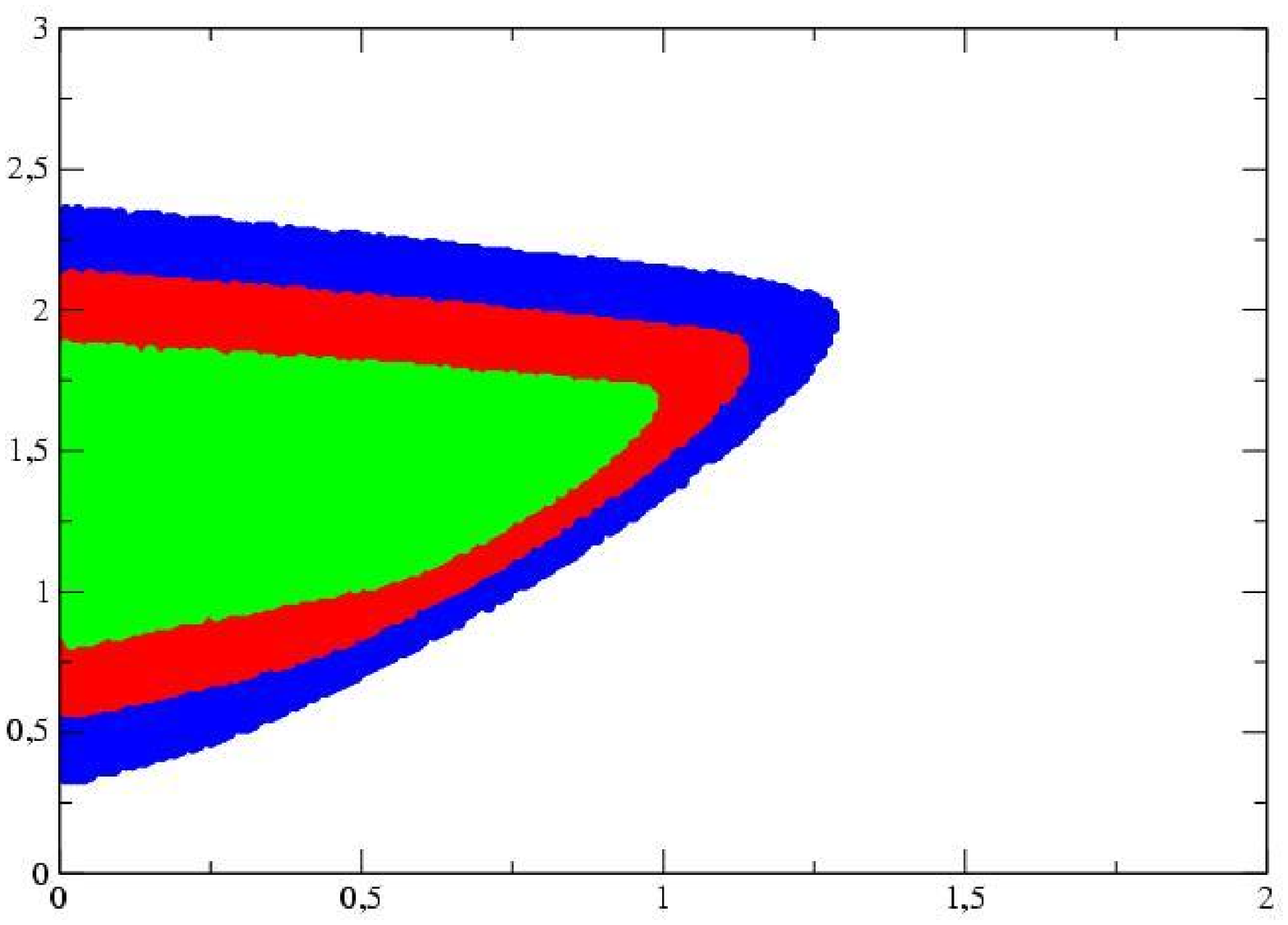,width=5cm,angle=-90}
\put(-35,-46){\tiny{$\Omega_{{\rm m} 0}$}}
\put(-65,-26){\tiny{$\Omega_{{\lambda} 0}$}}
\end{picture}&\phantom{42}&
\begin{picture}(60,55)
\epsfig{file=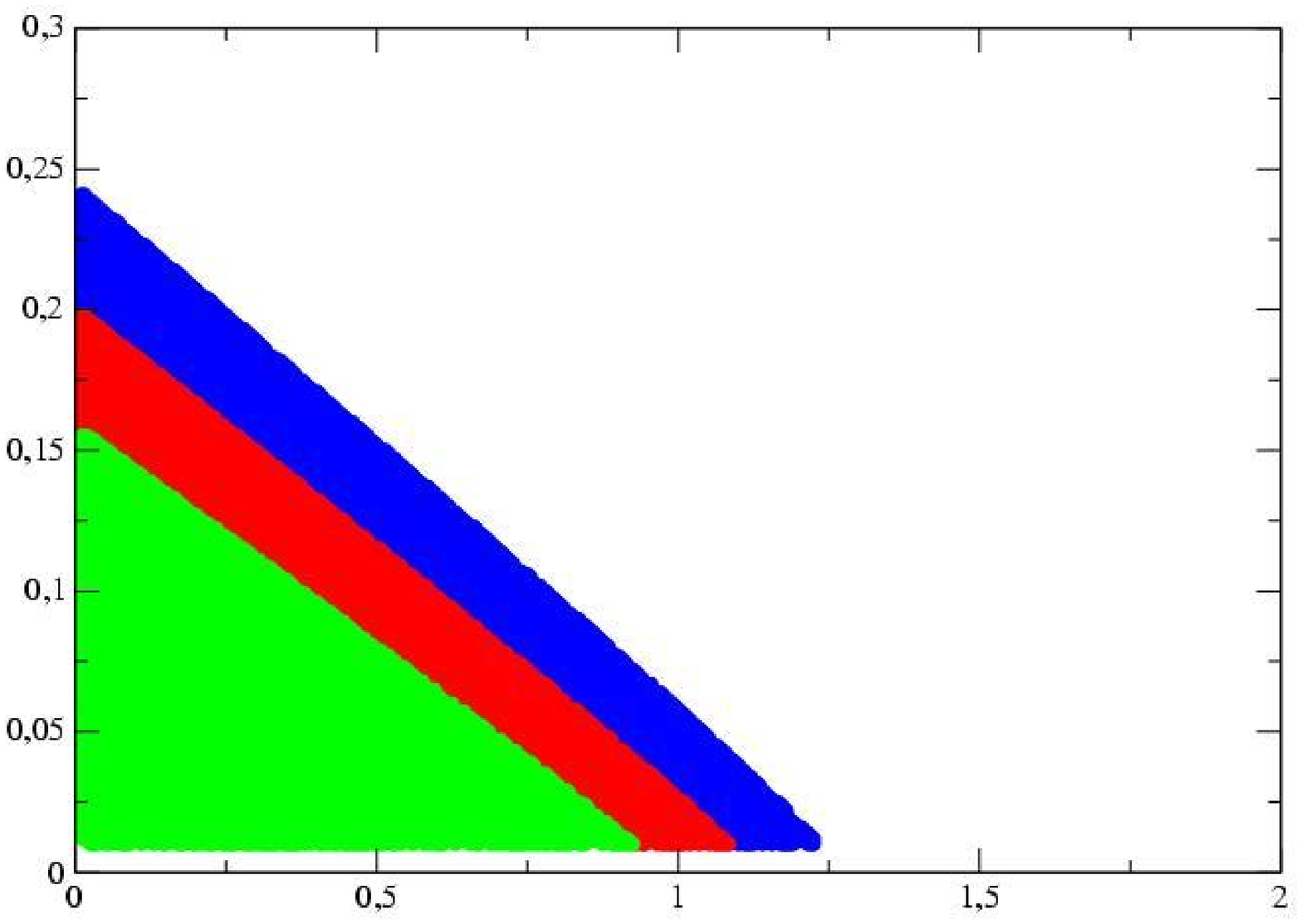,width=5cm,angle=-90}
\put(-35,-46){\tiny{$\Omega_{{\rm m} 0}$}}
\put(-65,-26){\tiny{$\Omega_{{\psi} 0}$}}
\end{picture}\\
\vspace{-1cm}\phantom{4242}\begin{picture}(60,55)
\epsfig{file=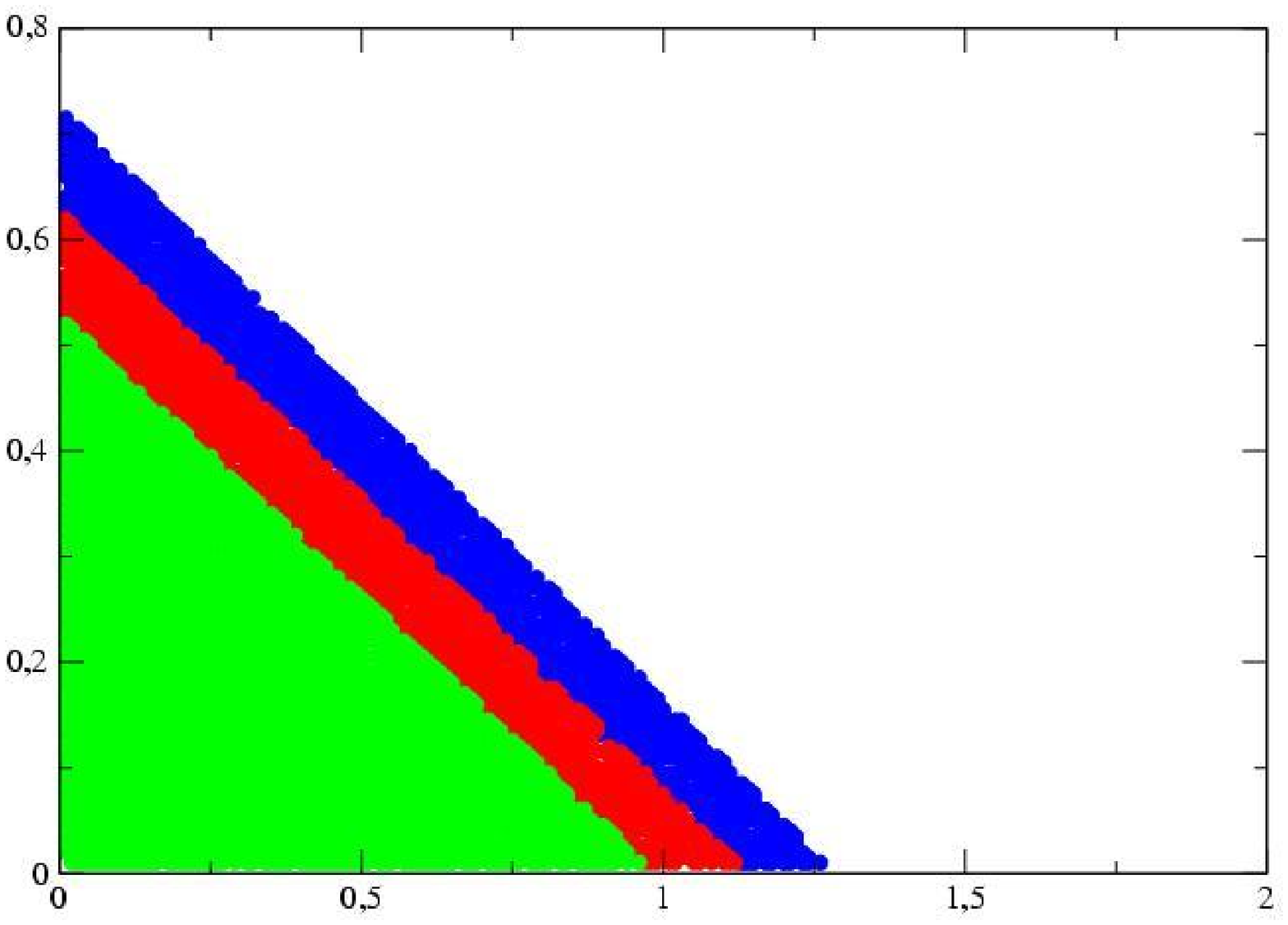,width=5cm,angle=-90}
\put(-35,-46){\tiny{$\Omega_{{\rm m} 0}$}}
\put(-65,-26){\tiny{$\Omega_{{\rm r} 0}$}}
\end{picture}&&
\begin{picture}(60,55)
\epsfig{file=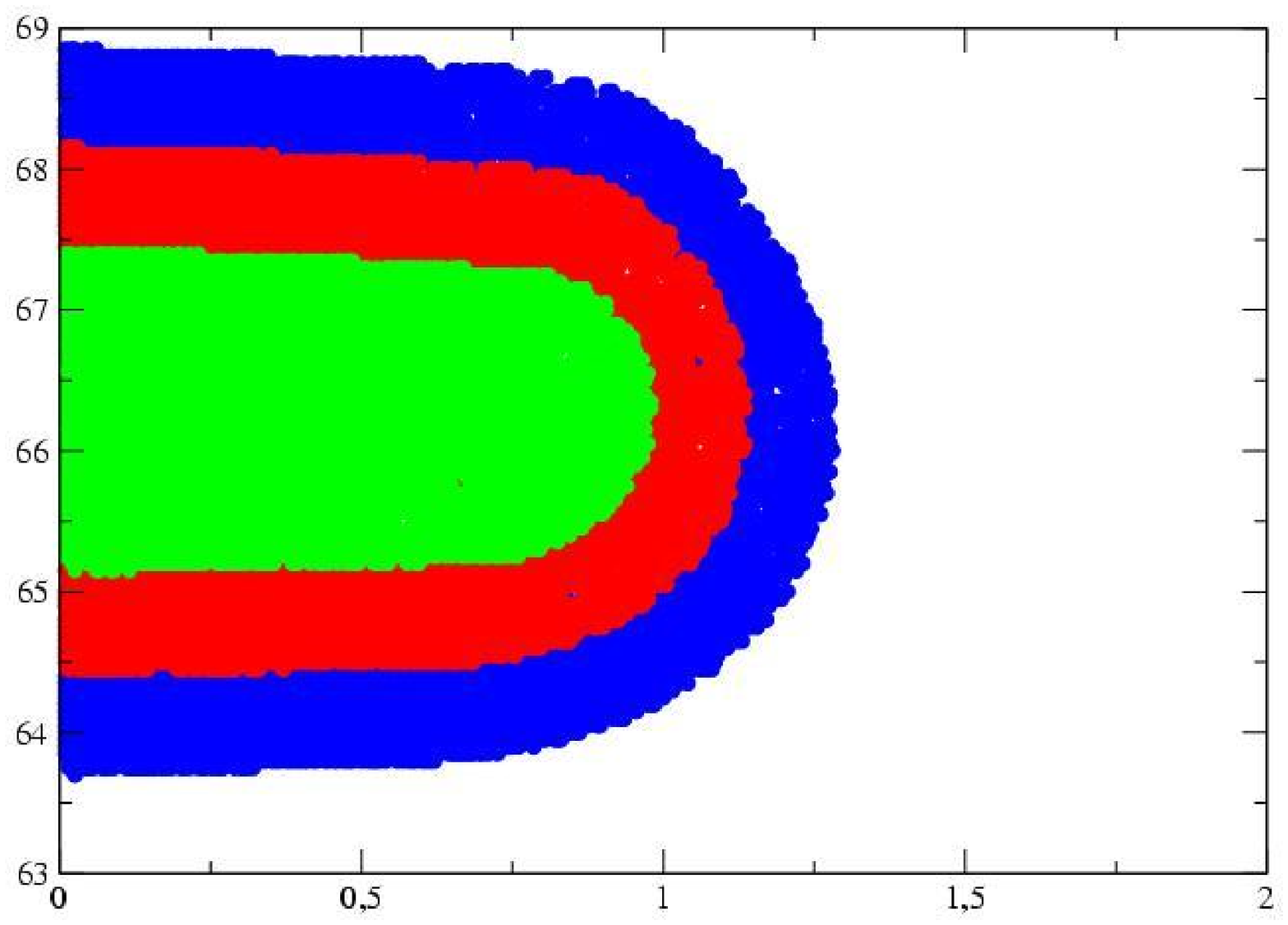,width=5cm,angle=-90}
\put(-35,-46){\tiny{$\Omega_{{\rm m} 0}$}}
\put(-65,-26){\tiny{$H_0$}}
\end{picture}\\
\vspace{-1cm}\phantom{4242}\begin{picture}(60,55)
\epsfig{file=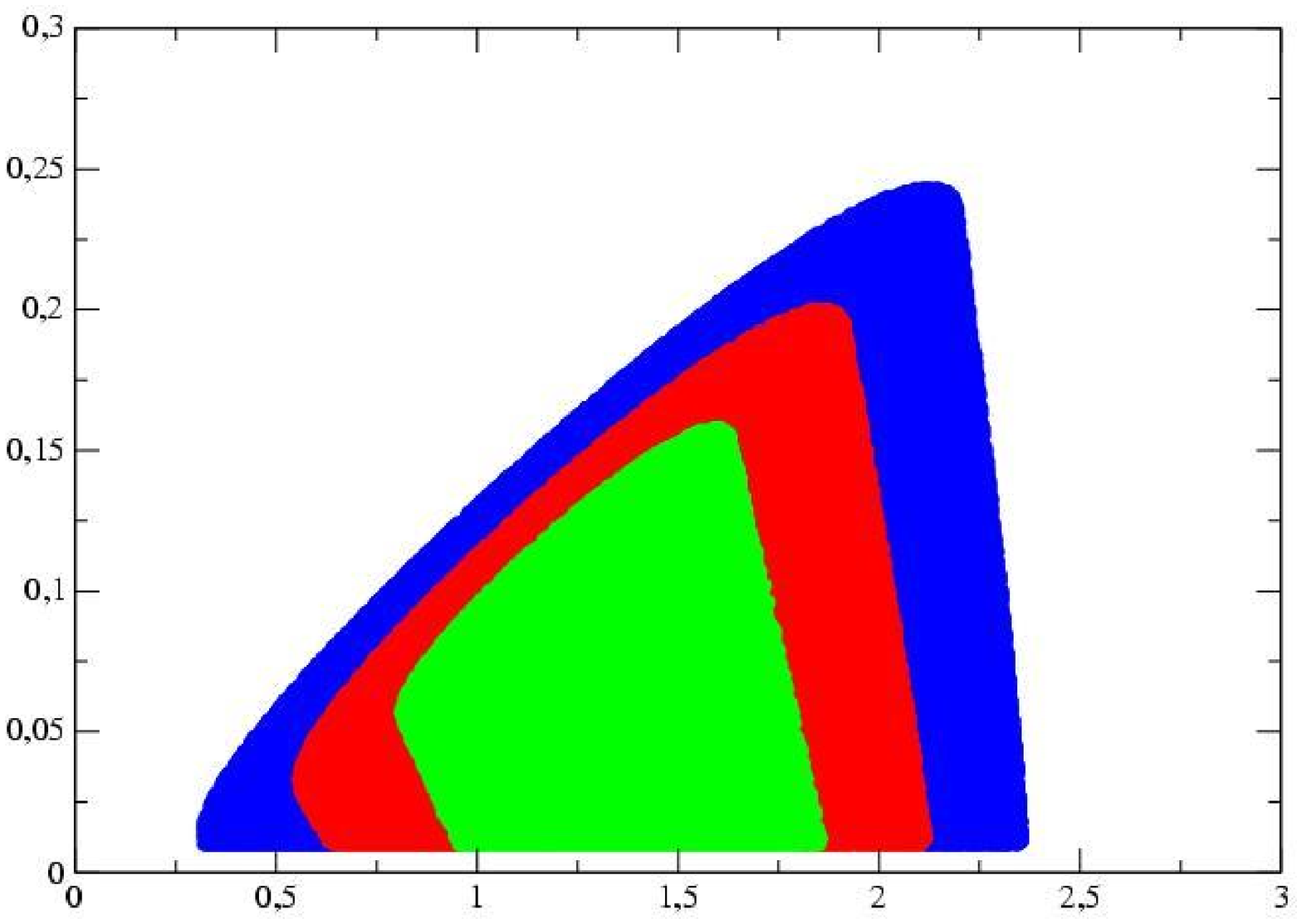,width=5cm,angle=-90}
\put(-35,-46){\tiny{$\Omega_{{\lambda} 0}$}}
\put(-65,-26){\tiny{$\Omega_{{\psi} 0}$}}
\end{picture}&&
\begin{picture}(60,55)
\epsfig{file=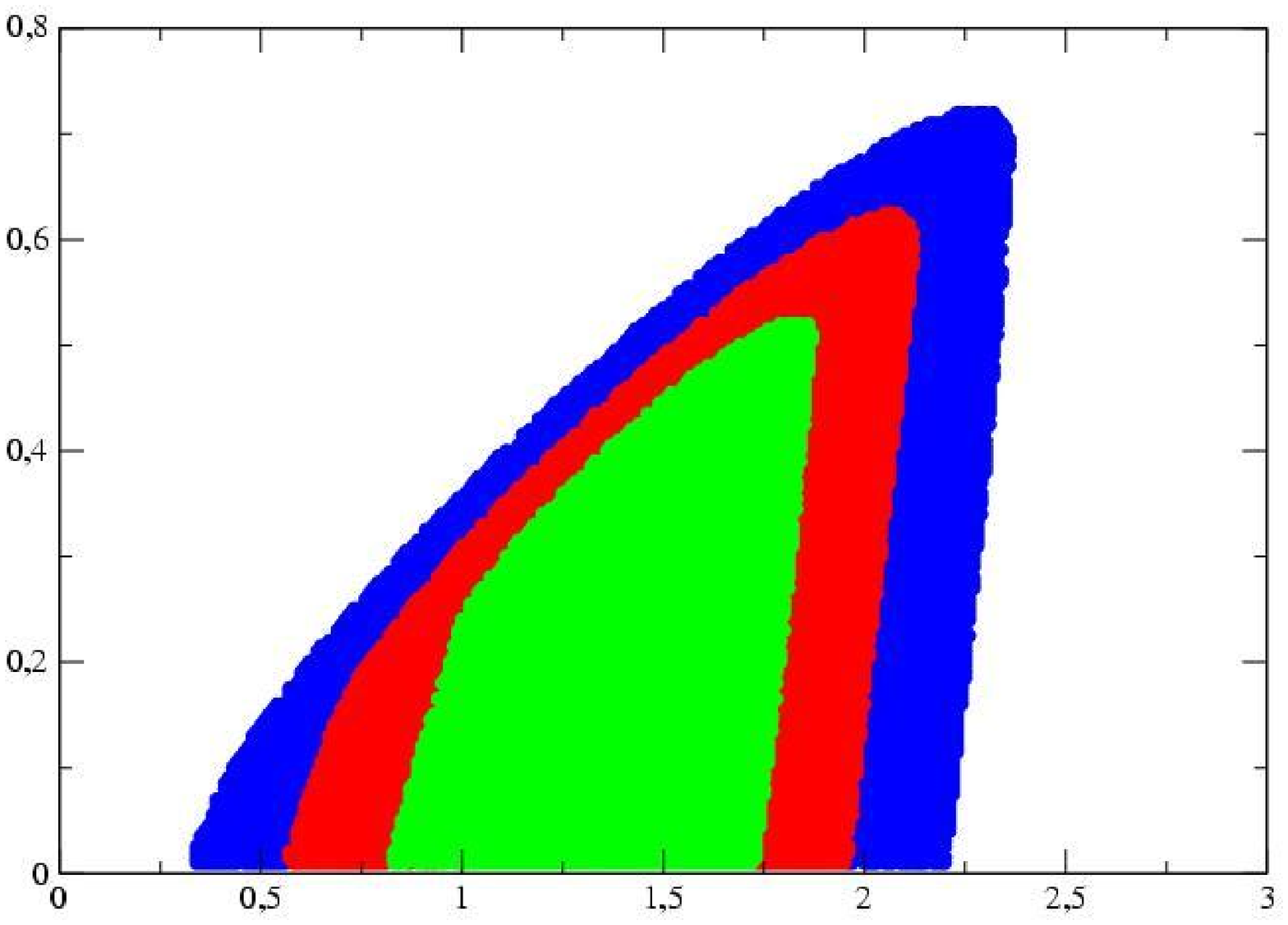,width=5cm,angle=-90}
\put(-35,-46){\tiny{$\Omega_{{\lambda} 0}$}}
\put(-65,-26){\tiny{$\Omega_{{\rm r} 0}$}}
\end{picture}\\
\vspace{-1cm}\phantom{4242}\begin{picture}(60,55)
\epsfig{file=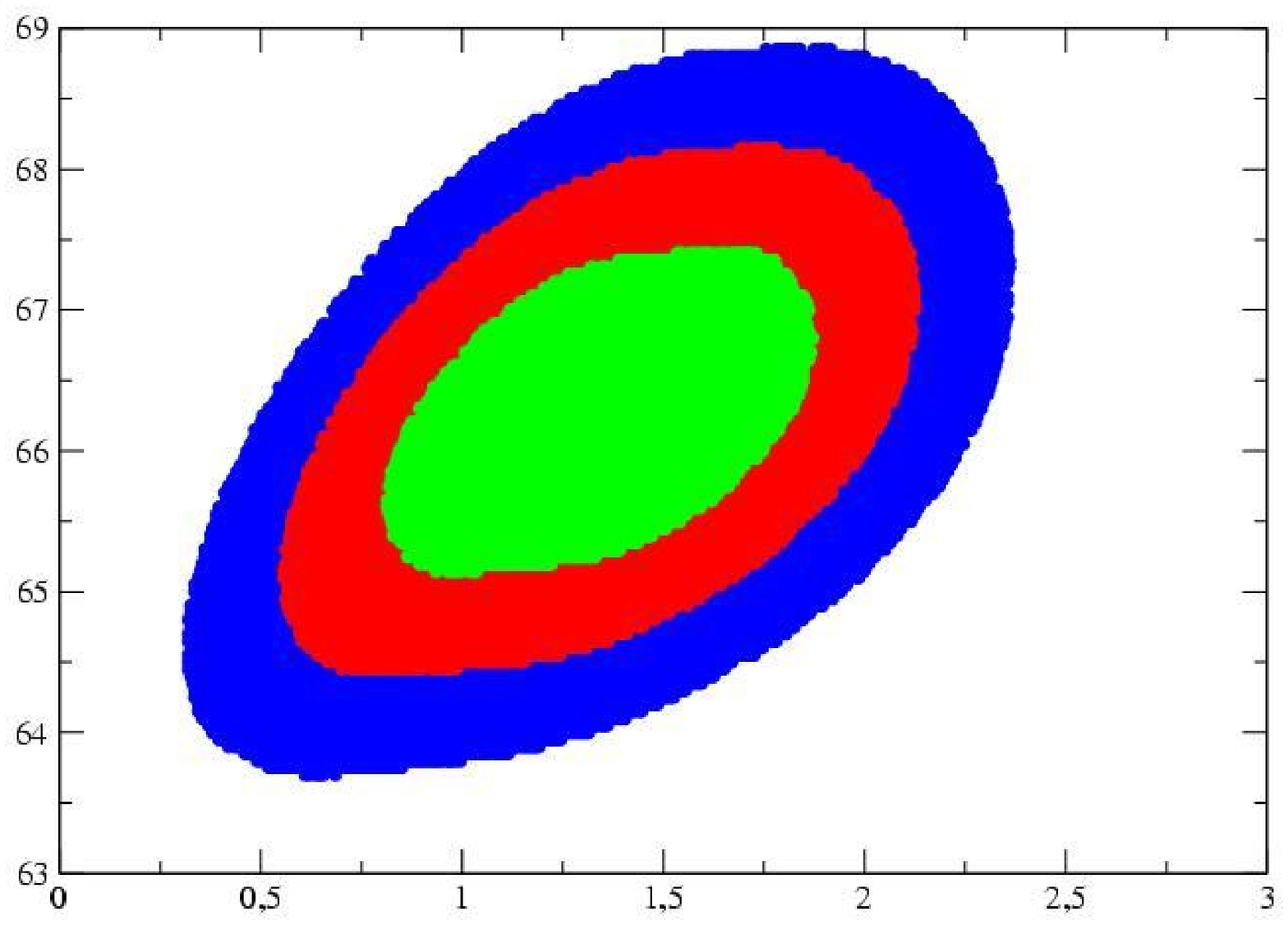,width=5cm,angle=-90}
\put(-35,-46){\tiny{$\Omega_{{\lambda} 0}$}}
\put(-65,-26){\tiny{$H_0$}}
\end{picture}&&
\begin{picture}(60,55)
\epsfig{file=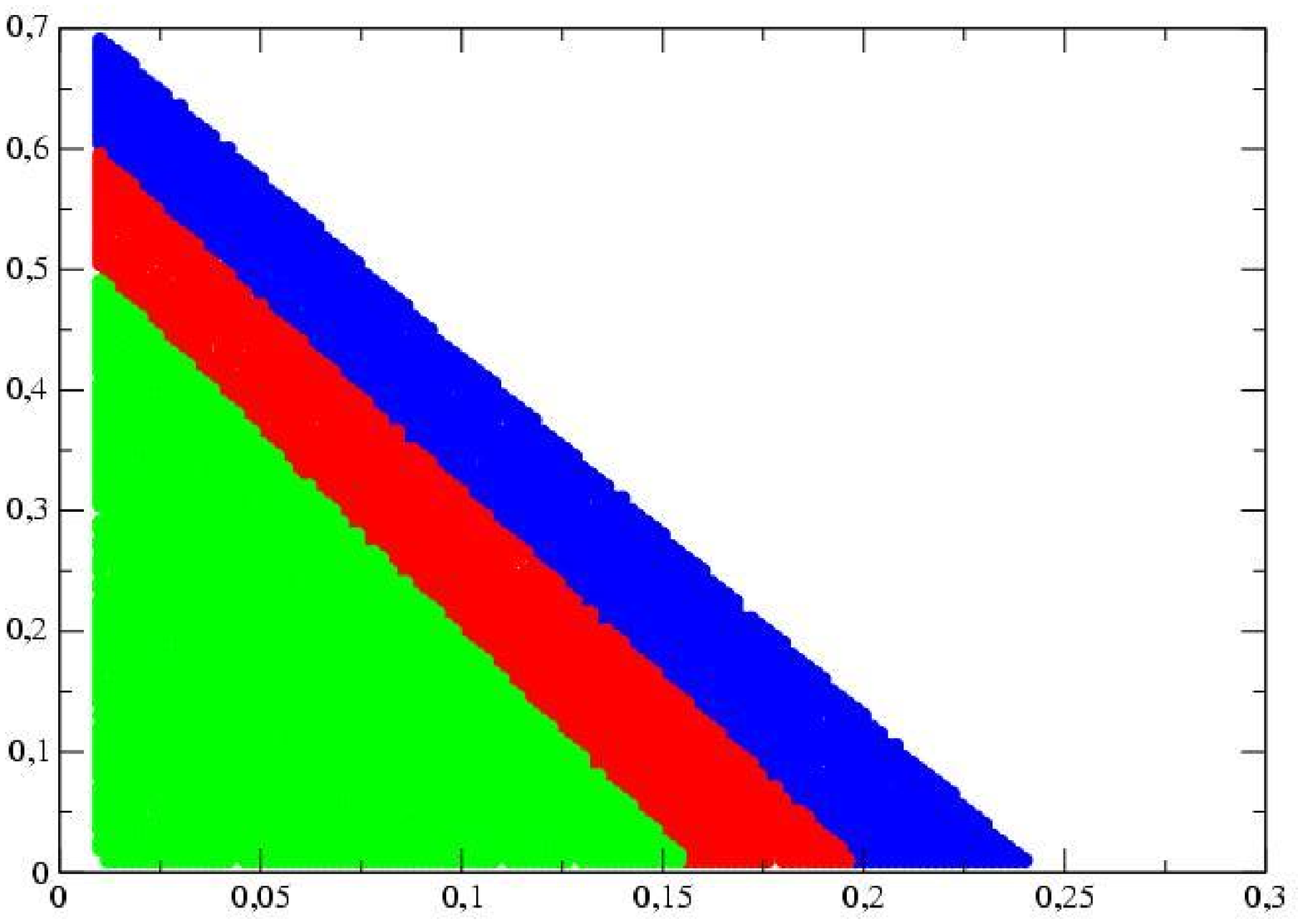,width=5cm,angle=-90}
\put(-35,-46){\tiny{$\Omega_{{\psi} 0}$}}
\put(-65,-26){\tiny{$\Omega_{{\rm r} 0}$}}
\end{picture}\\
\vspace{-1cm}\phantom{4242}\begin{picture}(60,55)
\epsfig{file=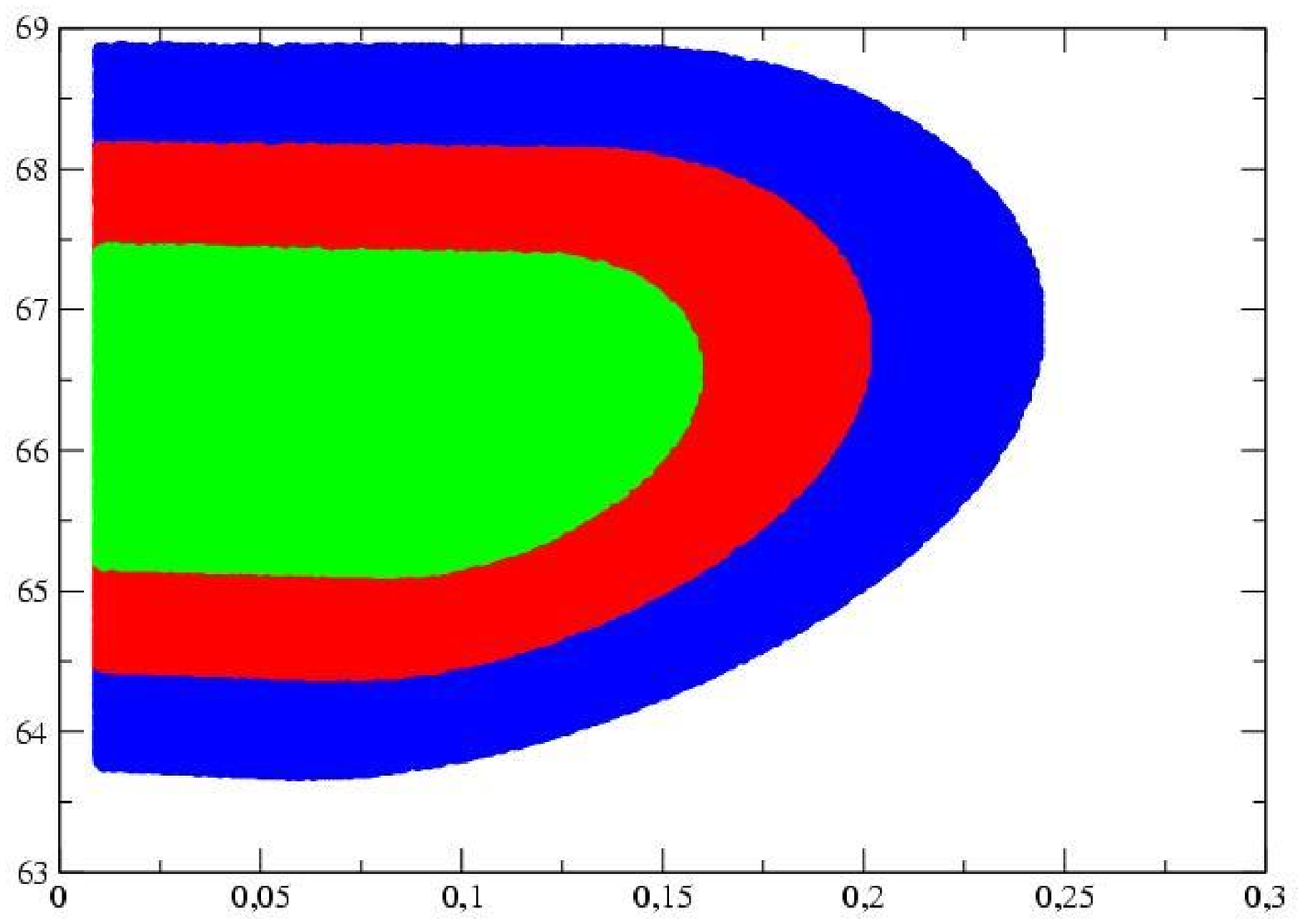,width=5cm,angle=-90}
\put(-35,-46){\tiny{$\Omega_{{\psi} 0}$}}
\put(-65,-26){\tiny{$H_0$}}
\end{picture}&&
\begin{picture}(60,55)
\epsfig{file=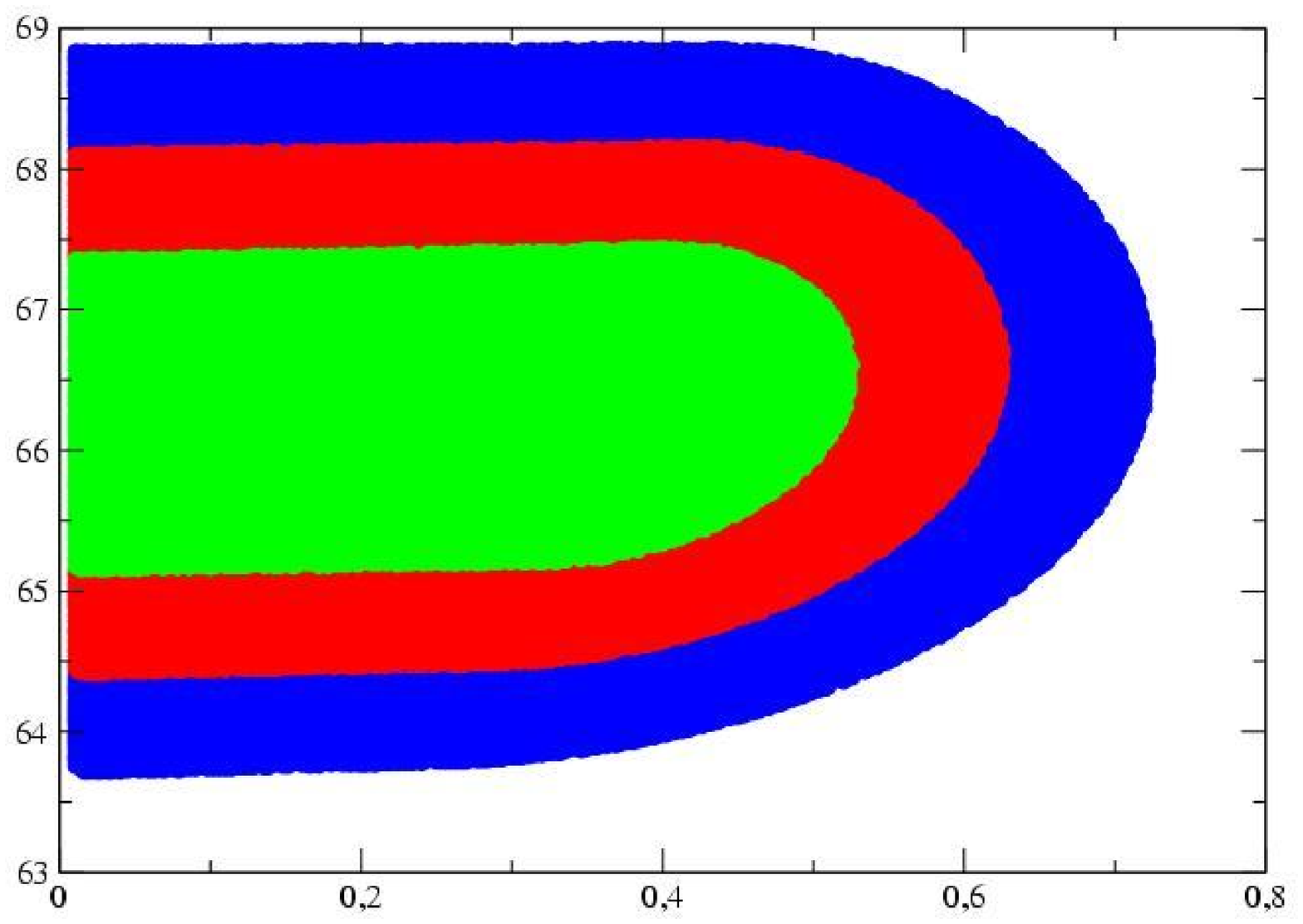,width=5cm,angle=-90}
\put(-35,-46){\tiny{$\Omega_{{\rm r} 0}$}}
\put(-65,-26){\tiny{$H_0$}}
\end{picture}\\\vspace{0.5cm}
\end{tabular}
\vspace{5cm}
\caption[Confidence contours for different parameter planes]{Confidence contours for all parameter planes of the Weyl-Cartan model for the unbinned data set containing 230 SN Ia.}
\label{FIG_triplet_contours_lambda230}
\end{figure}

The overall best-fit parameters for our model are displayed in table \ref{TABLE_triplet_best_unbinned}\footnote{Best-fit parameters are marked with ``WC'' in order to indicate that they  belong to a model which is based on Weyl-Cartan spacetime. This labeling will become handy in a forthcoming paper.}. Additionally, we worked out the 1$\sigma$, 2$\sigma$, and 3$\sigma$ confidence contours for all parameter planes and both data sets in figure \ref{FIG_triplet_contours_lambda} and \ref{FIG_triplet_contours_lambda230}. The overall best-fit for the data set containing 92 SNe of type Ia has $\chi^2 \approx 134.4$ which corresponds to  $\chi^2_\nu \approx 1.54$. The quality of the fit is improved for the unbinned data set with 230 SNe which has $\chi^2_\nu \approx 1.11$. As becomes clear from figure \ref{FIG_triplet_contours_lambda230} the new density parameter $\Omega_{\psi 0}$ is constrained to values smaller than 0.25 at the 3$\sigma$ level if we make use of T230 data set of Tonry \etal.    

We did not allow for negative values of the parameters in our fitting procedure. For parameters like $\Omega_{{\rm m}}$ such a choice would be clearly unphysical. On the other hand  negative values for $\Omega_{\psi}$ are {\it not} ruled out a priori, but lead to a limiting redshift $z_{\rm max}$ at which expression (\ref{EXT_obuk_Hubble_rate_with_density}) becomes negative, which of course is not allowed and forces us to limit our model to a specific redshift range. Since this limiting redshift is rather low even for moderate values of $\Omega_\psi$, remember the $\sim z^6$ scaling behavior of this new density parameter, we discard the possibility of introducing negative values for $\Omega_\psi$.

\begin{table}
\caption{Overall best-fit parameters for the unbinned data sets. }
\label{TABLE_triplet_best_unbinned}
\begin{indented}
\item[]\begin{tabular}{@{}llllllllll}
\br
\textbf{Symbol}&$\sharp$ SN &$\Omega_{{\rm m}0}$&$\Omega_{\lambda 0}$&$\Omega_{\psi 0}$&$\Omega_{{\rm r} 0}$&$H_0$&$\chi^2$&$\chi^2_\nu$&$q_0$\\
\mr
WC92 &92 &0.076&  1.218&  0.001&  0.295&  65.334&  134.446&1.54&-0.88\\
WC230&230&0.013&  1.446&  0.001&  0.363&  66.300&  249.112&1.11&-1.07\\
\br
\end{tabular}
\item[] $[H_0]={\rm km \, s}^{-1}{\rm Mpc}^{-1}$. 
\end{indented}
\end{table}

\subsection{Flux-averaging\label{NUMERICAL_binning}}

\begin{figure}
\setlength{\unitlength}{1mm}
\vspace{-5cm}
\begin{tabular}{lll}
\vspace{-1cm}\phantom{4242}\begin{picture}(60,55)
\epsfig{file=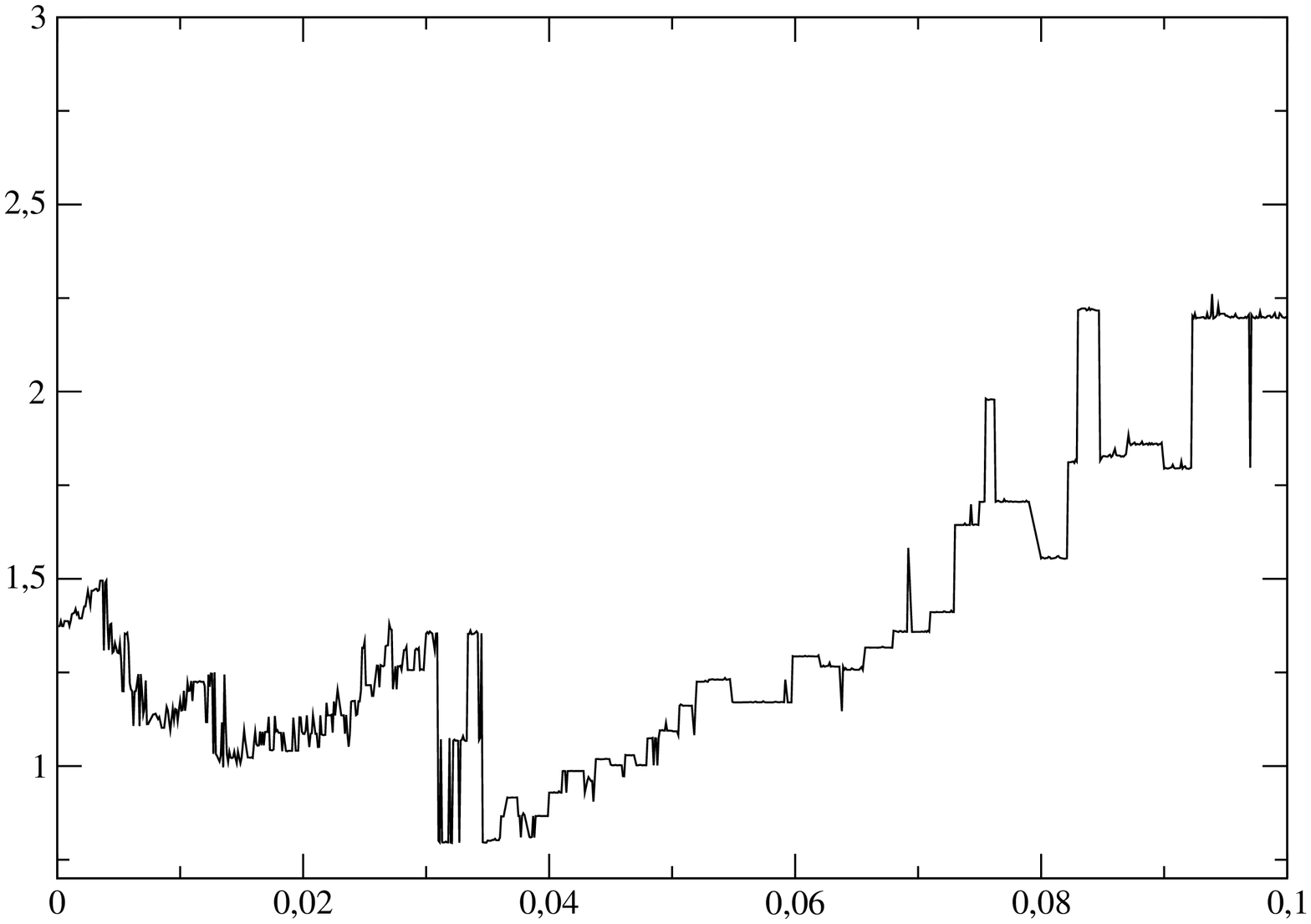,width=5cm,angle=-90}
\put(-35,-46){\tiny{$\Delta z$}}
\put(-65,-26){\tiny{$\chi^2_{\rm pdf}$}}
\end{picture}&\phantom{42}&
\begin{picture}(60,55)
\epsfig{file=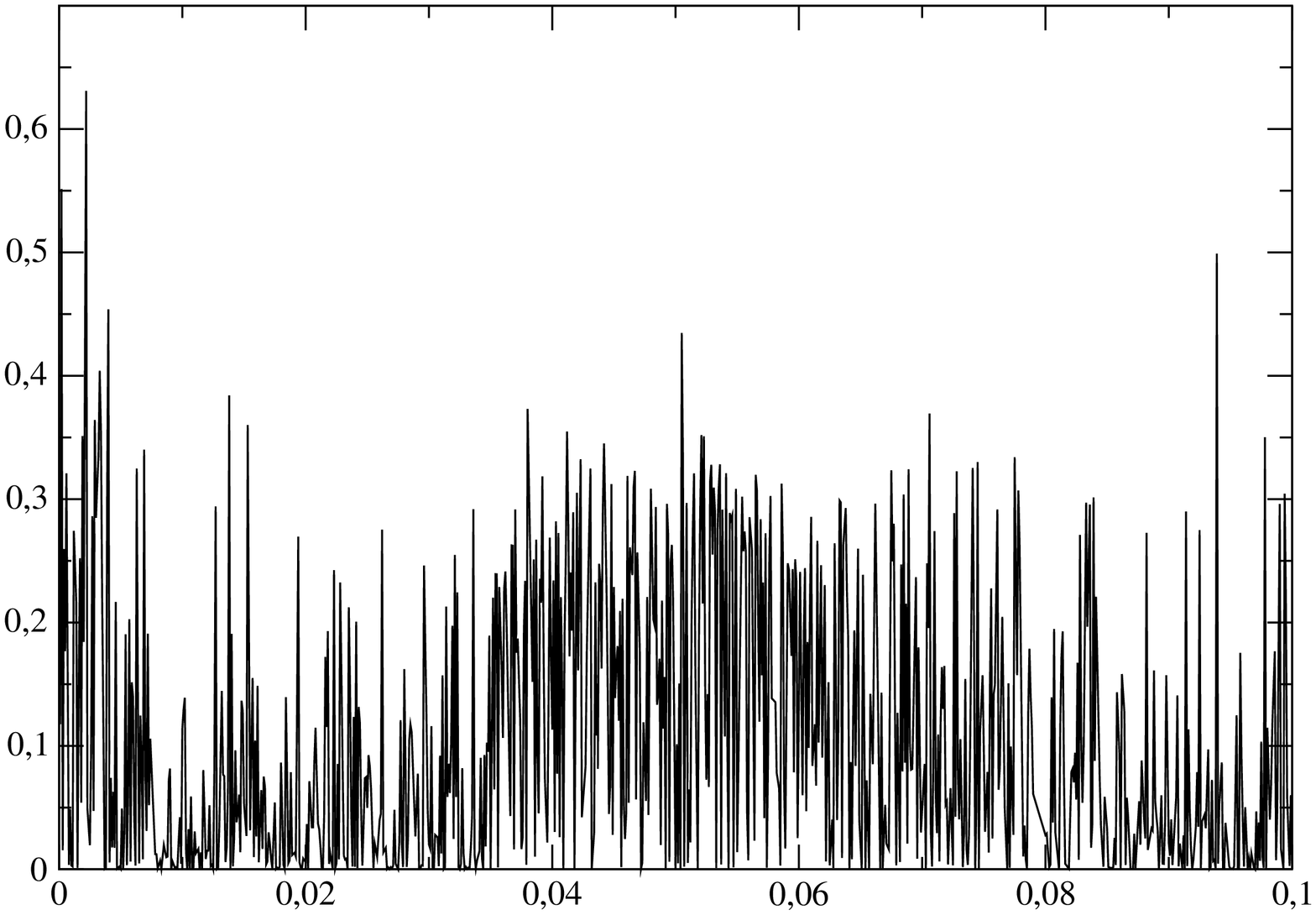,width=5cm,angle=-90}
\put(-35,-46){\tiny{$\Delta z$}}
\put(-65,-26){\tiny{$\Omega_{{\rm m} 0}$}}
\end{picture}\\
\vspace{-1cm}\phantom{4242}\begin{picture}(60,55)
\epsfig{file=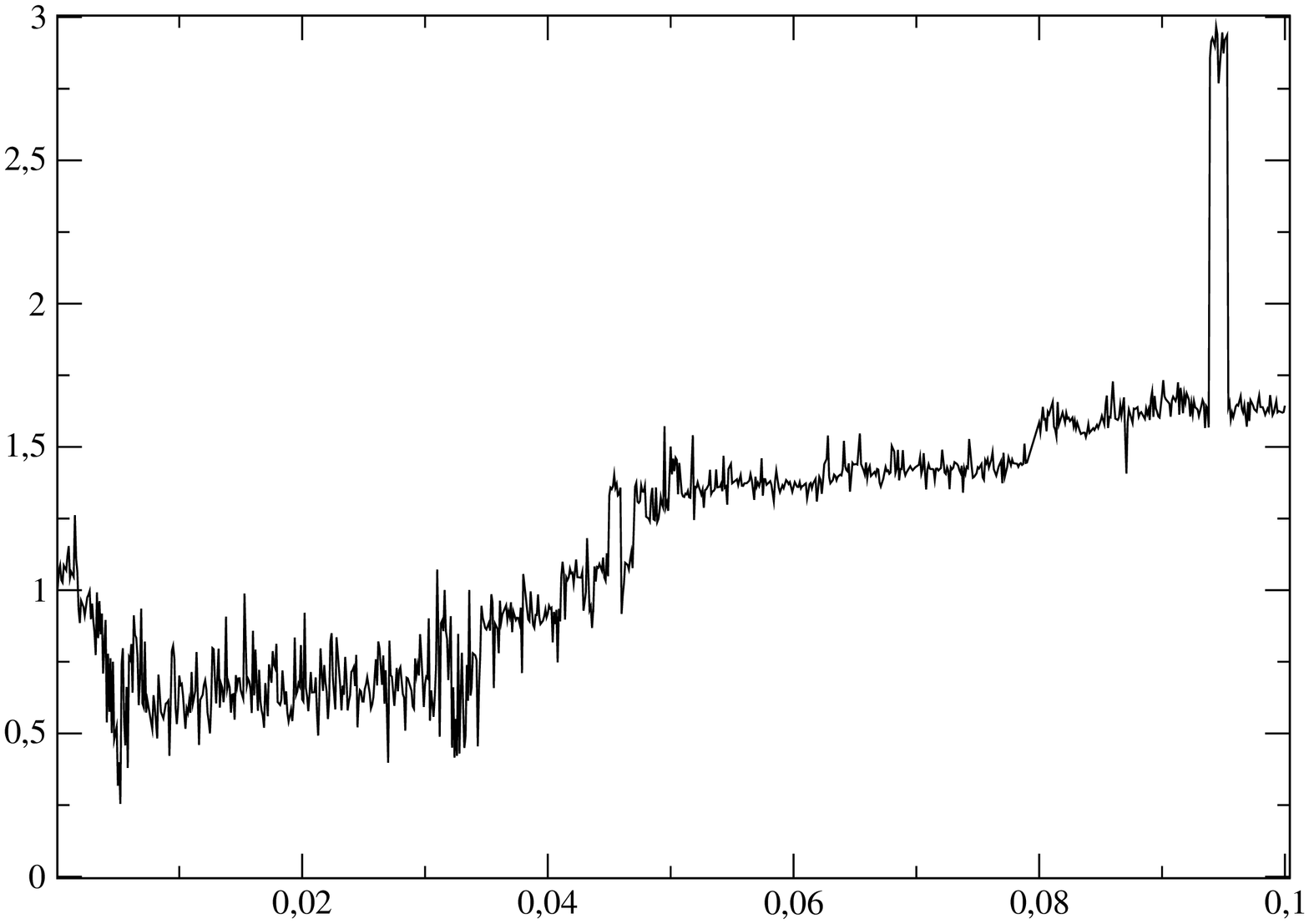,width=5cm,angle=-90}
\put(-35,-46){\tiny{$\Delta z$}}
\put(-65,-26){\tiny{$\Omega_{{\lambda} 0}$}}
\end{picture}&&
\begin{picture}(60,55)
\epsfig{file=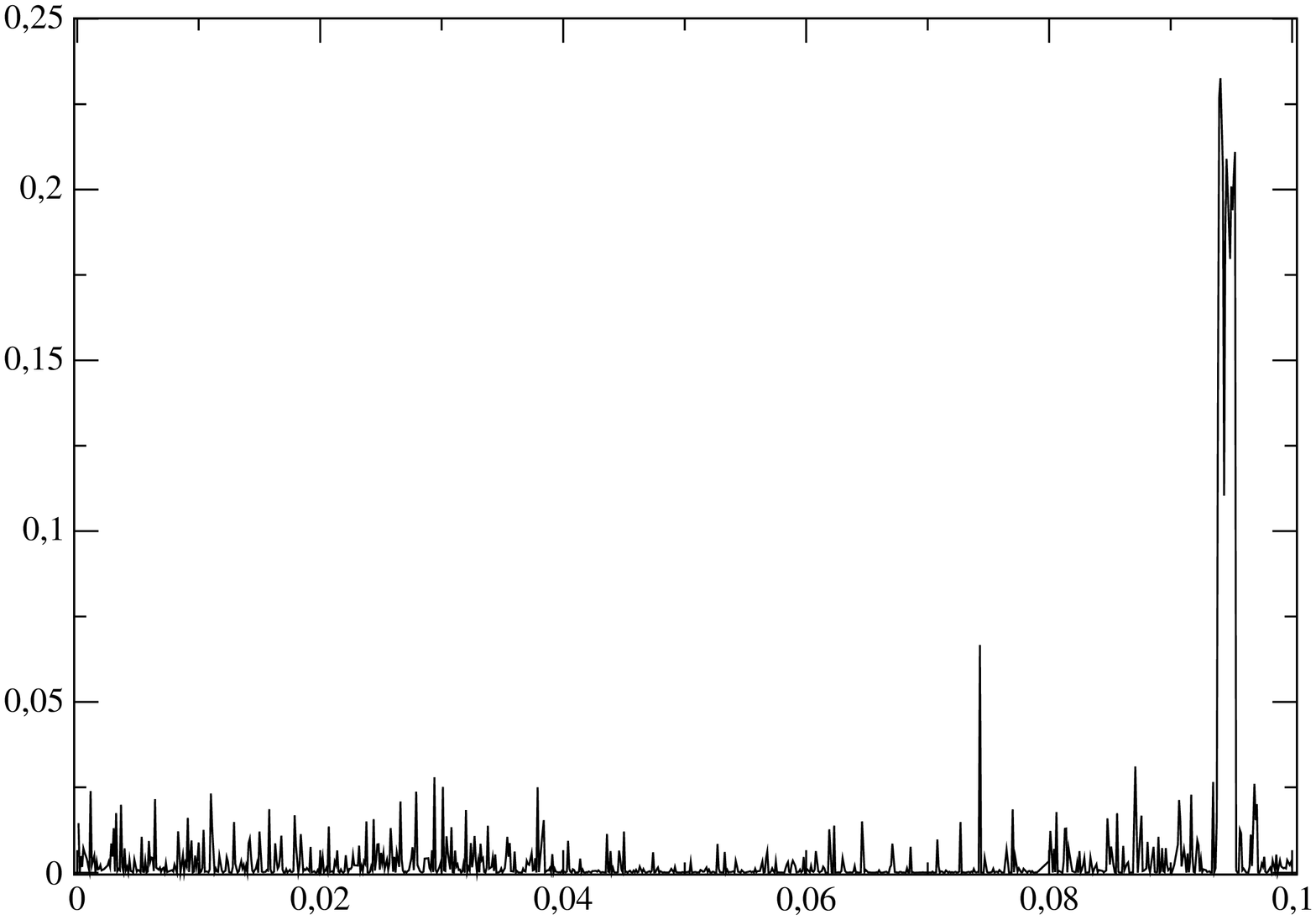,width=5cm,angle=-90}
\put(-35,-46){\tiny{$\Delta z$}}
\put(-65,-26){\tiny{$\Omega_{{\psi} 0}$}}
\end{picture}\\
\vspace{-1cm}\phantom{4242}\begin{picture}(60,55)
\epsfig{file=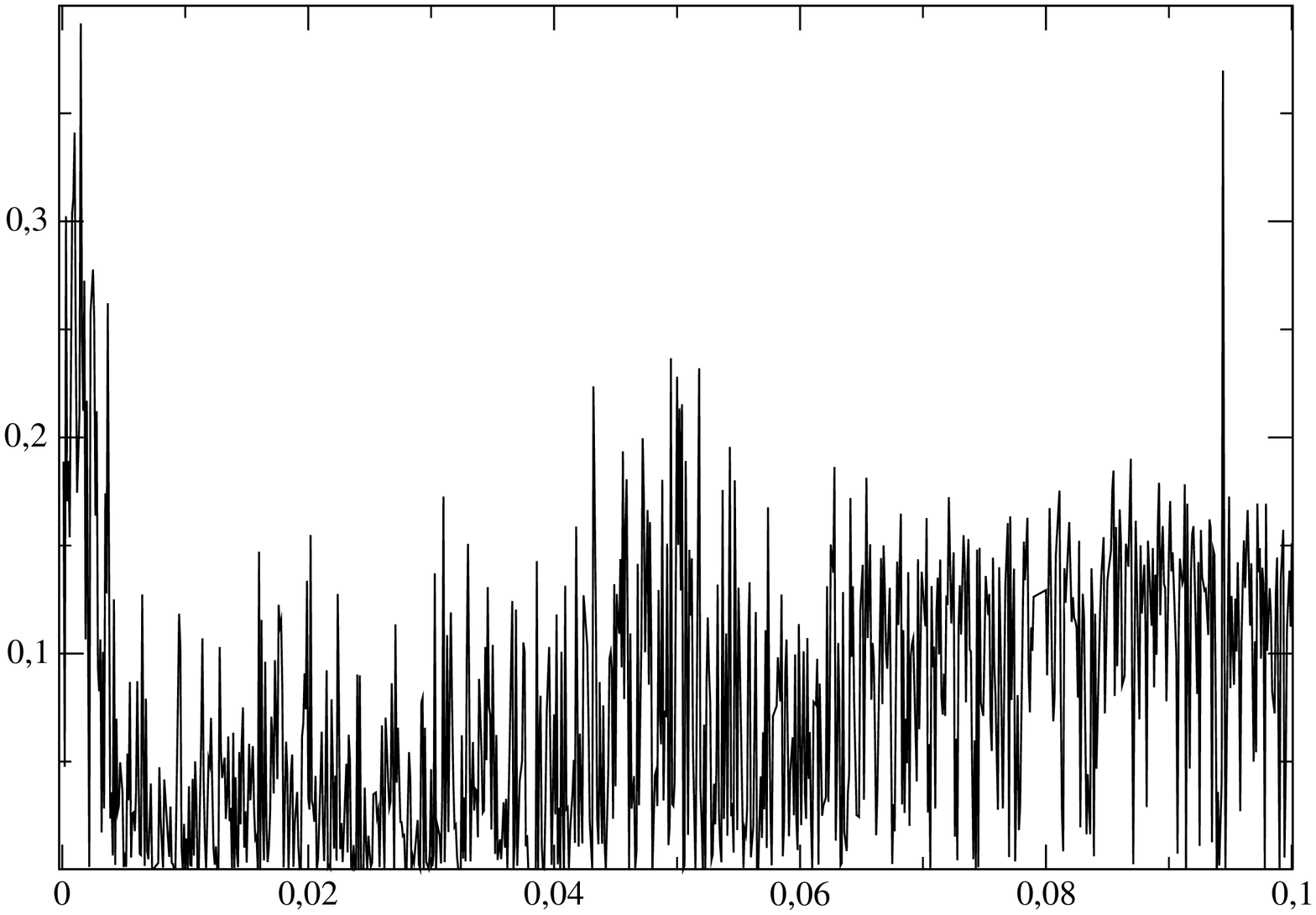,width=5cm,angle=-90}
\put(-35,-46){\tiny{$\Delta z$}}
\put(-65,-26){\tiny{$\Omega_{{\rm r} 0}$}}
\end{picture}&&
\begin{picture}(60,55)
\epsfig{file=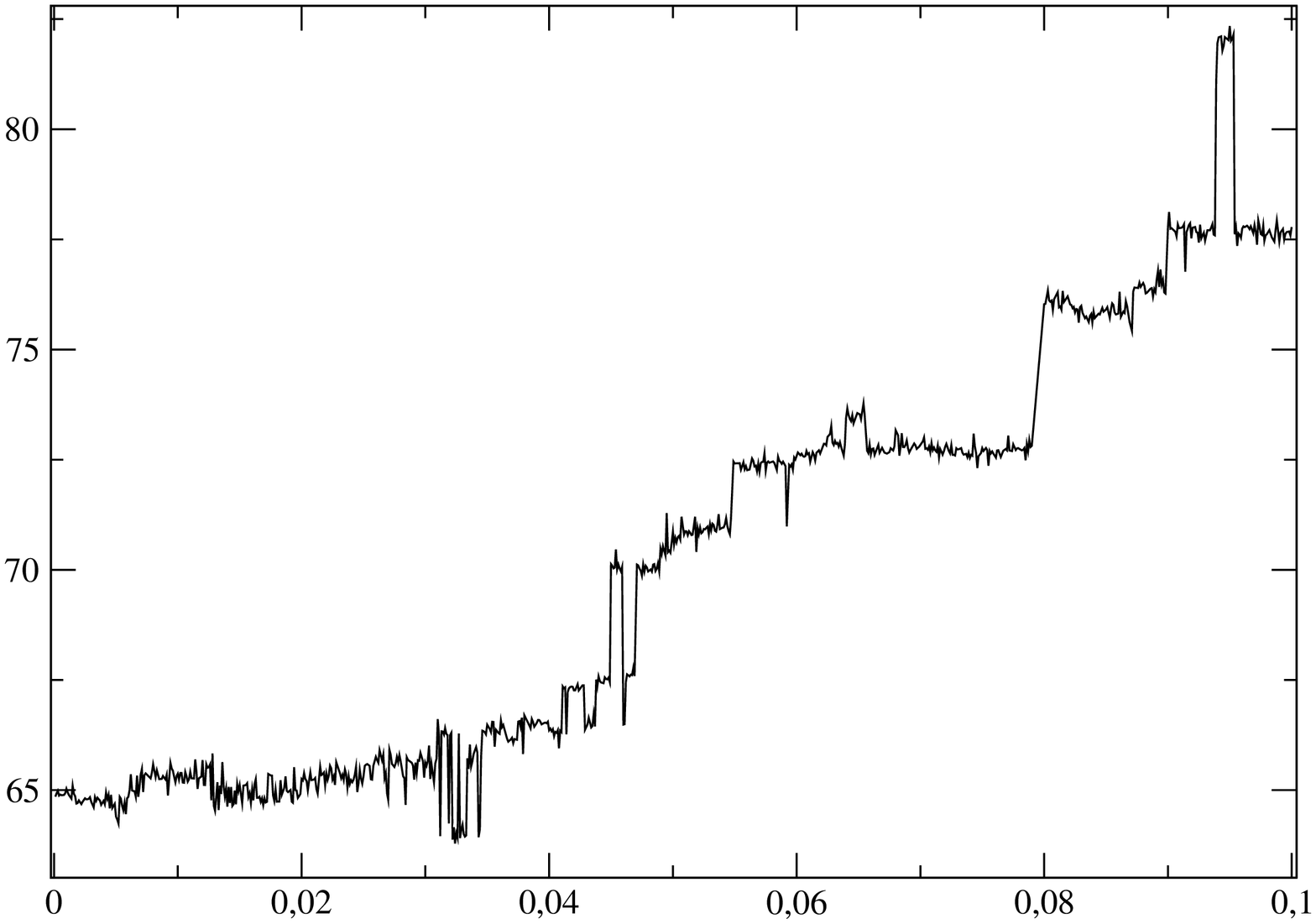,width=5cm,angle=-90}
\put(-35,-46){\tiny{$\Delta z$}}
\put(-65,-26){\tiny{$H_0$}}
\end{picture}\\
\end{tabular}
\vspace{6cm}
\caption[]{Variation of the best-fit parameters for different bin sizes for the data set of Wang which originally contained 92 SNe.}
\label{FIG_triplet_averaged92}
\end{figure}

In addition to the search for the best-fit parameters for the data sets described in section \ref{DATA_sets_section} we investigate the impact of placing the data in redshift bins of finite size. A similar analysis was carried out by Wang for the FLRW model in \cite{Wang,Wang2000}. Due to the lumpiness of the matter distribution in our universe one expects changes of the peak luminosity of single supernovae due to weak gravitational lensing. In order to reduce the impact of lensing one flux-averages the data by placing it in redshift intervals of finite size. Additionally, such a smoothing procedure should reduce the variation of the luminosity due to intrinsic dispersions. For an investigation covering the use of lensing shear maps to correct for the signal due to weak-gravitational lensing see \cite{Dalal}.  

\begin{figure}
\setlength{\unitlength}{1mm}
\vspace{-5cm}
\begin{tabular}{lll}
\vspace{-1cm}\phantom{4242}\begin{picture}(60,55)
\epsfig{file=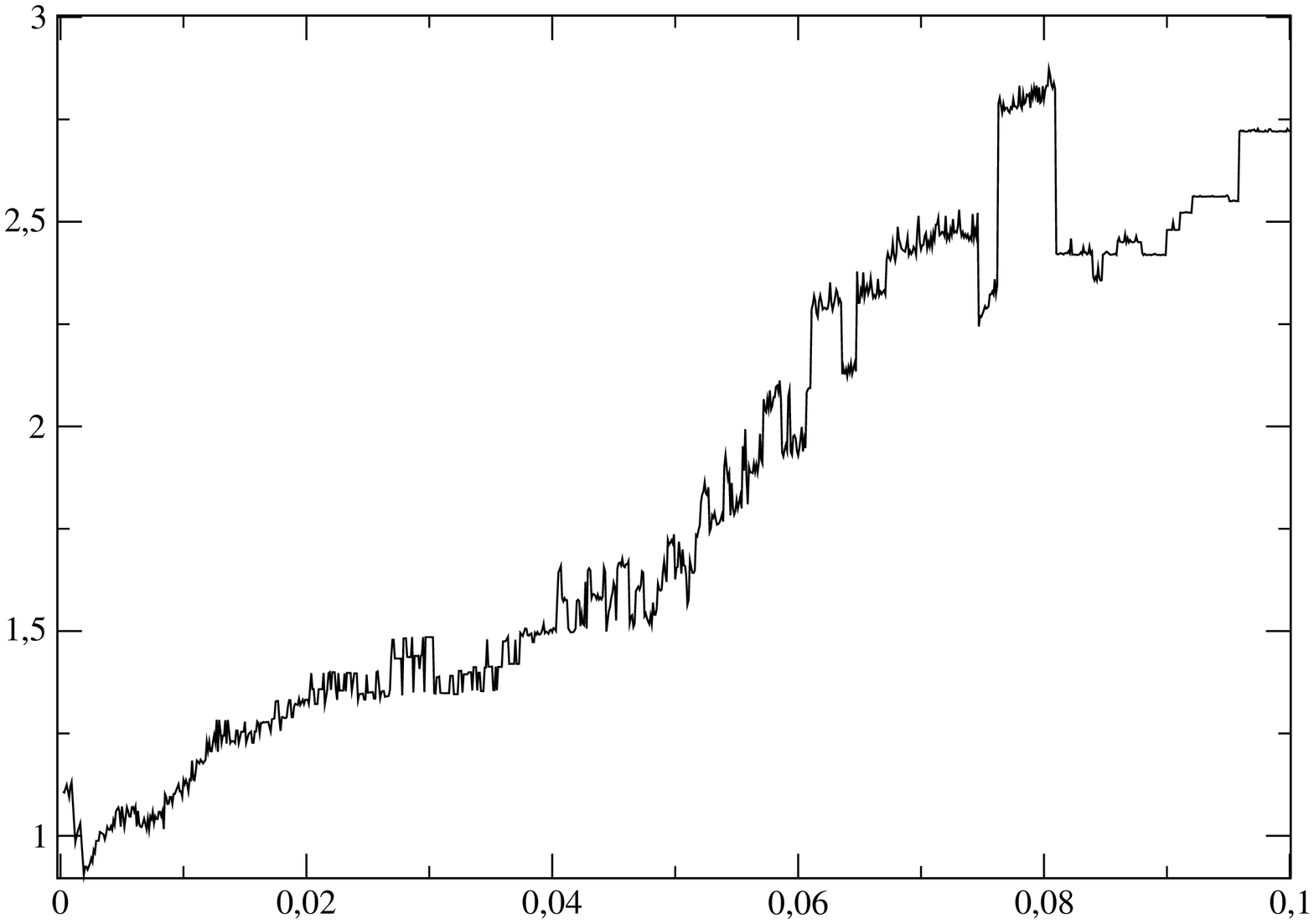,width=5cm,angle=-90}
\put(-35,-46){\tiny{$\Delta z$}}
\put(-65,-26){\tiny{$\chi^2_{\rm pdf}$}}
\end{picture}&\phantom{42}&
\begin{picture}(60,55)
\epsfig{file=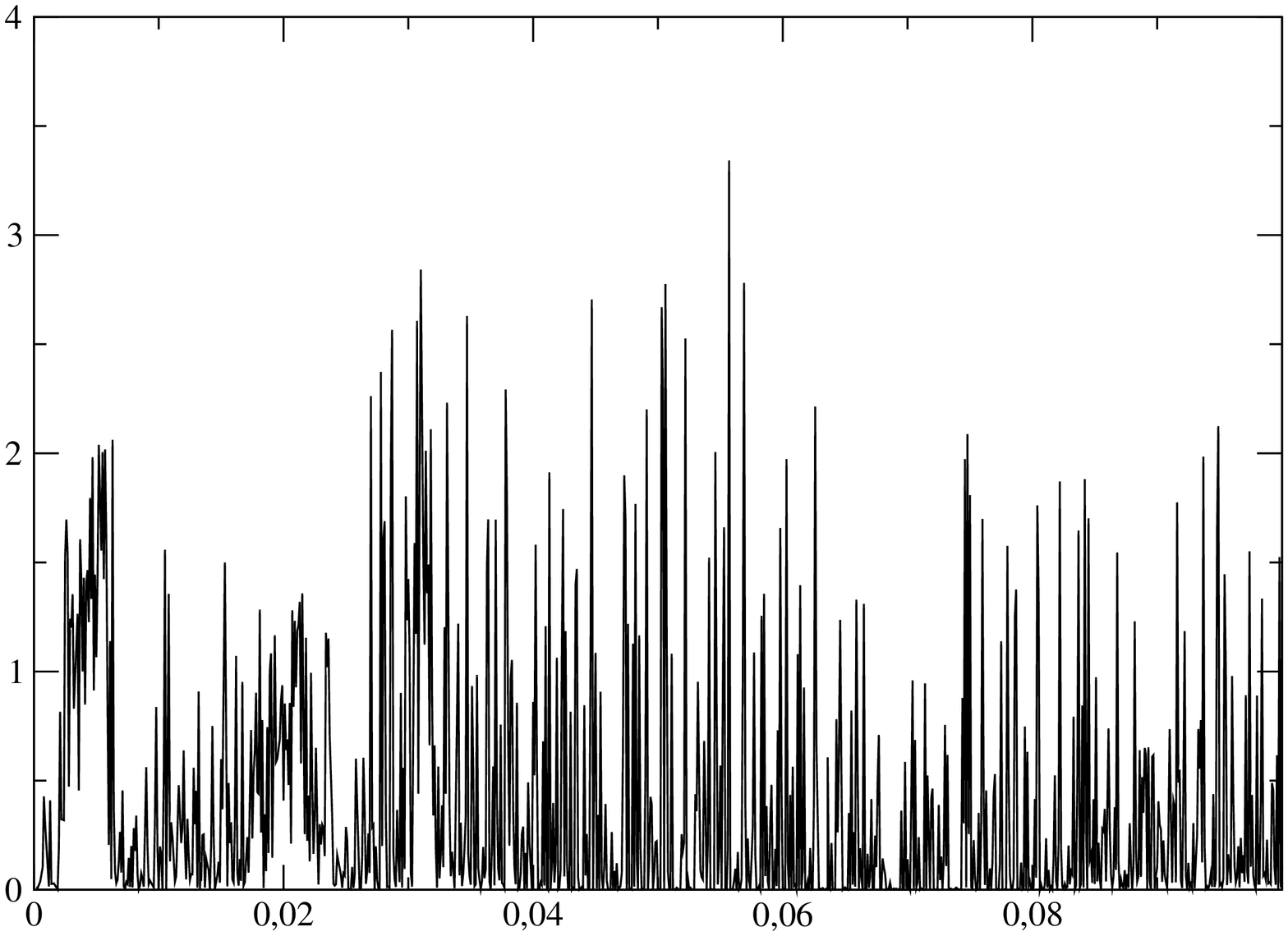,width=5cm,angle=-90}
\put(-35,-46){\tiny{$\Delta z$}}
\put(-65,-26){\tiny{$\Omega_{{\rm m} 0}$}}
\end{picture}\\
\vspace{-1cm}\phantom{4242}\begin{picture}(60,55)
\epsfig{file=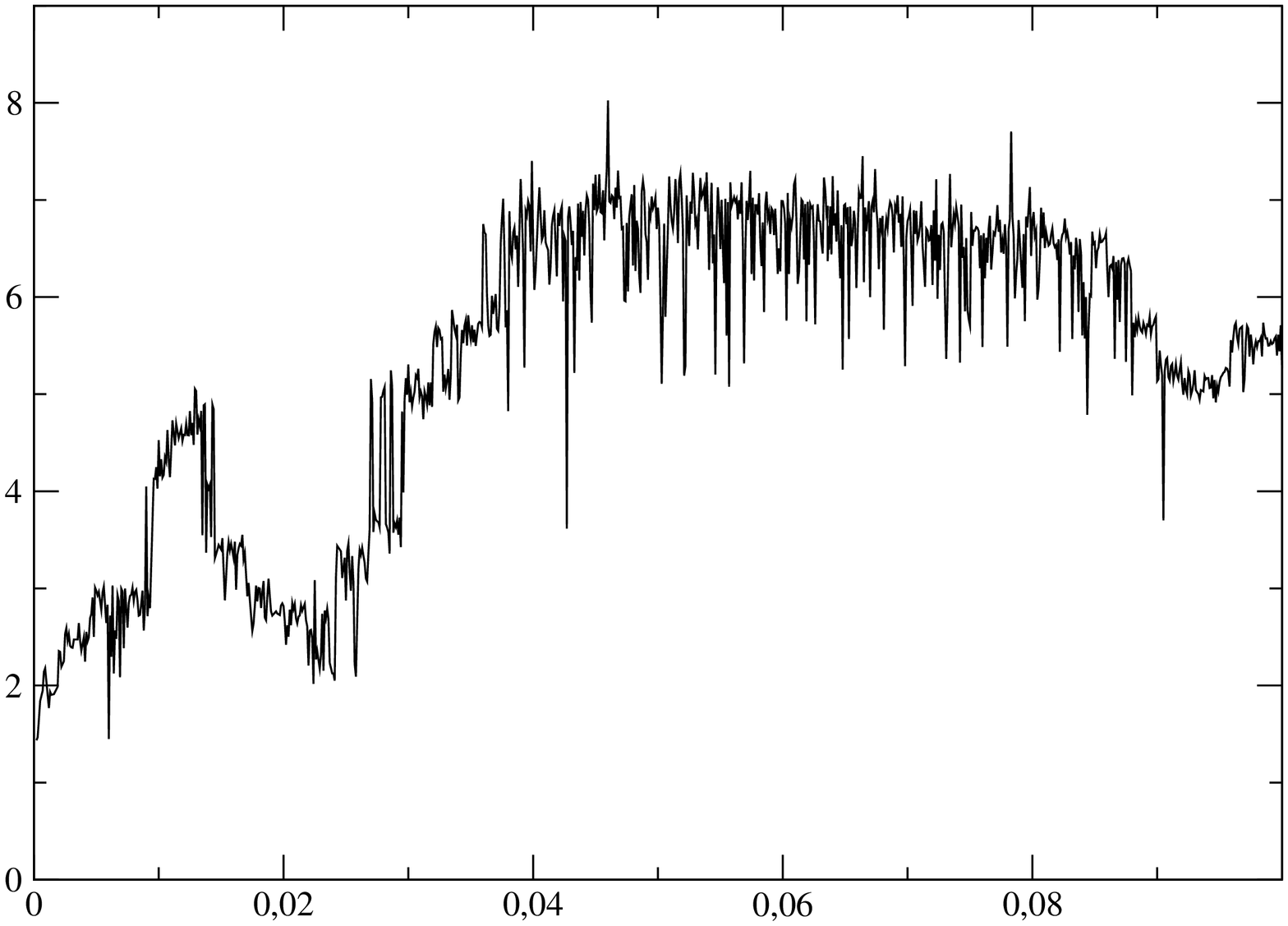,width=5cm,angle=-90}
\put(-35,-46){\tiny{$\Delta z$}}
\put(-65,-26){\tiny{$\Omega_{{\lambda} 0}$}}
\end{picture}&&
\begin{picture}(60,55)
\epsfig{file=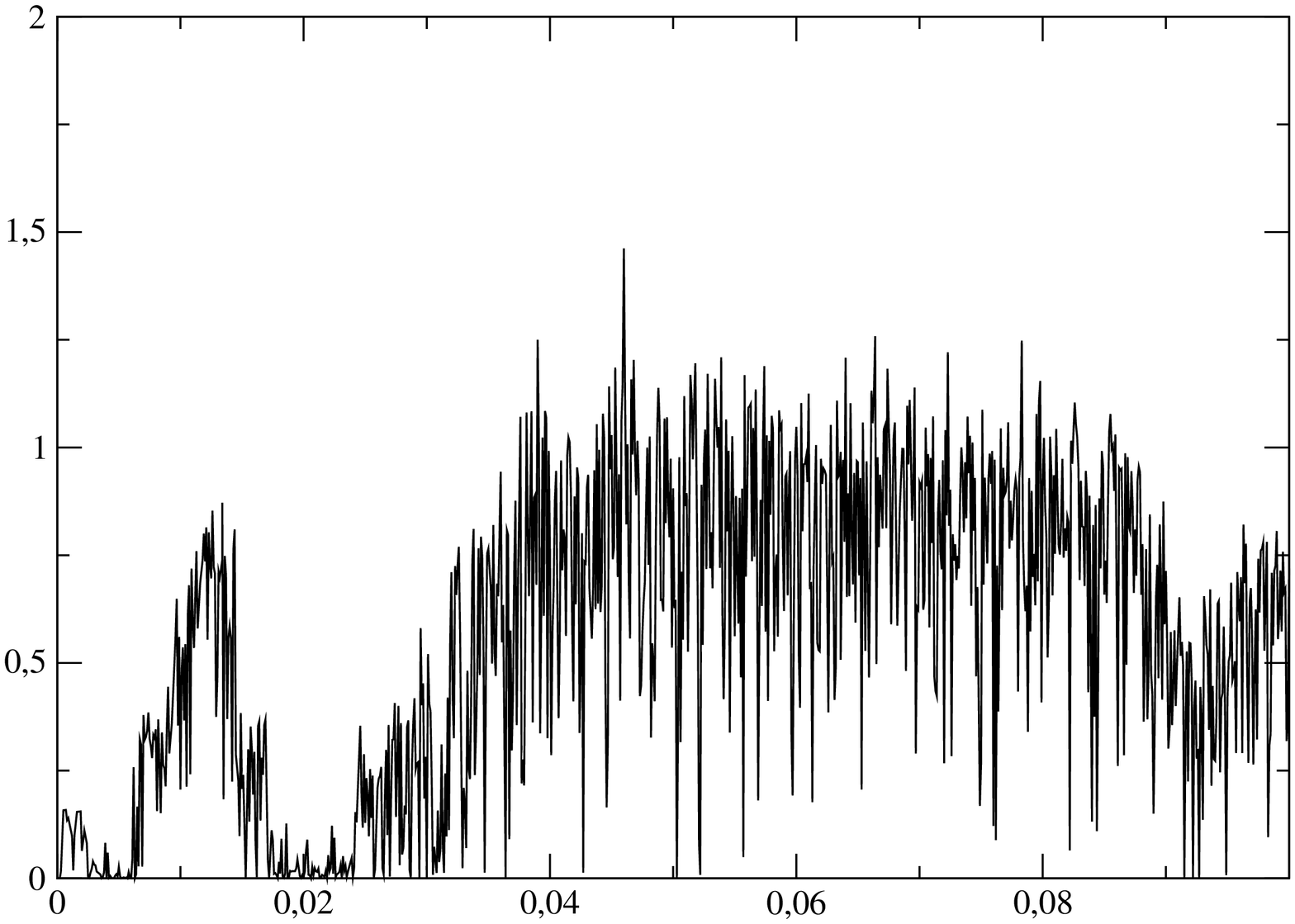,width=5cm,angle=-90}
\put(-35,-46){\tiny{$\Delta z$}}
\put(-65,-26){\tiny{$\Omega_{{\psi} 0}$}}
\end{picture}\\
\vspace{-1cm}\phantom{4242}\begin{picture}(60,55)
\epsfig{file=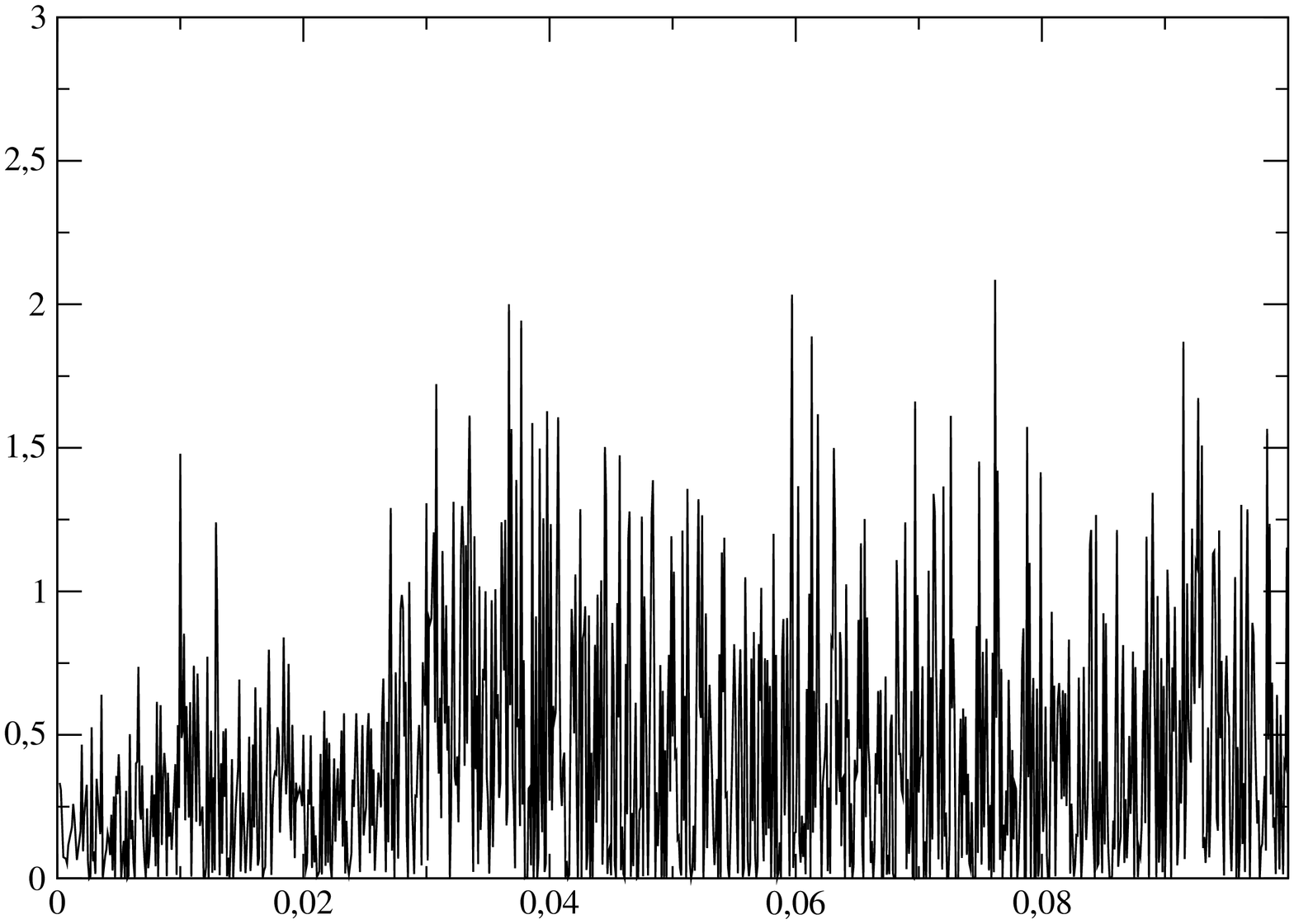,width=5cm,angle=-90}
\put(-35,-46){\tiny{$\Delta z$}}
\put(-65,-26){\tiny{$\Omega_{{\rm r} 0}$}}
\end{picture}&&
\begin{picture}(60,55)
\epsfig{file=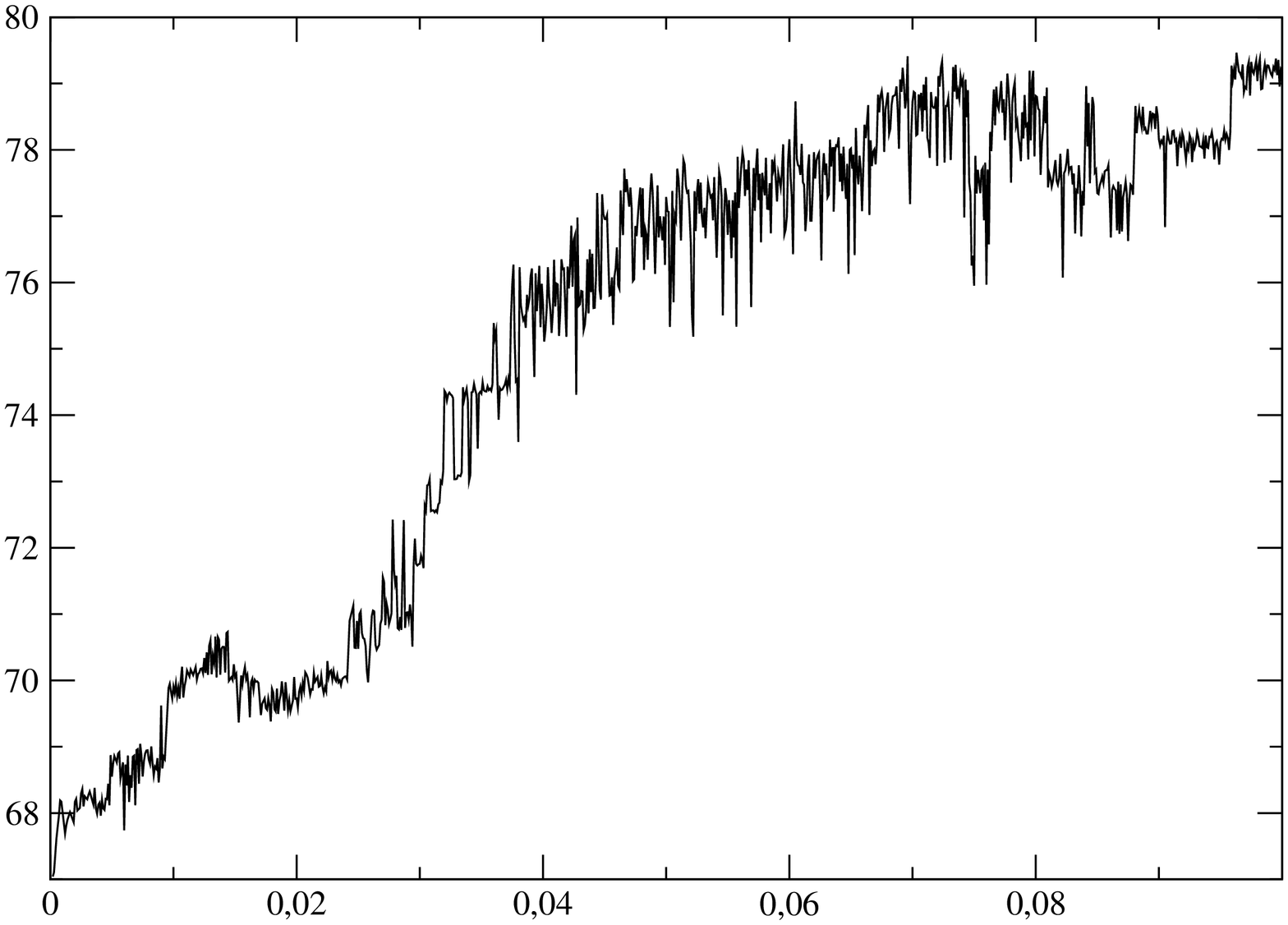,width=5cm,angle=-90}
\put(-35,-46){\tiny{$\Delta z$}}
\put(-65,-26){\tiny{$H_0$}}
\end{picture}\\
\end{tabular}
\vspace{6cm}
\caption[]{Variation of the best-fit parameters for different bin sizes for the data set of Tonry \etal which originally contained 230 SNe.}
\label{FIG_triplet_averaged230}
\end{figure}

In figure \ref{FIG_triplet_averaged92} we plotted the variation of the overall best-fit parameters for different bin sizes in the interval $[0,0.1]$ for the W92 data set. As becomes clear from the plot on the upper lhs we get an acceptable fit for bin sizes up to 0.08. The optimal bin size lies somewhere around 0.04. We also performed the binning procedure for the T230 data set. The results of this calculation are displayed in figure \ref{FIG_triplet_averaged230}. In contrast to the results for the W92 data set the optimal bin size is shifted to a lower bin width of about 0.002. For the T230 data set the fit quality is acceptable up to a bin size of about 0.04. 

\begin{table}
\caption{Overall best-fit parameters for binned data sets. }
\label{TABLE_triplet_best_binned}
\begin{indented}
\item[]\begin{tabular}{@{}lllllllllll}
\br
\textbf{Symbol}&$\sharp$ SN &$\Delta z$&$\Omega_{{\rm m}0}$&$\Omega_{\lambda 0}$&$\Omega_{\psi 0}$&$\Omega_{{\rm r} 0}$&$H_0$&$\chi^2$&$\chi^2_\nu$&$q_0$\\
\mr
WC114&114&0.002&0.001&  1.986&  0.155&  0.165&  67.861&  98.620 &0.90&-1.51\\
WC25 &25 &0.04 &0.860&  6.567&  0.754&  0.270&  75.663&  29.954 &1.49&-4.35\\
\br
\end{tabular}
\item[] $[H_0]={\rm km \, s}^{-1}{\rm Mpc}^{-1}$.
\end{indented}
\end{table}

\begin{figure}
\setlength{\unitlength}{1mm}
\vspace{-5cm}
\begin{tabular}{lll}
\vspace{-1cm}\phantom{4242}\begin{picture}(60,55)
\epsfig{file=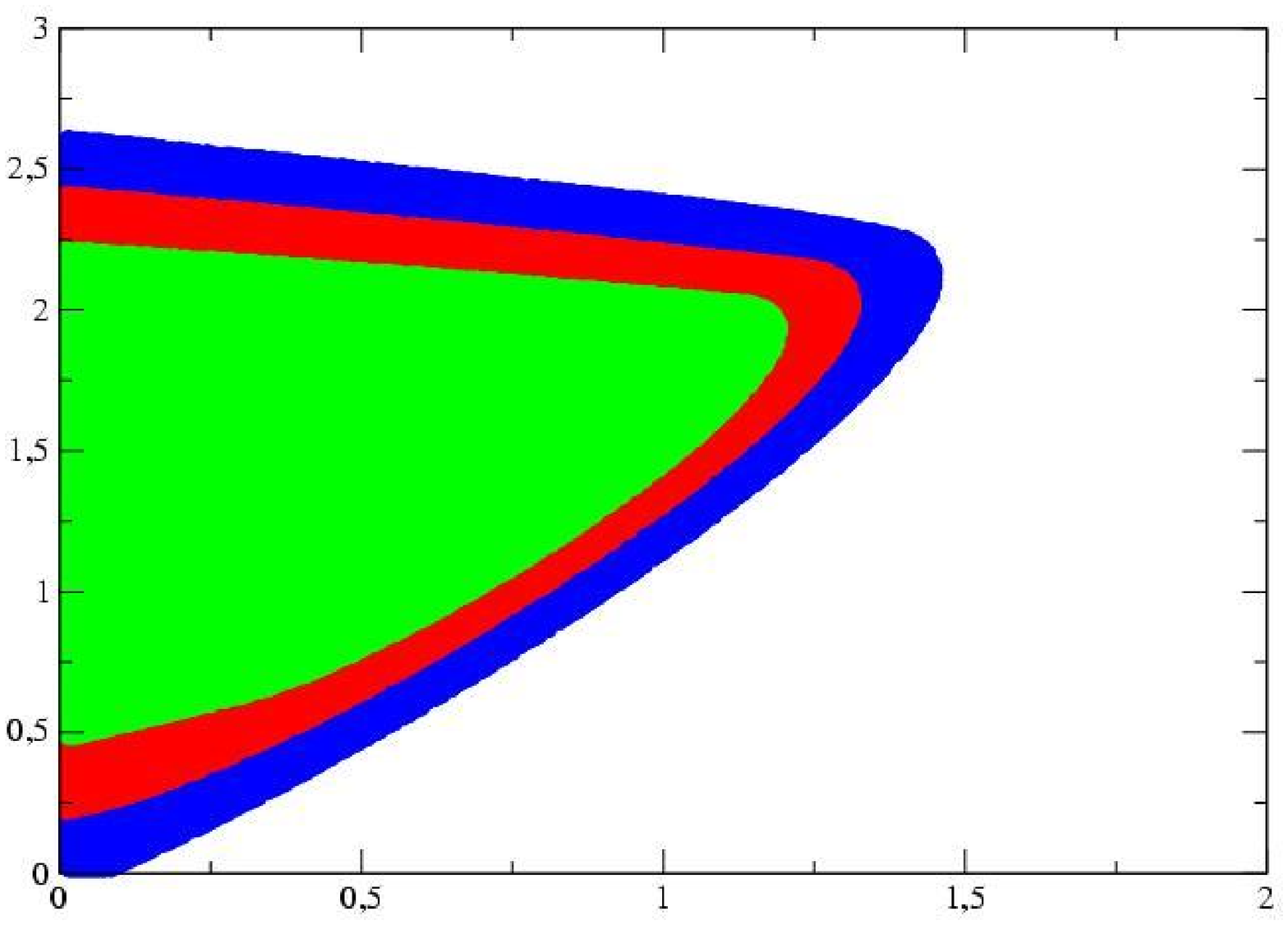,width=5cm,angle=-90}
\put(-35,-46){\tiny{$\Omega_{{\rm m} 0}$}}
\put(-65,-26){\tiny{$\Omega_{{\lambda} 0}$}}
\end{picture}&\phantom{42}&
\begin{picture}(60,55)
\epsfig{file=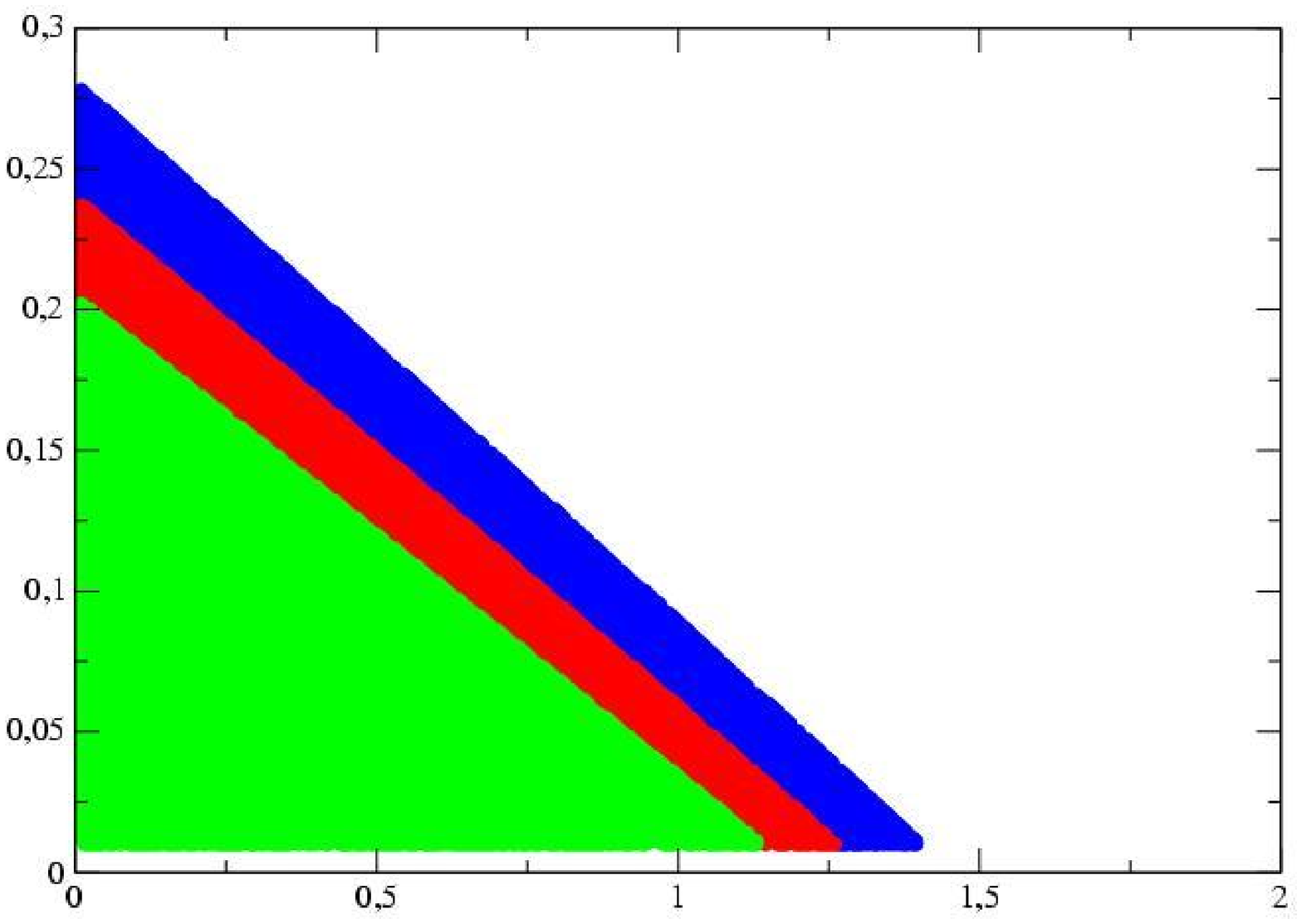,width=5cm,angle=-90}
\put(-35,-46){\tiny{$\Omega_{{\rm m} 0}$}}
\put(-65,-26){\tiny{$\Omega_{{\psi} 0}$}}
\end{picture}\\
\vspace{-1cm}\phantom{4242}\begin{picture}(60,55)
\epsfig{file=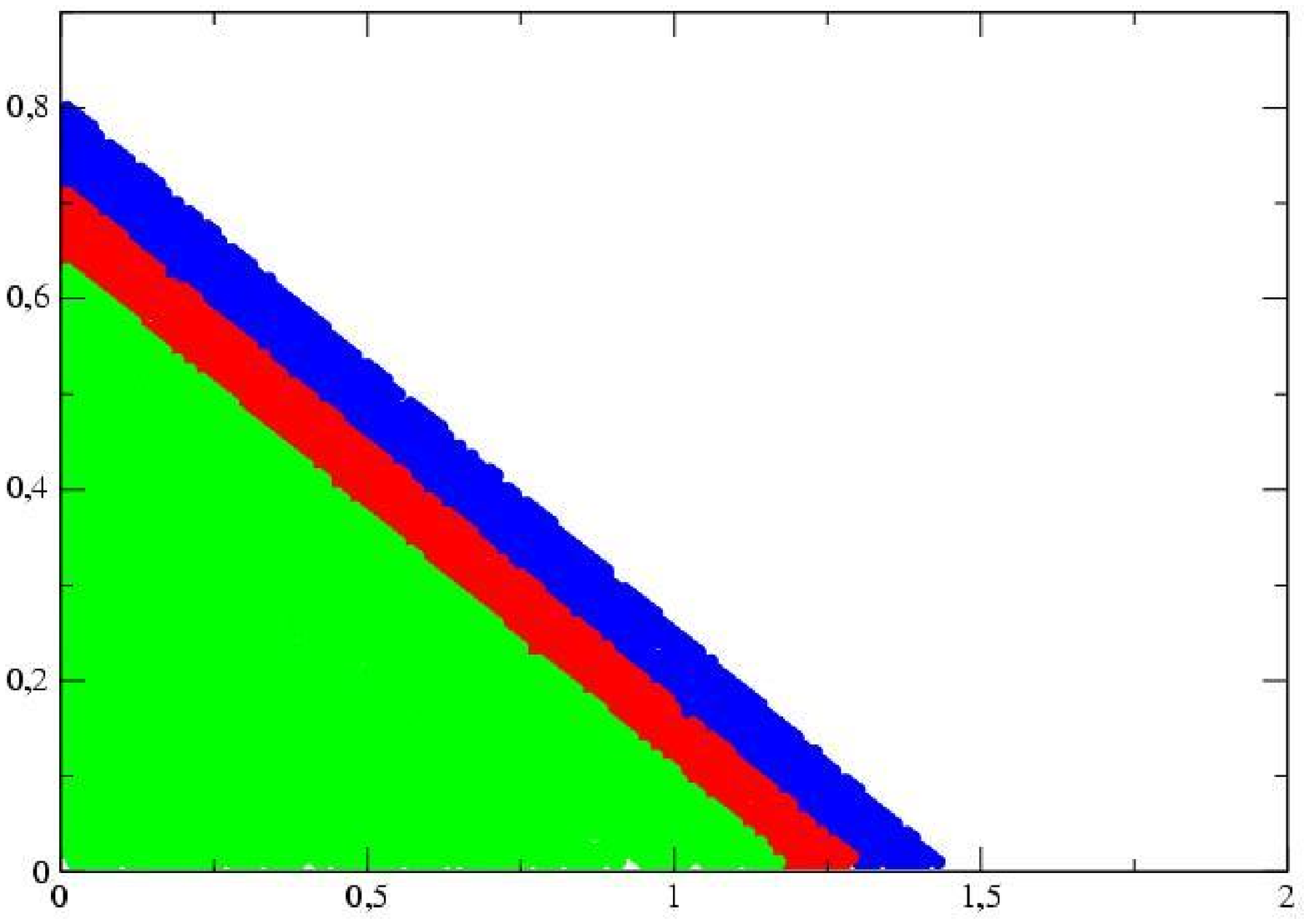,width=5cm,angle=-90}
\put(-35,-46){\tiny{$\Omega_{{\rm m} 0}$}}
\put(-65,-26){\tiny{$\Omega_{{\rm r} 0}$}}
\end{picture}&&
\begin{picture}(60,55)
\epsfig{file=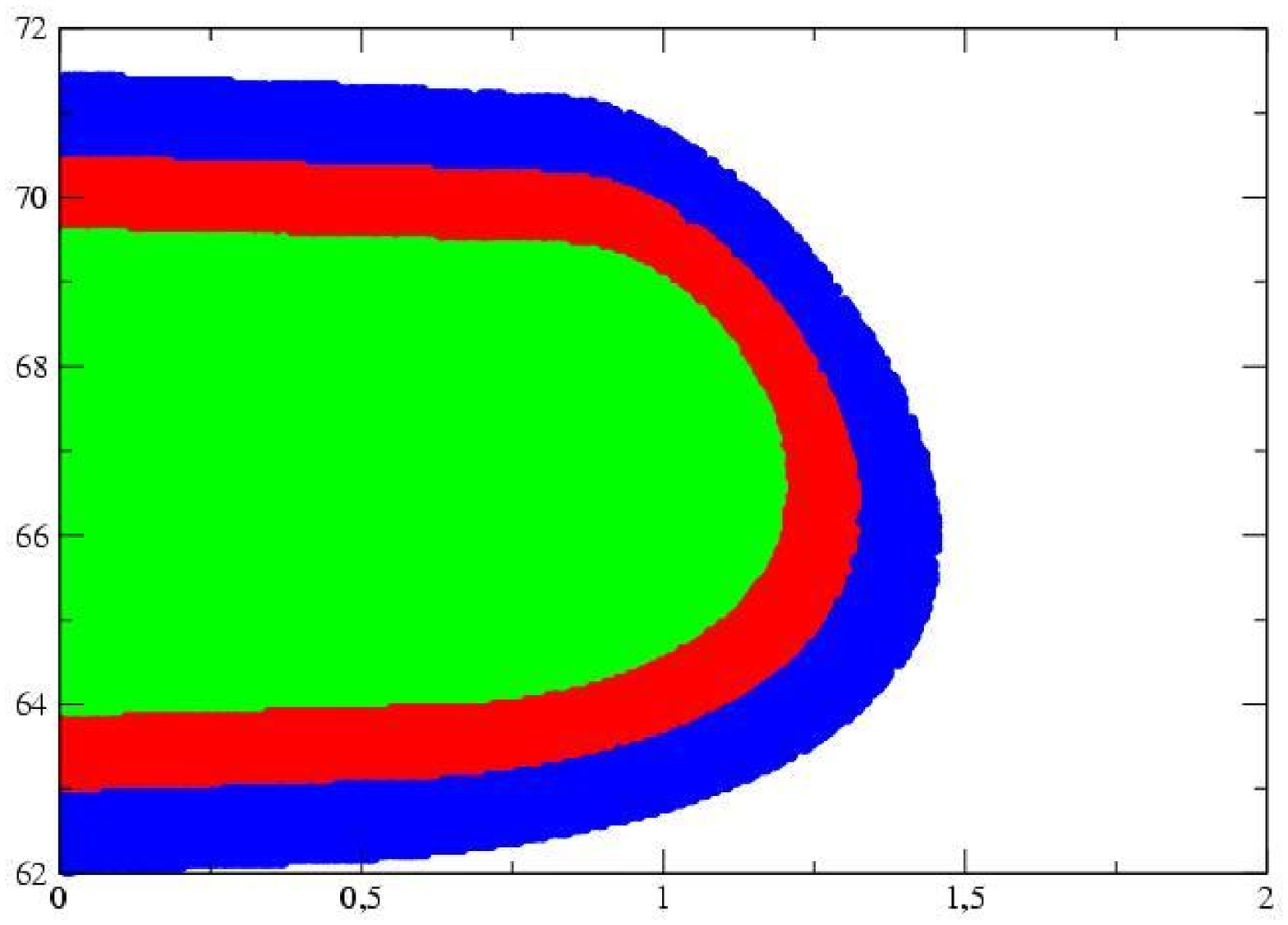,width=5cm,angle=-90}
 \put(-35,-46){\tiny{$\Omega_{{\rm m} 0}$}}
\put(-65,-26){\tiny{$H_0$}}
\end{picture}\\
\vspace{-1cm}\phantom{4242}\begin{picture}(60,55)
\epsfig{file=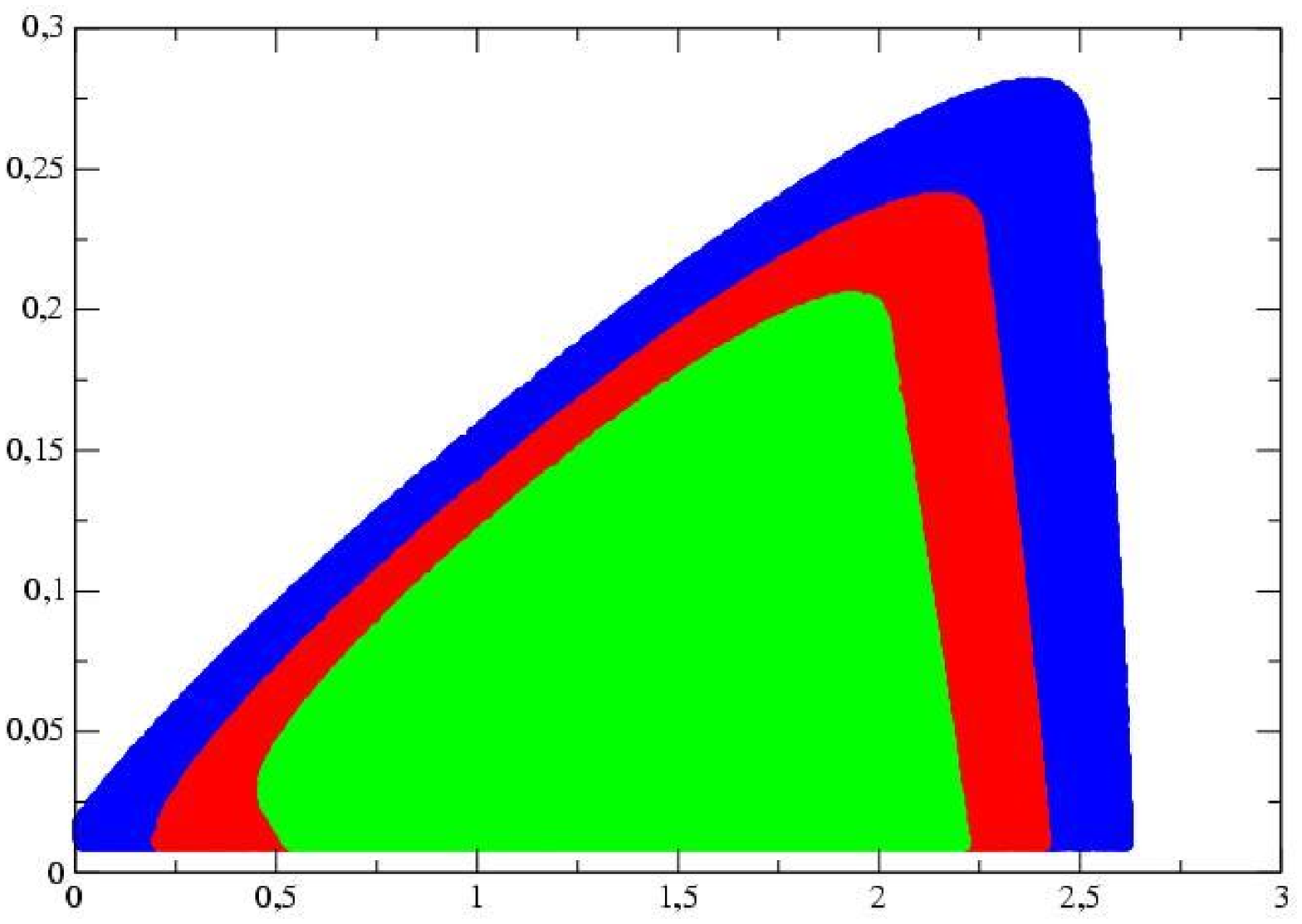,width=5cm,angle=-90}
\put(-35,-46){\tiny{$\Omega_{{\lambda} 0}$}}
\put(-65,-26){\tiny{$\Omega_{{\psi} 0}$}}
\end{picture}&&
\begin{picture}(60,55)
\epsfig{file=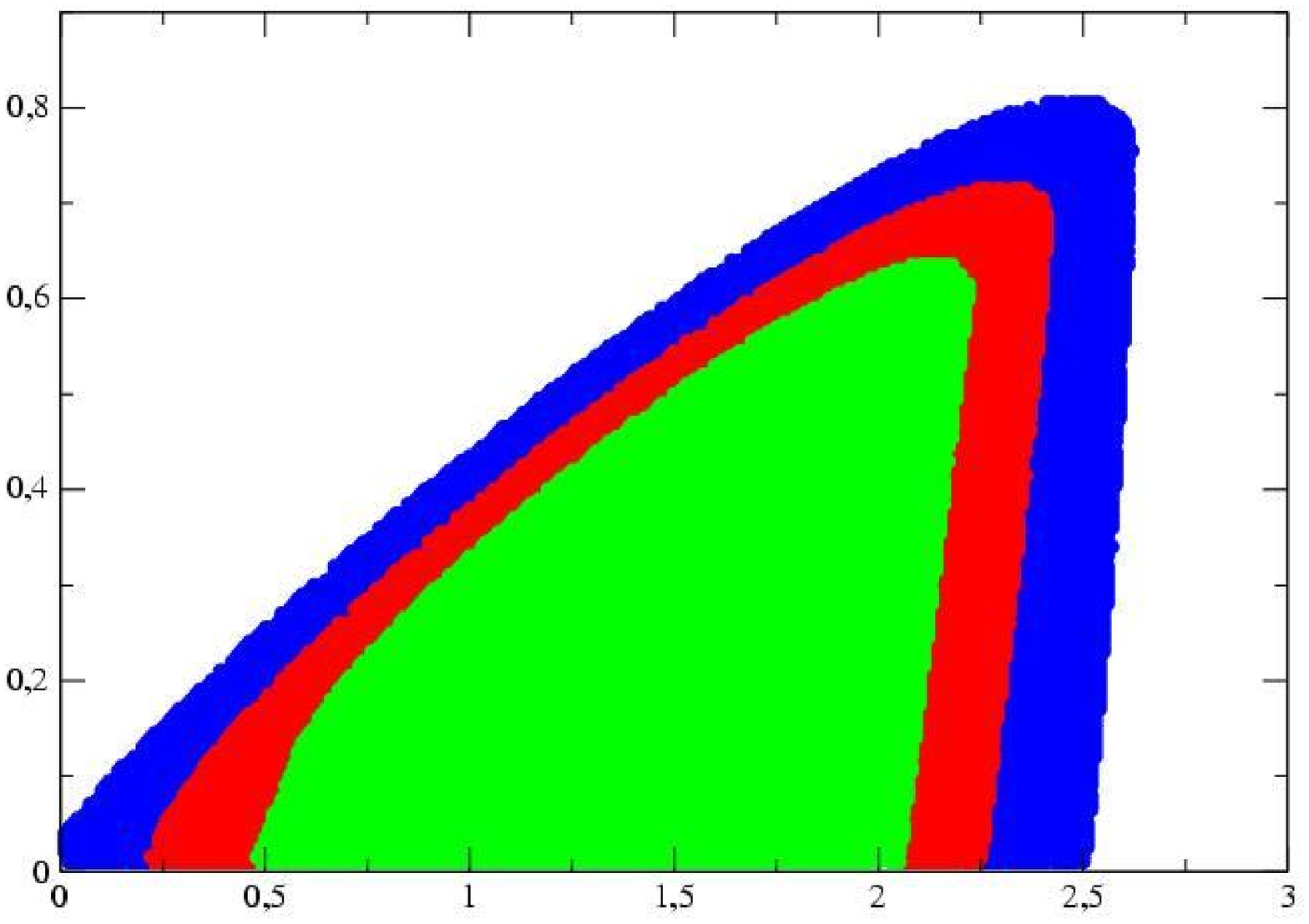,width=5cm,angle=-90}
\put(-35,-46){\tiny{$\Omega_{{\lambda} 0}$}}
\put(-65,-26){\tiny{$\Omega_{{\rm r} 0}$}}
\end{picture}\\
\vspace{-1cm}\phantom{4242}\begin{picture}(60,55)
\epsfig{file=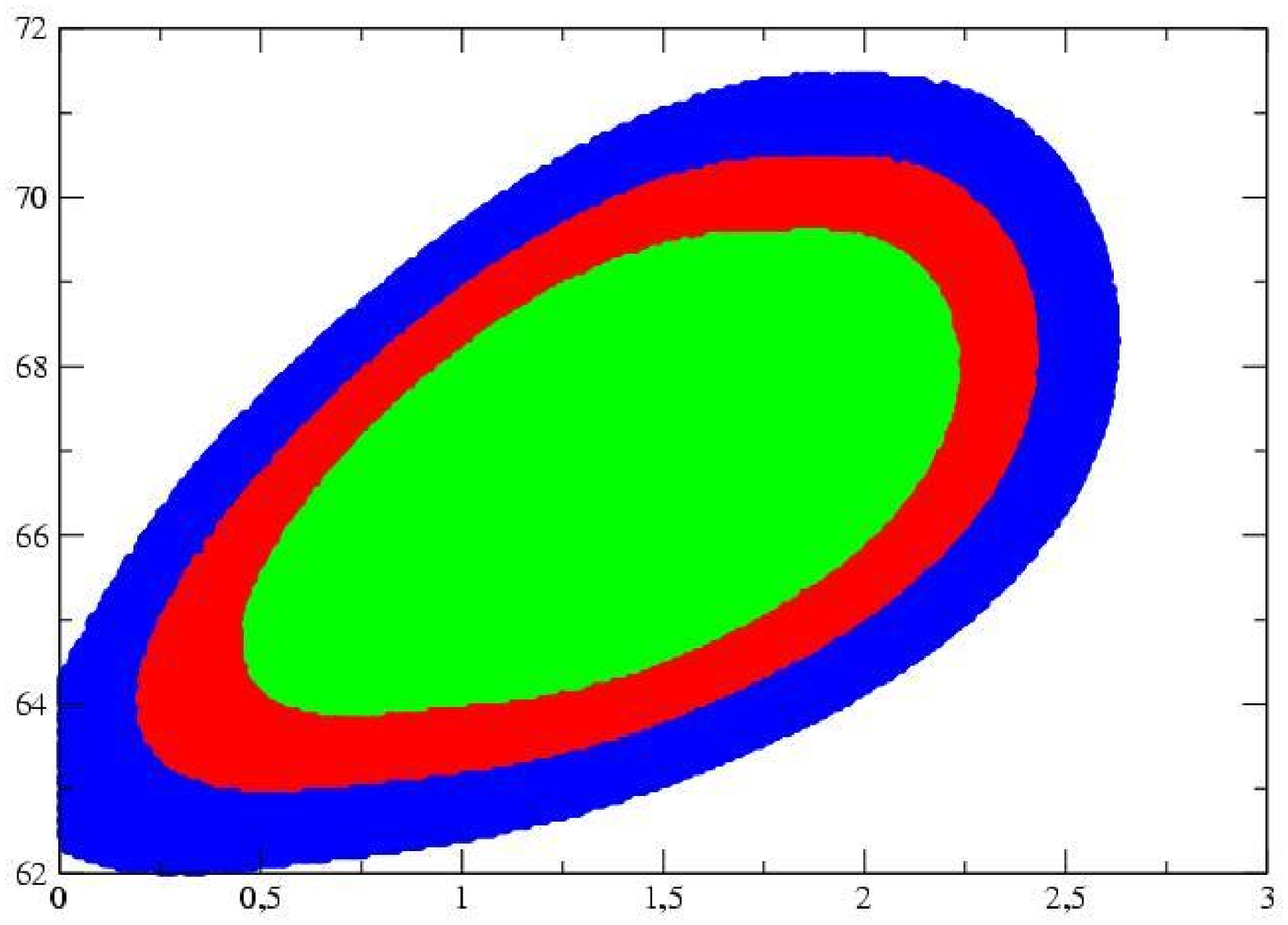,width=5cm,angle=-90}
\put(-35,-46){\tiny{$\Omega_{{\lambda} 0}$}}
\put(-65,-26){\tiny{$H_0$}}
\end{picture}&&
\begin{picture}(60,55)
\epsfig{file=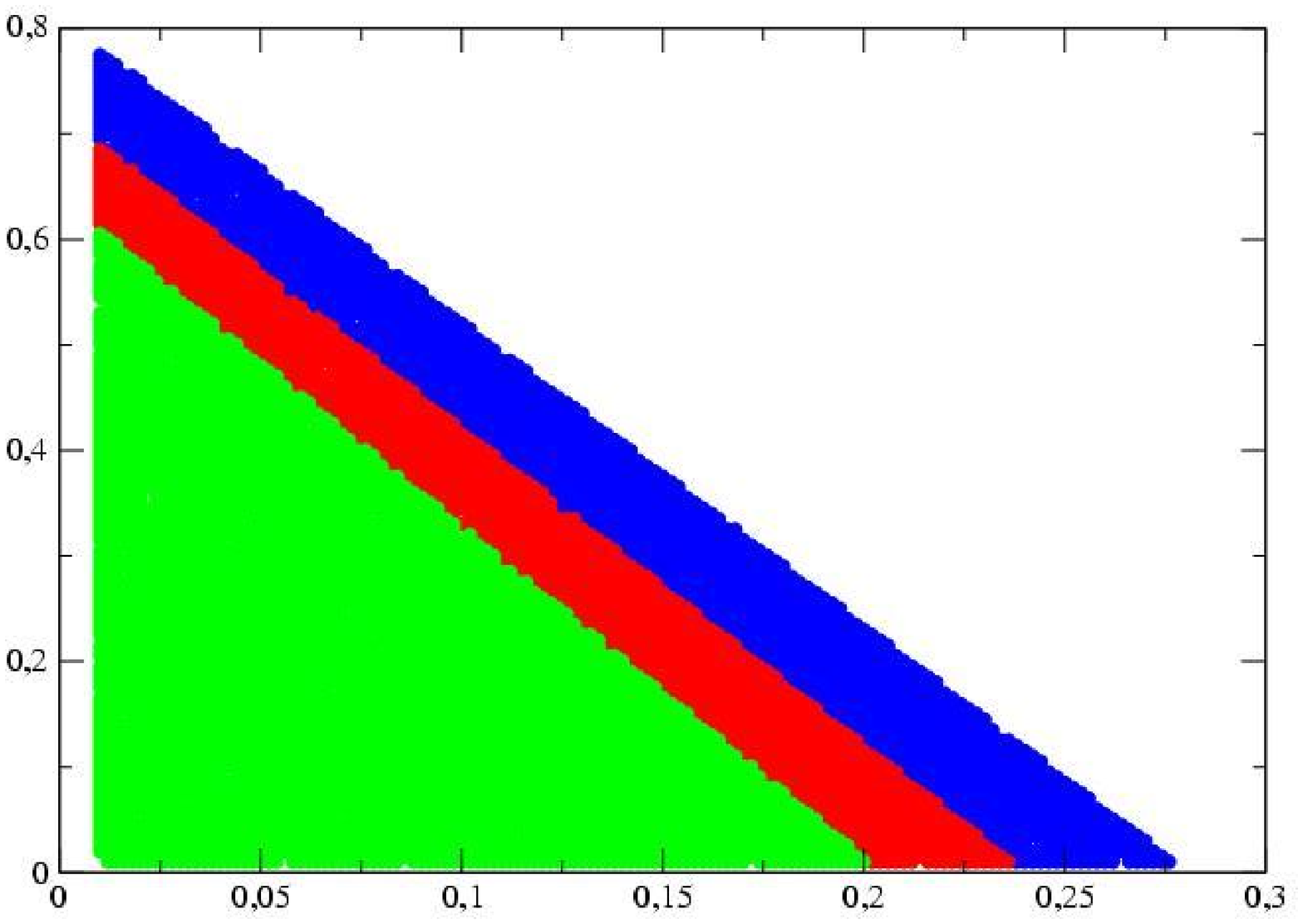,width=5cm,angle=-90}
\put(-35,-46){\tiny{$\Omega_{{\psi} 0}$}}
\put(-65,-26){\tiny{$\Omega_{{\rm r} 0}$}}
\end{picture}\\
\vspace{-1cm}\phantom{4242}\begin{picture}(60,55)
\epsfig{file=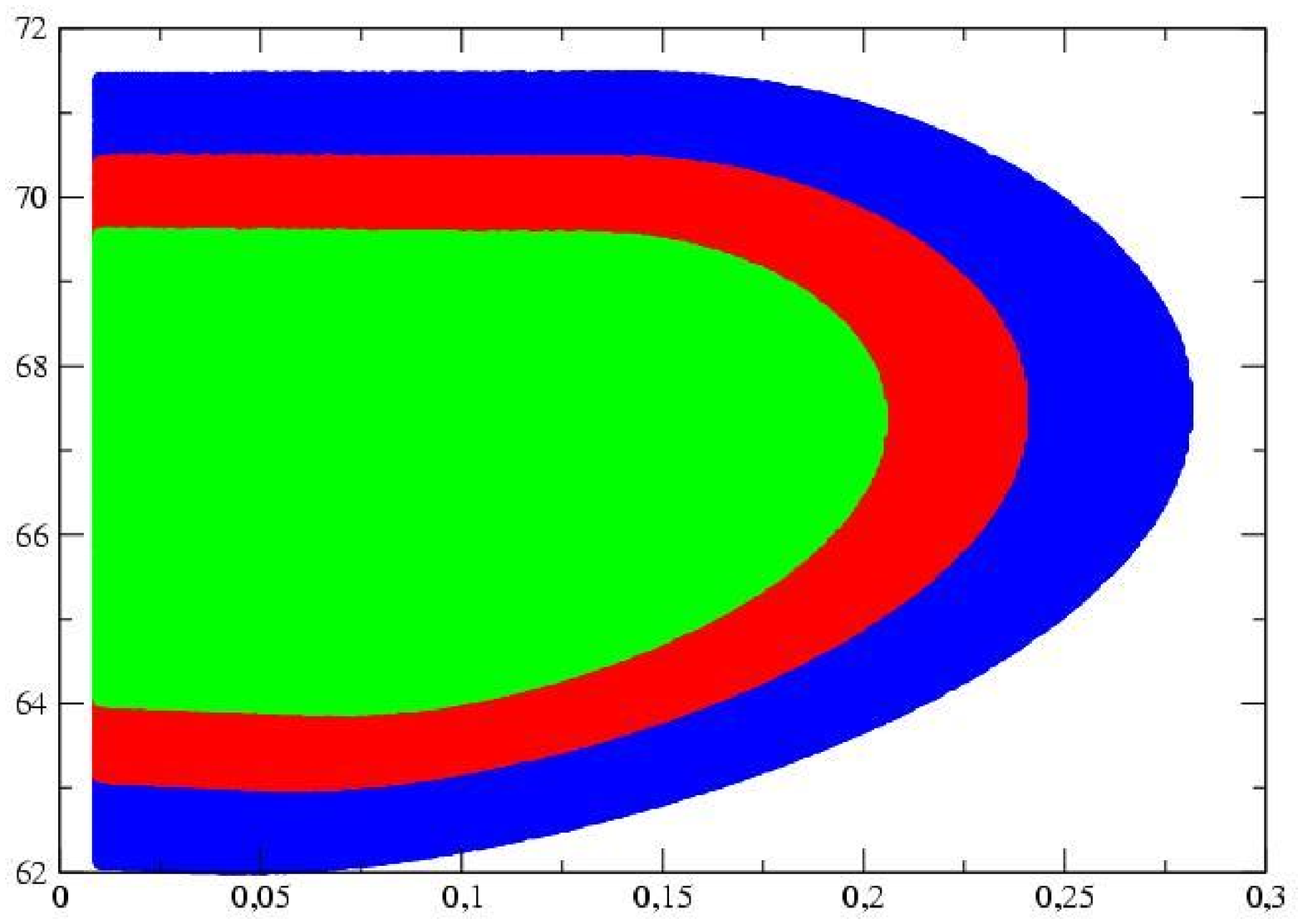,width=5cm,angle=-90}
\put(-35,-46){\tiny{$\Omega_{{\psi} 0}$}}
\put(-65,-26){\tiny{$H_0$}}
\end{picture}&&
\begin{picture}(60,55)
\epsfig{file=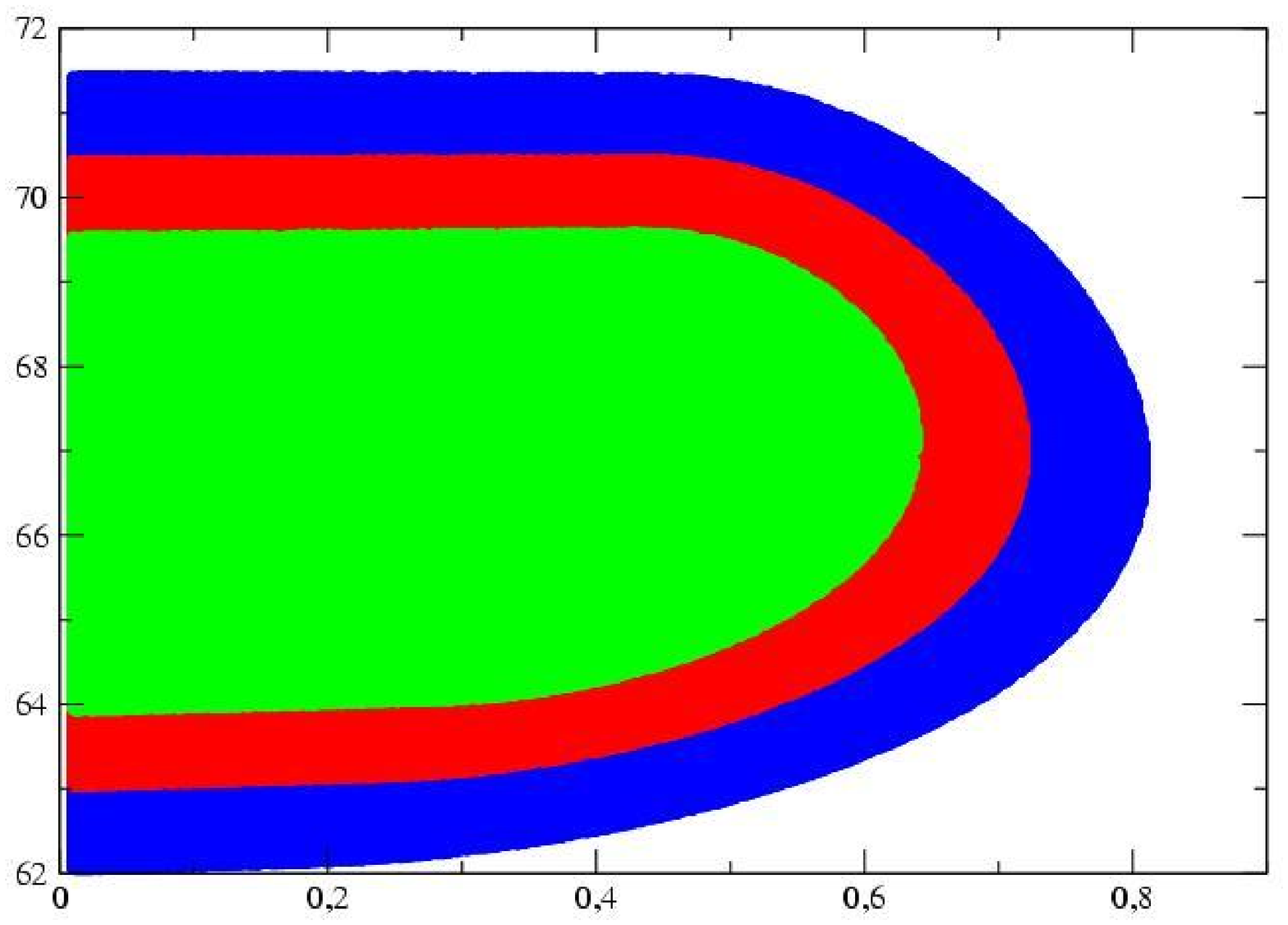,width=5cm,angle=-90}
\put(-35,-46){\tiny{$\Omega_{{\rm r} 0}$}}
\put(-65,-26){\tiny{$H_0$}}
\end{picture}\\\vspace{0.5cm}
\end{tabular}
\vspace{5cm}
\caption[]{Confidence contours for all parameter planes of the Weyl-Cartan model for the binned data set containing 114 SN Ia, which corresponds to bin-size $\Delta z \approx 0.002$.}
\label{FIG_triplet_contours_lambda230b}
\end{figure}

In figure \ref{FIG_triplet_contours_lambda230b} the confidence contours for the binned T230 data set with bin size $\Delta z \approx 0.002$ are displayed (the best-fit values for this set are also listed in table \ref{TABLE_triplet_best_binned}). The plots show that the averaging procedure leads to a slight broadening of the confidence contours. Of course the meaning of such a post-processed data set should not be overinterpreted. One has to keep in mind that the choice of the size of the redshift bin size is merely based on statistical and not on physical arguments. Nevertheless it is interesting to note that the estimates for $H_0$ exhibit a clear trend for varying bin size.

\section{Conclusions}\label{CONCLUSION_section}
\paragraph{Age}
For the best-fit parameters listed in table \ref{TABLE_triplet_best_unbinned} and table \ref{TABLE_triplet_best_binned} we obtain the following values for the age of the universe: (WC92, $11.77$ Gyrs), (WC230, $11.78$ Gyrs), (WC114, $9.48$ Gyrs), (WC25, $7.52$ Gyrs). Our best-fit parameters yield cosmological models with younger\footnote{Younger with respect to the standard FLRW model with $\Omega_{{\rm m} 0}=0.3$, $\Omega_{\lambda 0}= 0.7$, and $H_0=65 \, {\rm km}\, {\rm s}^{-1} {\rm Mpc}^{-1}$, for which one obtains an age of $14.5$ Gyrs.} age than the estimates from globular clusters and nuclear cosmochronology which range from 11 to 15 Gyrs \cite{Chaboyer,Truran}, but we should keep in mind that the uncertainties in the parameters allow a broad range of age of the Universe. When one considers the whole allowed interval as displayed in figures \ref{FIG_triplet_contours_lambda}, \ref{FIG_triplet_contours_lambda230}, and \ref{FIG_triplet_contours_lambda230b}, the age of the Universe is not a problem.

\begin{figure}
\begin{center}
\epsfig{file=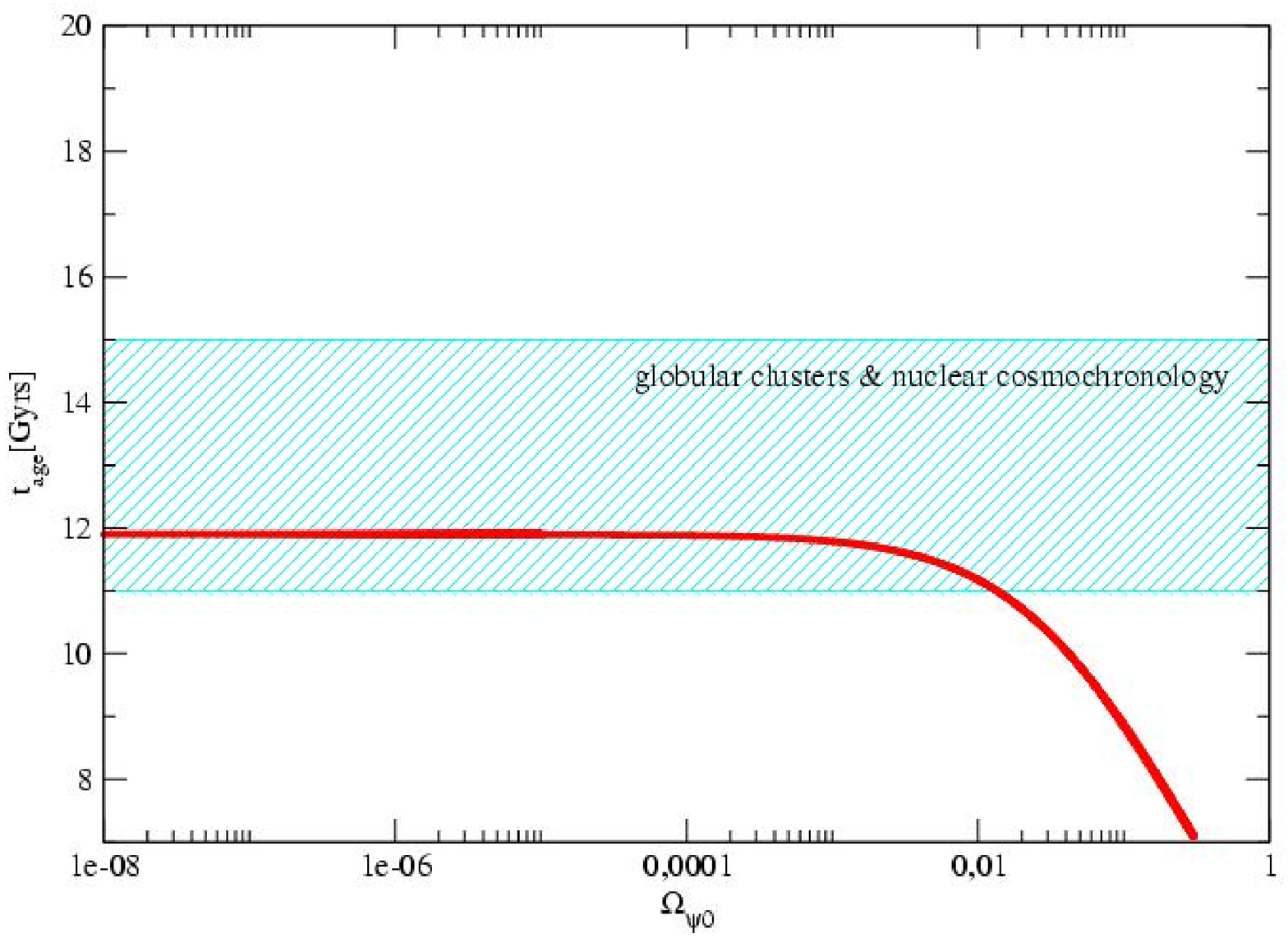,width=9.6cm,angle=-90}
\end{center}
\caption[Age function]{Age of the universe for varying $\Omega_{\psi 0}$ parameter. The other parameters are fixed to the values as listed in table \ref{TABLE_triplet_best_unbinned} for the WC230 fit. The shaded region corresponds to the age of the universe as inferred from globular clusters and nuclear cosmochronology \cite{Chaboyer,Truran}.} \label{FIG_age}
\end{figure}

In figure \ref{FIG_age} we plotted the variation the age for varying $\Omega_{\psi 0}$. Regarding to this plot there should be no conflict with the age of the oldest objects in the universe as long as $\Omega_{\psi 0}$ is smaller than $0.01$.  

\begin{figure}
\begin{center}
\epsfig{file=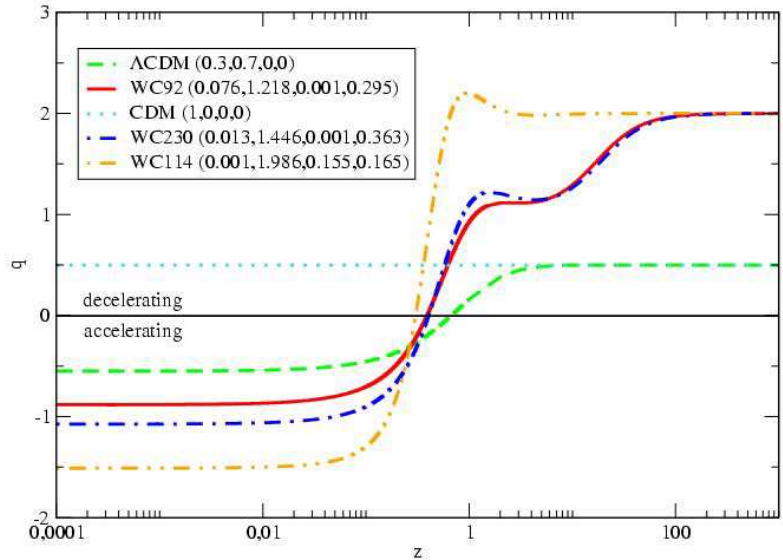, width=9.6cm,angle=-90}
\end{center}
\caption[Deceleration parameter]{Deceleration parameter versus redshift for different parameter choices. We make use of the notation $(\Omega_{{\rm m} 0},\Omega_{{\lambda} 0},\Omega_{{\psi} 0},\Omega_{{\rm r} 0})$.} 
\label{FIG_deceleration_parameter}
\end{figure}

\paragraph{Deceleration parameter}
For a model with pressure-less matter, radiation, cosmological constant, and the additional contribution from $\Omega_\psi$, the deceleration parameter is given by
\begin{equation}
\hspace{-2.5cm}q=\frac{\Omega _{{\rm m}0}\left( 1+z\right) ^{3}-2\Omega _{\lambda 0}+4\Omega _{\psi 0}\left( 1+z\right)^{6}+2 \Omega _{{\rm r}0}\left( 1+z\right) ^{4}}{2 \left[ z \Omega _{{\rm m}0}\left( 1+z\right) ^{2}-z \Omega _{\lambda 0 }\left( 2+z\right) +z\Omega _{\psi 0} \sum\limits_{i=2}^{5}\left( 1+z\right) ^{i}+z\Omega _{{\rm r} 0} \sum\limits_{i=2}^{3}\left( 1+z\right) ^{i}+ \left(
1+z\right) ^{2}\right]}.
\label{EXT_deceleration_parameter}
\end{equation}
The deceleration factor versus the redshift is plotted in figure \ref{FIG_deceleration_parameter}. Therefrom we can infer that the presence of an additional (positive) $\Omega _{\psi 0}$ contribution would lead to a later transition to the accelerating epoch than in the standard model with $\Omega _{{\rm m}0}\approx0.3$ and $\Omega _{{\lambda} 0}\approx 0.7$.

\paragraph{Future measurements}

\begin{figure}
\begin{center}
\epsfig{file=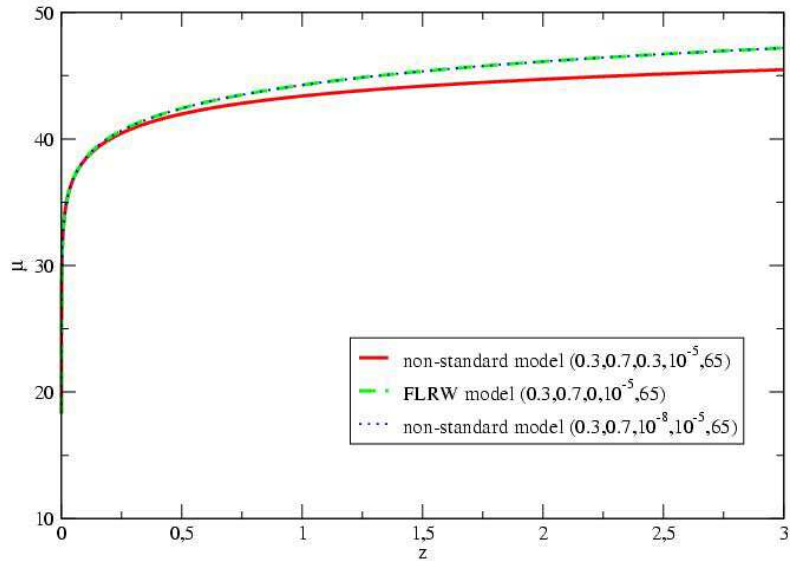,width=9.6cm,angle=-90}
\end{center}
\caption[$\mu - z $ relation]{Distance modulus versus redshift for the FLRW and our non-standard model. We make use of the notation $(\Omega_{{\rm m} 0},\Omega_{{\lambda} 0},\Omega_{{\psi} 0},\Omega_{{\rm r} 0},H_0)$. For very small values of the $\Omega_{\psi 0}$ density parameter our model is virtually indistinguishable from the FLRW at low redshifts. The red line corresponds to the upper limit on $\Omega_{\psi 0}$ of about $0.3$ as inferred from our previous fits.} \label{FIG_mu_z_both_models}
\end{figure}

\begin{figure}
\begin{center}
\epsfig{file=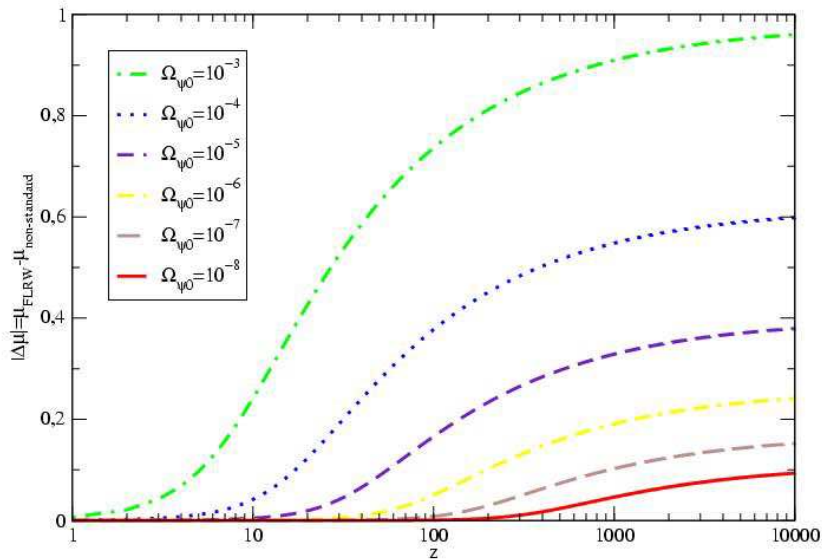,width=9.6cm,angle=-90}
\end{center}
\caption[$\mu - z $ relation]{Absolute difference between the distance modulus of the FLRW and our new model for different values of the $\Omega_{\psi 0}$ parameter. The rest of the cosmological parameters is fixed to the following values: $\Omega_{{\rm m} 0}=0.3,$ $\Omega_{{\lambda} 0}=0.7,$ $\Omega_{{\rm r} 0}=0.3,$ $H_0=65$ ${\rm km \, s}^{-1}{\rm Mpc}^{-1}$.} \label{FIG_mu_z_abs_both_models}
\end{figure}

In figure \ref{FIG_mu_z_both_models} we plotted the distance modulus versus the redshift for the FLRW model as well as for our new model. As becomes clear from this plot the new model would not be distinguishable from the FLRW model if we fix the new $\Omega_{\psi 0}$ density parameter to very small values. If one sticks to the upper limit for $\Omega_{\psi 0}$ of about $0.3$, which we obtained by fitting to the data set of Wang, it should be possible to distinguish our model from the FLRW scenario as soon as more data at redshifts between $z=1$ and $z=2$ becomes available. In order to quantify the difference in $\mu$ between the FLRW and our model, we displayed $|\Delta \mu|$ in figure \ref{FIG_mu_z_abs_both_models} for several values of the $\Omega_{\psi 0}$ density parameter. 

\paragraph{Summary}

Most interestingly there exist Weyl-Cartan based scenarios in which the usual factorization of the density factors is applicable, in contrast to the more general scenario discussed in \cite{Puetzfeld1,Puetzfeld2} the model presented in this work allows for such a simple relation between the density parameters. We worked out the magnitude-redshift relation and managed to pin-down the cosmological parameters by using recent SN Ia data. Moreover we have shown that our results are also applicable to the models of two other groups. The model discussed here turns out to be a viable cosmological model from the viewpoint of the SN Ia observations. 

The best-fit parameter sets WC92, WC230, and WC114 reveal a clear trend to low values for the pressure-less matter component $\Omega_{{\rm m} 0}$. This is a rather desirable behavior since one would prefer a model which predicts rather small values for $\Omega_{{\rm m} 0}$ and thereby lies closer to the results for the baryonic matter component as inferred from nucleosynthesis \cite{Olive2}. 

Unfortunately our model does also favor a non-vanishing cosmological constant, as it is also favored by the standard FLRW model. Hence, in order to get rid of the cosmological constant one would have to change either the Lagrangian in (\ref{Obukhov_Lagrangian}) or the ansatz for non-Riemannian field strengths, i.e.\ torsion and nonmetricity, which in this model are determined by the one form in (\ref{EXT_obukhov_triplet_1_form}). Nevertheless it should be possible to work out a modified model which removes the need for the cosmological constant and thereby the need for a rather mysterious dark energy component.    

We have to stress at this point that our goal was to perform fits to the available SN Ia data sets without any prerequisites. Future work will concentrate on the determination of the cosmological parameters \cite{Peebles1,Lineweaver} with the help of other independent cosmological tests like nucleosynthesis \cite{Olive2} and the fluctuations within the cosmic microwave background \cite{Jaffe,WMAP1}. 

In addition one could investigate the impact of the recently released SNe data set from the Deep Survey \cite{Barris} on the fit result.  

Most interestingly Daly \etal suggested a test on the basis of recent radio galaxy measurements reaching out to $z \approx 1.8$ which is currently not covered by the supernova data. As we have shown before the availability of data at higher redshifts should increase our capability of discriminating between our model and the usual FLRW scenario. Hence the analysis of this data set might lead to some interesting new insights.

\ack
The authors are grateful to Y.\ Wang (Oklahoma) for providing her SNe data set. Additionally, one of the authors (DP) wants to thank C.\ Raas (Cologne) for his patient and enduring advice on several computer related issues. XC is supported by the NSF grant PHY99-07949.\bigskip

\appendix

\section{Metric-affine gravity \label{TRIPLET_KAPITEL}}

In MAG we have the metric $g_{\alpha \beta }$, the coframe $\vartheta^{\alpha }$, and the connection 1-form $\Gamma _{\alpha}{}^{\beta}$ (with values in the Lie algebra of the four-dimensional linear group $GL(4,R)$) as new independent field variables.\ Here $\alpha ,\beta ,\ldots =0,1,2,3$ denote (anholonomic) frame indices. Spacetime is described by a metric-affine geometry with the gravitational field strengths nonmetricity $Q_{\alpha \beta }:=-Dg_{\alpha \beta }$, torsion $T^{\alpha }:=D\vartheta^{\alpha }$, and curvature $R_{\alpha }{}^{\beta }:=d\Gamma _{\alpha}{}^{\beta }-\Gamma _{\alpha }\,^{\gamma }\wedge \Gamma _{\gamma }{}^{\beta} $. A Lagrangian formalism for a matter field $\Psi $ minimally coupled to the gravitational potentials $g_{\alpha \beta }$, $\vartheta ^{\alpha }$, $\Gamma_{\alpha }{}^{\beta }$ has been set up in \cite{PhysRep}. The dynamics of an ordinary MAG theory is specified by a total Lagrangian 
\begin{equation}
L=V_{{\rm MAG}}(g_{\alpha \beta },\vartheta ^{\alpha },Q_{\alpha \beta},T^{\alpha },R_{\alpha }{}^{\beta })+L_{{\rm mat}}(g_{\alpha \beta},\vartheta ^{\alpha },\Psi ,D\Psi ).
\end{equation}
The variation of the action with respect to the independent gauge potentials leads to the field equations: 
\begin{eqnarray}
\frac{\delta L_{{\rm mat}}}{\delta \Psi } &=&0,  \label{matter} \\
DM^{\alpha \beta }-m^{\alpha \beta } &=&\sigma ^{\alpha \beta },\label{zeroth} \\
DH_{\alpha }-E_{\alpha } &=&\Sigma _{\alpha ,}  \label{first} \\
DH^{\alpha }{}_{\beta }-E^{\alpha }{}_{\beta } &=&\Delta ^{\alpha }{}_{\beta}.  \label{second}
\end{eqnarray}
Equations (\ref{zeroth}) and (\ref{first}) are the generalized Einstein equations with the symmetric energy-momentum 4-form $\sigma^{\alpha \beta}$ and the canonical energy-momentum 3-form $\Sigma _{\alpha }$ as sources. Equation (\ref{second}) is an additional field equation which takes into account other aspects of matter, such as spin, shear and dilation currents, represented by the hypermomentum $\Delta ^{\alpha }{}_{\beta }$. We made use of the definitions of the gauge field excitations, 
\begin{equation}
H_{\alpha }:=-\frac{\partial V_{{\rm MAG}}}{\partial T^{\alpha }},\quad H^{\alpha }{}_{\beta }:=-\frac{\partial V_{{\rm MAG}}}{\partial R_{\alpha}{}^{\beta }},\quad M^{\alpha \beta }:=-2\frac{\partial V_{{\rm MAG}}}{\partial Q_{\alpha \beta }},  \label{exications}
\end{equation}
of the canonical energy-momentum, the metric stress-energy, and the hypermomentum current of the gauge fields, 
\begin{equation}
E_{\alpha }:=\frac{\partial V_{{\rm MAG}}}{\partial \vartheta ^{\alpha }},\quad m^{\alpha \beta }:=2\frac{\partial V_{{\rm MAG}}}{\partial g_{\alpha \beta }},\quad E^{\alpha }{}_{\beta }=-\vartheta ^{\alpha }\wedge H_{\beta}-g_{\beta \gamma }M^{\alpha \gamma },  \label{gauge_currents}
\end{equation}
and of the canonical energy-momentum, the metric stress-energy, and the hypermomentum currents of the matter fields, 
\begin{equation}
\Sigma _{\alpha }:=\frac{\delta L_{{\rm mat}}}{\delta \vartheta ^{\alpha }},\quad \sigma ^{\alpha \beta }:=2\frac{\delta L_{{\rm mat}}}{\delta g_{\alpha \beta }},\quad \Delta ^{\alpha }{}_{\beta }:=\frac{\delta L_{{\rm mat}}}{\delta \Gamma _{\alpha }{}^{\beta }}.  \label{matter_currents}
\end{equation}
Provided the matter equations (\ref{matter}) are fulfilled, the following Noether identities hold: 
\begin{eqnarray}
D\Sigma _{\alpha } &=&\left( e_{\alpha }\rfloor T^{\beta }\right) \wedge \Sigma _{\beta }-\frac{1}{2}\left( e_{\alpha }\rfloor Q_{\beta \gamma}\right) \sigma ^{\beta \gamma }+\left( e_{\alpha }\rfloor R_{\beta}{}^{\gamma }\right) \wedge \Delta ^{\beta }{}_{\gamma },
\label{noether_ident_1} \\
D\Delta ^{\alpha }{}_{\beta } &=&g_{\beta \gamma }\sigma ^{\alpha \gamma}-\vartheta ^{\alpha }\wedge \Sigma _{\beta }.  \label{noether_ident_2}
\end{eqnarray}
They show that the field equation (\ref{zeroth}) is redundant, thus we only need to take into account (\ref{first}) and (\ref{second}).

As suggested in \cite{Exact2}, the most general parity conserving quadratic Lagrangian expressed in terms of the irreducible pieces of the nonmetricity $Q_{\alpha \beta }$, torsion $T^{\alpha }$, and curvature $R_{\alpha \beta }$ reads 
\begin{eqnarray}
V_{{\rm MAG}}=&\frac{1}{2\kappa }\biggl[ &-a_{0}R^{\alpha \beta }\wedge \eta_{\alpha \beta }-2\lambda \eta +T^{\alpha }\wedge \,^{\star }\left(\sum_{I=1}^{3}a_{I}\,^{(I)}T_{\alpha }\right)   \nonumber \\
&&+Q_{\alpha \beta }\wedge \,^{\star }\left(\sum_{I=1}^{4}b_{I}\,^{(I)}Q^{\alpha \beta }\right)\nonumber \\  
&&+b_{5}\left(^{(3)}Q_{\alpha \gamma }\wedge \vartheta ^{\alpha }\right) \wedge \,^{\star}\left( \,^{(4)}Q^{\beta \gamma }\wedge \vartheta _{\beta }\,\right) \nonumber \\
&&+2\left( \sum_{I=2}^{4}c_{I}\,^{(I)}Q_{\alpha \beta }\right) \wedge \vartheta ^{\alpha }\wedge \,^{\star }T^{\beta }\biggr]  \nonumber \\
-\frac{1}{2\rho }\, R^{\alpha \beta }\wedge \,^{\star}&&\hspace{-0.5cm}\biggl[\sum_{I=1}^{6}w_{I}\,^{(I)}W_{\alpha \beta}+\sum_{I=1}^{5}z_{I}\,^{(I)}Z_{\alpha \beta }+w_{7}\vartheta _{\alpha}\wedge \left( e_{\gamma }\rfloor \,^{(5)}W^{\gamma }{}_{\beta }\right)\nonumber \\
&&\hspace{-0.5cm}+z_{6}\vartheta _{\gamma }\wedge \left( e_{\alpha }\rfloor \,^{(2)}Z^{\gamma }{}_{\beta }\right) +\sum_{I=7}^{9}z_{I}\vartheta _{\alpha}\wedge \left( e_{\gamma }\rfloor \,^{(I-4)}Z^{\gamma }{}_{\beta }\right) \biggr].
\label{general_v_mag}
\end{eqnarray}
The constants entering (\ref{general_v_mag}) are the cosmological constant $\lambda $, the weak and strong coupling constant $\kappa $ and $\rho $, and the 28 dimensionless parameters 
\begin{equation}
a_{0},\dots ,a_{3},b_{1},\dots ,b_{5},c_{2},\dots ,c_{4},w_{1},\dots,w_{7},z_{1},\dots ,z_{9}.  \label{general_coupling}
\end{equation}
This Lagrangian and the presently known exact solutions in MAG have been reviewed in \cite{Exact2}. In table \ref{tabelle_defs} we collected some of symbols defined within this section.

\subsection{Triplet ansatz \label{triplet_ansatz_section}}

One way to obtain field equations of manageable size is to constrain the general Lagrangian in equation (\ref{general_v_mag}). A special case that has received much attention over the last years is the so-called triplet ansatz, which was firstly investigated by Obukhov \etal in \cite{Obukhov}. Within this ansatz one usually chooses:
\begin{equation}
w_{1},\dots ,w_{7}=0, \quad z_{1},\dots ,z_{3},z_{5},\dots,z_{9}=0,\quad z_{4}\neq 0.  \label{triplet_coupling}
\end{equation}
Thus, one considers a general weak part but only a very constrained strong part of (\ref{general_v_mag}). Additionally one makes the following ansatz for the nonmetricity and torsion in terms of a single 1-form $\omega$.
\index{triplet!ansatz} 
\begin{eqnarray}
Q &=&k_{0}\omega ,\quad \Lambda =k_{1}\omega ,\quad T=k_{2}\omega,\label{triplet_allg} \\
T^{\alpha } &=&\,^{(2)}T^{\alpha }=\frac{1}{3}\vartheta ^{\alpha }\wedge T,  \label{torsion_triplet} \\
Q_{\alpha \beta } &=&\,^{(3)}Q_{\alpha \beta }+\,^{(4)}Q_{\alpha \beta }=\frac{4}{9}\left( \vartheta _{(\alpha }e_{\beta )}\rfloor \Lambda -\frac{1}{4}g_{\alpha \beta }\Lambda \right) +g_{\alpha \beta }Q.
\label{nonmet_triplet}
\end{eqnarray}

\begin{table}
\caption{Summary of definitions made in \ref{TRIPLET_KAPITEL}.}
\label{tabelle_defs}
\begin{indented}
\item[]\begin{tabular}{@{}lllll}
\br
Potentials&Field strengths&Excitations&Gauge currents&\\ 
\mr
{$g_{\alpha\beta}$}&{$Q_{\alpha \beta }:=-Dg_{\alpha \beta}$}& $M^{\alpha \beta}:=-2\frac{\partial V}{\partial Q_{\alpha \beta}}$
&$m^{\alpha \beta}:=2\frac{\partial V}{\partial g_{\alpha \beta}}$  & \\
{$\vartheta^\alpha$}&{$T^{\alpha }\hspace{0.2cm}:=D\vartheta^\alpha$}&
$H_\alpha\hspace{0.22cm}:=-\frac{\partial V}{\partial T^{\alpha}}$ &
$E_\alpha\hspace{0.21cm}:=\frac{\partial V}{\partial \vartheta^{\alpha}}$ & \\
{$\Gamma_{\alpha}{}^{\beta}$}&{$R_{\alpha}{}^{\beta}:=$''$D$''$\Gamma_{\alpha}{}^{\beta}$}&
$H^{\alpha}{}_{\beta}:=-\frac{\partial V}{\partial R_{\alpha}{}^{\beta}}$ &
$E^{\alpha}{}_{\beta}:=\frac{\partial V}{\partial \Gamma_{\alpha}{}^{\beta}}$ &\\
\br
\end{tabular}
\end{indented}
\end{table}

\section{Differential geometric formalism\label{OPERATIONS_SECTION}}

We assume a connected $n$-dimensional differential manifold $Y_n$ as
underlying structure throughout the paper. A vector basis of its tangent space
$T_p Y_n$ is denoted by $e_\alpha$, which is dual (i.e. $e_\alpha \rfloor
\vartheta^\beta=\delta_\alpha^\beta$) to the basis
$\vartheta^{\alpha}$ of the cotangent space $T_p^{\ast} Y_n$. A $p$-form
$\Xi$ can be expanded with respect to this basis as follows
\begin{equation}
\Xi=\frac{1}{p!}\, \Xi_{\beta_1 \dots \beta_p} \, \vartheta^{\beta_1} \wedge
\dots \wedge \vartheta^{\beta_p}.
\end{equation}
Table \ref{tabelle_8} provides a rough overview of the operators used
throughout the paper. For a more comprehensive treatment the reader should
consult \cite{Nakahara} or section 3, and Appendix A of \cite{PhysRep}.   
\begin{table}
\caption{Operators.}
\label{tabelle_8}
\begin{indented}
\item[]\begin{tabular}{@{}lllll}
\br
Operation&Symbol&Input&&Output\\ 
\mr
Exterior multiplication&$\wedge$&$p$-form $\wedge$ $q$-form&$\rightarrow$&$ (p+q)$-form\\ 
Interior multiplication&$\rfloor$&vector $\rfloor$ $p$-form&$\rightarrow$&$ (p-1)$-form\\
Exterior derivative&$d$&$d$ $p$-form&$\rightarrow$&$ (p+1)$-form\\
Hodge star in a $n$-dimen. space&$^{\star}$&$^{\star} p$-form&$\rightarrow$&$ (n-p)$-form\\

\br
\end{tabular}
\end{indented}
\end{table}

\section{Units\label{NATURAL_UNITS}}

In this work we made use of \textit{natural units}, i.e. $\hbar=c=k_B=1$ (cf table \ref{tabelle_8}).
\begin{table}
\caption{Natural units.}
\label{tabelle_8}
\begin{indented}
\item[]\begin{tabular}{@{}ccccc}
\br
[energy] & [mass] & [time] & [length] & [temperature]\\ 
\mr
length$^{-1}$ & length$^{-1}$ & length & length & length$^{-1}$\\ 
\br
\end{tabular}
\end{indented}
\end{table}
Additionally, we have to be careful with the coupling constants and the coordinates within the coframe. In order to keep things as clear as possible, we provide a list of the quantities emerging throughout all sections in table \ref{tabelle_9}.
\begin{table}
\caption{Dimensions of quantities.}
\label{tabelle_9}
\begin{indented}
\item[]\begin{tabular}{@{}lll}
\br
Quantities & \textit{1} & \textit{Length} \\ \mr
Gauge potentials & $[g_{\alpha \beta }],[\Gamma_{\alpha \beta}]$& $[\vartheta ^{\alpha }]$ \\ 
Field strengths & $[Q_{\alpha \beta }],[R_{\alpha \beta}]$ & $[T^\alpha]$\\ 
Gauge field excitations & $[M^{\alpha \beta }],[H^{\alpha }{}_{\beta }]$ & $[H_{\alpha }]^{-1}$ \\ 
Gauge field currents & $[E^{\alpha }{}_{\beta }],[m^{\alpha \beta }]$ & $[E_{\alpha }]^{-1}$ \\ 
Matter currents & $[\Delta_{\alpha \beta}],[\sigma^{\alpha \beta}]$ & $[\Sigma _{\alpha }]^{-1}$ \\ \mr
Coordinates & $[\theta ],[\phi ],[r]$ & $[t]$ \\\mr 
Functions & $[\Omega_w],[\Omega_k],[\Omega_{\rm m}],[\Omega_\psi]$ & $[S],[\mu]^{-\frac{1}{4}},[p]^{-\frac{1}{4}},[H]^{-1}$ \\ \mr
Miscellany & $[z],[q],[w],$ & $[\Sigma _{\alpha \beta }]^{-\frac{1}{4}},[d_{\tiny \rm luminosity}]$
\\
&$[m],[M],[{\cal M}],[\mu]$& \\\mr 
Constants & $[\chi ],[a_{I}],[b_{I}],[c_{I}],[k],[\psi]$ & $[\kappa ]^{\frac{1}{2}},[\lambda]^{-\frac{1}{2}},[G]^{\frac{1}{2}},[\upsilon]^{\frac{1}{4}}$ \\ 
\br 
\end{tabular}
\end{indented}
\end{table}
Note that $[d]=1$ and $[\,^{\star }]=$ length$^{n-2p}$, where $n=$ dimension of the spacetime, $p=$ degree of the differential form on which $^{\star}$ acts.


\bigskip


\begin{thebibliography}{9}

\bibitem{KolbTurner}  E.W. Kolb, Michael S. Turner: \textit{The early universe.} Addison-Wesley (1990)

\bibitem{PadmanabhanAstro3} T. Padmanabhan: \textit{Theoretical astrophysics Volume III: Galaxies and Cosmology.} Cambridge University Press, 1st edition (2002)

\bibitem{Recipes} W.H. Press, S.A. Teukolsky, W.T. Vetterling, B.P. Flannery: \textit{Numerical recipes in C - The Art of scientific computing.} Cambridge,
  2nd edition (1992)

\bibitem{Bevington} P.R. Bevington: \textit{Data reduction and error analysis for the physical sciences.} McGraw-Hill (1969)

\bibitem{Martin} B.R. Martin: \textit{Statistics for physicists.} Academic Press (1971)

\bibitem{Nakahara}  M. Nakahara: \textit{Geometry, Topology and Physics.} Adam Hilger, Bristol (1990)

\bibitem{Puetzfeld1}  D. Puetzfeld: \textit{A cosmological model in Weyl-Cartan spacetime: I. Field equations and solutions.} Class. Quantum Grav. \textbf{19} (2002) 3363-3280 Los Alamos e-Print Archive \texttt{gr-qc/0111014}  

\bibitem{Puetzfeld2}  D. Puetzfeld: \textit{A cosmological model in Weyl-Cartan spacetime: II. Magnitude-redshift relation.} Class. Quantum Grav. \textbf{19} (2002) 4463-4482 Los Alamos e-Print Archive \texttt{gr-qc/0205052}  

\bibitem{PhysRep}  F.W. Hehl, J.D. McCrea, E.W. Mielke, Y. Ne\'{}eman: \textit{Metric-affine gauge theory of gravity: Field equations, Noether identities, world spinors, and breaking of dilation invariance.} Phys. Rep. \textbf{258} (1995) 1-171

\bibitem{Exact2}  F.W. Hehl, A. Mac\'{\i}as: \textit{Metric-affine gauge theory of gravity: II. Exact solutions.} Int. J. Mod. Phys. D, Vol.\textbf{8}, No. 4 (1999) 399-416

\bibitem{Obukhov}  Y.N. Obukhov, E.J. Vlachynsky, W. Esser, F.W. Hehl: \textit{Effective Einstein theory from metric-affine gravity models via irreducible decompositions.} Phys. Rev. \textbf{D56} 12 (1997) 7769-7778

\bibitem{Obukhov2}  Y.N. Obukhov, R. Tresguerres: \textit{Hyperfluid - a
model of classical matter with hypermomentum.} Phys. Lett. \textbf{A184} 17-22 (1993)

\bibitem{Babourova} O.V. Babourova, B.N. Frolov: \textit{Matter with dilaton charge in Weyl-Cartan spacetime and evolution of the universe.} Class. Quantum Grav. \textbf{20} (2003) 1423-1442 Los Alamos e-print Archive \texttt{gr-qc/0209077}
\bibitem{Hamuy} M. Hamuy et al.: {\it The absolute luminosities of the Cal\'an/Tololo type Ia supernovae.} Los Alamos e-print Archive \texttt{astro-ph/9609059}

\bibitem{Perlmutter} S. Perlmutter et al.: \textit{Measurements of $\Omega$ and $\Lambda$ from 42 high-redshift supernovae.} Astrophys.\ J.\ \textbf{517} (1999) 565

\bibitem{Perlmutter2} S. Perlmutter et al.: \textit{Measurements of the cosmological parameters $\Omega$ and $\Lambda$ from the first seven supernovae at $z \ge 0.35$.} Astrophys. J. \textbf{483} (1997) 565-581

\bibitem{Schmidt} B.P. Schmidt et al.: \textit{The high-z supernova search: Measuring cosmic deceleration and global curvature of the universe using type Ia supernovae.} Astrophys. J. \textbf{507} (1998) 46-63

\bibitem{Garnavich1} P.M. Garnavich et al.: \textit{Constraints on cosmological models from Hubble space telescope observations of high-z supernovae.} Astrophys. J. \textbf{493} L53-57 (1998)

\bibitem{Riess1} A.G. Riess et al.: \textit{The farthest known supernova: Support for an accelerating universe and a glimpse of the epoch of deceleration.} Astrophys. J. \textbf{560} (2001) 49-71 Los Alamos e-print Archive \texttt{astro-ph/0104455}

\bibitem{Riess2} A.G. Riess et al.: \textit{Observational evidence from supernovae for an accelerating universe and a cosmological constant.} Astrophys. J. \textbf{116} (1998) 1009-1038

\bibitem{Tonry} J.L. Tonry et al.: \textit{Cosmological results from high-z supernovae.}  Astrophys. J. \textbf{594} (2003) 1-24 Los Alamos e-print Archive \texttt{astro-ph/0305008}

\bibitem{Barris} B.J. Barris et al.: \textit{23 high redshift supernovae from the IfA Deep Survey: Doubling the SN sample at z$\ge$0.7.} Los Alamos e-print Archive \texttt{astro-ph/0310843}

\bibitem{Peebles1} P.J.E. Peebles: \textit{The cosmological tests.} Int. J. Mod. Phy. A \textbf{16} 4223-4233 (2001)

\bibitem{Lineweaver} C.H. Lineweaver: \textit{Cosmological parameters} Los Alamos e-print Archive \texttt{astro-ph/0112381}

\bibitem{Olive2} K.A. Olive, G. Steigman, T.P. Walker: \textit{Primordial nucleosynthesis: Theory and observations.} Phys. Rept. \textbf{333} 389-407 (2000) Los Alamos e-print Archive \texttt{astro-ph/9905320}

\bibitem{Jaffe} A.H. Jaffe et al.: \textit{Cosmology from MAXIMA-1, BOOMERANG \& COBE/DMR CMB observations.} Phys. Rev. Lett. \textbf{86} (2001) 3475-3479 Los Alamos e-print Archive \texttt{astro-ph/0007333}

\bibitem{WMAP1} D.N. Spergel et al.: \textit{First year Wilkinson microwave anisotropy probe (WMAP) observations: Determination of cosmological parameters.}  Astrophys. J. Suppl. \textbf{148} (2003) 175 Los Alamos e-print Archive \texttt{astro-ph/0302209}

\bibitem{Wang} Y. Wang: \textit{Flux-averaging analysis of type Ia supernova data.} Astrophys. J. \textbf{536} 531-539 (2000)

\bibitem{Wang2000} Y. Wang: \textit{Supernova pencil beam survey.} Astrophys. J. \textbf{531} 676-683 (2000)

\bibitem{Dalal} N. Dalal, D.E. Holz, X. Chen, J.A. Frieman: \textit{Corrective lenses for high redshift supernovae.} Astrophys. J. \textbf{585} L11-L14 (2003) Los Alamos e-print Archive \texttt{astro-ph/0206339}

\bibitem{Chaboyer} B. Chaboyer et al.: \textit{The age of globular clusters in the light of HIPPARCOS: Resolving the age problem?} Astrophys. J. \textbf{494} (1998) 96-110

\bibitem{Truran} J.W. Truran et al.: \textit{Nucleosynthesis clocks and the age of the galaxy.} ASP Conference Series, Vol. TBD, 2001, Eds. T. von Hippel, N. Manset, C. Simpson Los Alamos e-print Archive \texttt{astro-ph/0109526}

\bibitem{KamionTurn} M. Kamionkowski, M.S. Turner: \textit{Thermal relics: Do we know their abundances?} Phys. Rev. \textbf{D42} 3310-3320 (1990)

\bibitem{Khalatnikov} I.M. Khalatnikov, A.Y. Kamenshchik: \textit{A generalisation of the Heckmann - Schuecking cosmological solution.} Phys. Lett. \textbf{B553} 119-125 (2003) Los Alamos e-print Archive \texttt{gr-qc/0301022} 


\end{thebibliography}
\end{document}